\documentclass[a4paper,fleqn,usenatbib]{mnras}
\usepackage{mathptmx}

\usepackage[T1]{fontenc}
\usepackage{ae,aecompl}

\usepackage{graphicx}	
\usepackage{amsmath}	
\usepackage{amssymb}	
\usepackage{longtable}
\usepackage{color}
\usepackage{times}
\usepackage[dvipsnames]{xcolor}

\newcommand{\us}{$\mu$s}
\newcommand{\dg}{$^{\circ}$ }
\newcommand{\vt}{$\overline{V}_T$ }

\title[EPTA timing of 42 MSPs]{High-precision timing of 42 millisecond pulsars with the European Pulsar Timing Array}

\author[G. Desvignes et al.]{\parbox{\textwidth}{G.~Desvignes,$^{1}$\thanks{E-mail:gdesvignes@mpifr-bonn.mpg.de}
 R.~N.~Caballero,$^1$
 L.~Lentati,$^2$
 J.~P.~W.~Verbiest,$^{3,1}$
 D.~J.~Champion,$^1$
 B.~W.~Stappers,$^4$
 G.~H.~Janssen,$^{5,4}$
 P.~Lazarus,$^1$
 S.~Os{\l}owski,$^{3,1}$
 S.~Babak,$^{6}$
 C.~G.~Bassa,$^{5,4}$
 P.~Brem,$^{6}$
 M.~Burgay,$^{7}$
 I.~Cognard,$^{8,9}$
 J.~R.~Gair,$^{10}$
 E.~Graikou,$^{1}$
 L.~Guillemot,$^{8,9}$
 J.~W.~T.~Hessels,$^{5,11}$
 A.~Jessner,$^1$
 C.~Jordan,$^4$
 R.~Karuppusamy,$^{1}$
 M.~Kramer,$^{1,4}$
 A.~Lassus,$^{1,}$
 K.~Lazaridis,$^{1}$
 K.~J.~Lee,$^{1,12}$
 K.~Liu,$^{1}$
 A.~G.~Lyne,$^{4}$
 J.~McKee,$^4$
 C.~M.~F.~Mingarelli,$^{13,1,14}$
 D.~Perrodin,$^{7}$
 A.~Petiteau,$^{15}$
 A.~Possenti,$^{7}$
 M.~B.~Purver,$^{4}$
 P.~A.~Rosado,$^{16,17}$
 S.~Sanidas,$^{11,4}$
 A.~Sesana,$^{14,6}$
 G.~Shaifullah,$^{3,1}$
 R.~Smits,$^{5}$
 S.~R.~Taylor,$^{18,10}$
 G.~Theureau,$^{8,9,19}$
 C.~Tiburzi,$^{1,3}$ 
 R.~van~Haasteren$^{13}$
 and A~.Vecchio$^{14}$ }
\vspace{0.4cm} \\ 
\parbox{\textwidth}{
$^{1}$ Max-Planck-Institut f\"ur Radioastronomie, Auf dem H\"ugel, 69 D-53121 Bonn, Germany\\
$^{2}$ Institute of Astronomy / Battcock Centre for Astrophysics, University of Cambridge, Madingley Road, Cambridge CB3 0HA, United Kingdom\\
$^{3}$ Fakult\"at f\"ur Physik, Universit\"at Bielefeld, Postfach 100131, 33501 Bielefeld, Germany\\
$^{4}$ Jodrell Bank Centre for Astrophysics, School of Physics and Astronomy, The University of Manchester, Manchester M13 9PL, UK\\
$^{5}$ ASTRON, the Netherlands Institute for Radio Astronomy, Postbus 2, 7990 AA, Dwingeloo, The Netherlands\\
$^{6}$ Max-Planck-Institut f{\"u}r Gravitationsphysik, Albert-Einstein-Institut, Am M\"uhlenberg 1, 14476, Golm, Germany \\
$^{7}$ INAF - ORA - Osservatorio Astronomico di Cagliari, via della Scienza 5, I-09047 Selargius (CA), Italy\\
$^{8}$ Laboratoire de Physique et Chimie de l'Environnement et de l'Espace LPC2E CNRS-Universit{\'e} d'Orl{\'e}ans, F-45071 Orl{\'e}ans, France\\
$^{9}$ Station de radioastronomie de Nan{\c c}ay, Observatoire de Paris, CNRS/INSU F-18330 Nan{\c c}ay, France\\
$^{10}$ Institute of Astronomy, University of Cambridge, Madingley Road, Cambridge, CB3 0HA, UK \\
$^{11}$ Anton Pannekoek Institute for Astronomy, University of Amsterdam, Science Park 904, 1098 XH Amsterdam, The Netherlands\\
$^{12}$ Kavli institute for astronomy and astrophysics, Peking University, Beijing 100871, P.R. China\\
$^{13}$ TAPIR (Theoretical Astrophysics), California Institute of Technology, Pasadena, California 91125, USA \\
$^{14}$ School of Physics and Astronomy, University of Birmingham, Edgbaston, Birmingham, B15 2TT, United Kingdom \\
$^{15}$ Universit\'e Paris-Diderot-Paris7 APC - UFR de Physique, B\^atiment Condorcet, 10 rue Alice Domont et L\'eonie Duquet 75205 PARIS CEDEX 13, France  \\
$^{16}$ Centre for Astrophysics \& Supercomputing, Swinburne University of Technology, PO Box 218, Hawthorn VIC 3122, Australia \\
$^{17}$ Max-Planck-Institut f\"ur Gravitationsphysik, Albert-Einstein-Institut, Callinstra\ss e 38, 30167, Hannover, Germany \\
$^{18}$ Jet Propulsion Laboratory, California Institute of Technology, Pasadena, California 91109, USA\\
$^{19}$ Laboratoire Univers et Th\'eories, Observatoire de Paris, CNRS/INSU,
Universit\'e Paris Diderot, 5 place Jules Janssen, 92190 Meudon, France \\
}
}

\date{Accepted XXX. Received YYY; in original form ZZZ}

\pubyear{2015}

\begin{document}
\label{firstpage}
\pagerange{\pageref{firstpage}--\pageref{lastpage}}
\maketitle

\begin{abstract}
We report on the high-precision timing of 42 radio millisecond pulsars
(MSPs) observed by the European Pulsar Timing Array (EPTA). This EPTA Data
Release 1.0 extends up to mid-2014 and 
baselines range from 7-18 years. It forms the basis for the stochastic
gravitational-wave background, anisotropic background, and continuous-wave
limits recently presented by the EPTA elsewhere. 
The Bayesian timing analysis performed with TempoNest
yields the detection of several new parameters:
 seven parallaxes, nine proper motions and, in the case of six binary pulsars,
an apparent change of the semi-major axis.
We find the NE2001 Galactic electron density model to be a better match to our parallax
distances (after correction from the Lutz-Kelker bias) than the M2 and M3 models by  \citet{sch12}. However, we measure
an average uncertainty of 80\% (fractional) for NE2001, three times larger than what is
typically assumed in the literature.
We revisit the transverse velocity distribution for a set of 19 isolated and 57 binary MSPs
and find no statistical difference between these two populations.
We detect  Shapiro delay in the timing residuals of PSRs
J1600$-$3053 and J1918$-$0642, implying pulsar and companion masses
$m_p=1.22_{-0.35}^{+0.5} \text{M}_{\odot}$, $m_c =
0.21_{-0.04}^{+0.06} \text{M}_{\odot }$ and $m_p=1.25_{-0.4}^{+0.6}
\text{M}_{\odot}$, $m_c = 0.23_{-0.05}^{+0.07}
\text{M}_{\odot }$, respectively.
Finally, we use the measurement of the orbital period derivative to set a
stringent constraint on the distance to PSRs J1012$+$5307 and J1909$-$3744, and
set limits on the longitude of ascending node through the search of the
annual-orbital parallax for PSRs J1600$-$3053 and J1909$-$3744.
\end{abstract}



\section{Introduction}

Three decades ago \citet{bkh+82} discovered the first millisecond pulsar (MSP), 
spinning at  642 Hz. Now over 300 MSPs have been found; see the Australia
Telescope National Facility (ATNF) pulsar
catalog\footnote{http://www.atnf.csiro.au/people/pulsar/psrcat/} \citep{mht+05}. MSPs are
thought to be neutron stars spun-up to rotation periods (generally) shorter than 30 ms via the
transfer of mass and angular momentum from a binary companion
\citep{acr+82,rs82}. We know that the vast majority of the MSP population ($ 
\simeq 80$\%) still
reside in binary systems and these objects have been shown to be incredible
probes for testing physical theories. Their applications range from
high-precision tests of general relativity (GR) in the quasi-stationary strong-field regime
\citep{ksm+06,fwe+12} to constraints on the equation of state of matter at supra-nuclear densities
\citep{dpr+10,afw+13}. Binary systems with a MSP and a white dwarf in wide orbits
offer the most stringent tests of the strong equivalence principle
\citep[e.g.][]{sfl+05,fkw12,rsa+14}.

 Most
of these applications and associated results mentioned above arise from the use of the pulsar
timing technique that relies on two properties of the radio MSPs: their
extraordinary rotational and average pulse profile stability. The pulsar
timing technique tracks the times of arrival (TOAs) of the pulses recorded at
the observatory and compares them to the prediction of a best-fit model. This
model, which is continuously improved as more observations are made available,
initially contains the pulsar's astrometric parameters, the rotational parameters
and the parameters describing the binary orbit, if applicable. With the recent
increase in timing precision due to e.g. improved receivers, larger available 
bandwidth and the use of coherent dedispersion \citep{hr75},
parameters that have a smaller effect on the TOAs have become measurable.

The first binary pulsar found, PSR B1913+16 \citep{ht75}, yielded the first evidence
for gravitational waves (GWs) emission. Since then, several ground-based detectors have been built
around the globe, e.g. Advanced  LIGO \citep{lig15} and Advanced Virgo
\citep{vir15}, to
directly detect GWs in the frequency range of 10-7000 Hz.
Also a space mission, eLISA \citep{lis+13}, is being designed to study GWs in the mHz regime.
Pulsars, on the other hand, provide a complementary probe for GWs
 by opening a new window in the nHz regime \citep{saz78,det79}.
Previous limits on the amplitude of the stochastic GW background (GWB) have been set by studying individual MSPs
\citep[e.g.][]{ktr94}.
However, an ensemble of pulsars spread over the sky (known as Pulsar Timing
Array; PTA) is required to ascertain the presence of a GWB and discriminate between
possible errors in the Solar System ephemeris or in the reference time
standards \citep{hd83,fb90}.

A decade ago, \citet{jhl+05} claimed that timing a set of a least 20 MSPs with a precision of
100 ns for five years would allow a direct detection of the GWB.
Such high timing precision has not yet been reached \citep{abb+15}.
Nonetheless, \citet{sej+13} recently argued that when a PTA enters a new signal
regime where the GWB signal starts to prevail over the low frequency pulsar
timing noise, the sensitivity of this PTA
depends more strongly on the number of pulsars than the cadence of the
observations or the timing precision. Hence, datasets consisting of many pulsars
with long observing baselines, even with timing precision of $\sim 1 \mu$s, constitute
an important step towards the detection of the GWB. In addition to the GWB
studies, such long and precise datasets allow additional timing parameters, and
therefore science, to be extracted from the same data.

Parallax measurements can contribute to the construction of Galactic electron
density models \citep{tc93,cl02}. Once built, these models can provide distance
 estimates for pulsars along generic lines-of-sight. New parallax measurements hence
allow a comparison and improvement of the current free electron distribution
models \citep{sch12}.
An accurate distance is also crucial to correct the spin-down rate of the
pulsar from the bias introduced by its proper motion \citep{shk70}. This same
correction has to be applied to the observed orbital period derivative before
any test of GR can be done with this parameter \citep{dt91}.

In binary systems, once the Keplerian parameters are known, it may be possible
to detect post-Keplerian (PK) parameters. These theory-independent parameters 
 describe the relativistic
deformation of a Keplerian orbit as a function of the  Keplerian parameters and
the {\it a priori} unknown pulsar
mass ($m_p$), companion mass ($m_c$) and inclination angle ($i$).
Measurement of the Shapiro delay, an extra propagation delay of the radio waves
due to the gravitational potential of the companion, gives 2 PK parameters
(range $r$ and shape $s\equiv \sin i$).
Other relativistic effects such as the advance of periastron $\dot{\omega}$
and the orbital decay $\dot{P_b}$ provide one extra PK parameter each.
In GR, any PK parameter can be described by the Keplerian parameters plus the
two masses of the system. Measuring three or more PK parameters therefore
overconstrains the masses, allowing one to perform tests of GR
\citep{tw89,ksm+06}.

The robustness of the detections of these parameters can be hindered
by the presence of stochastic influences like dispersion measure (DM)
variations and red (low-frequency) spin noise in the timing residuals \citep{chc+11,lah+14}.
Recent work by \citet{kcs+13} and \citet{lbj+14} discussed the modeling of the
DM variations while \citet{chc+11} used Cholesky decomposition of the covariance
matrix to properly estimate the parameters in the presence of red noise.
 Correcting for the DM
variations and the effects of red noise has often been done through an iterative
process. However, TempoNest, a Bayesian pulsar timing analysis software
\citep{lah+14} used in this work allows one to model these stochastic
influences
simultaneously while performing a non-linear timing analysis.

In this paper we report on the timing solutions of 42 MSPs observed by the
European Pulsar Timing Array (EPTA). The EPTA is a collaboration of European
research institutes and radio observatories that was established in 2006
\citep{kc13}. The EPTA makes use of the five largest (at decimetric
wavelengths) radio telescopes  in Europe: the Effelsberg Radio Telescope in
Germany (EFF), the Lovell Radio
Telescope at the Jodrell Bank Observatory (JBO) in England, the Nan\c cay Radio
Telescope (NRT) in France, the Westerbork Synthesis Radio Telescope (WSRT) in
the Netherlands and the Sardinia Radio Telescope (SRT) in Italy. As the SRT is
currently being commissioned, no data from this telescope are included in this paper.
The EPTA also operates the Large European Array for Pulsars (LEAP), where
data from the EPTA telescopes are coherently combined to form a tied-array telescope with
an equivalent diameter of 195 meters, providing a significant improvement in
the sensitivity of pulsar timing observations \citep{bjk+15}.

This collaboration has already led to previous publications. Using multi-telescope
data on PSR J1012$+$5307, \citet{lwj+09} put a limit on the gravitational
dipole radiation and the variation of the gravitational constant $G$.
\citet{jsb+10} presented long-term timing results of four MSPs, two of which are
updated in this work. More recently, \citet{hlj+11} set the first EPTA upper
limit on the putative GWB. Specifically for a GWB formed by circular, GW-driven
supermassive black-hole binaries, they measured the amplitude $A$ of the
characteristic strain level at a frequency of 1/yr, $A < 6 \times
10^{-15}$, using a subset of the EPTA data from only 5 pulsars.


Similar PTA efforts are ongoing around the globe with the Parkes Pulsar
Timing Array (PPTA; \citet{mhb+13}) and the NANOGrav collaboration
\citep{mac13}, also setting limits on the GWB \citep{dfg+13, src+13}.


The EPTA dataset introduced here, referred to as the EPTA Data
Release 1.0, serves as the reference dataset for the
following studies: an analysis of the DM variations (Janssen et al., in prep.),
 a modeling of the red noise in each pulsar \citep{cll+15}, a limit on the stochastic GWB
\citep{ltm+15} and the anisotropic background \citep{tmg+15} as well as a
search for continuous GWs originating from single sources
\citep{bps+15}.  The organization of this paper is as follows. The
instruments and methods to extract the TOAs at each observatory are described
in Section~\ref{sec:obs}. The combination and timing procedures are detailed
in Section~\ref{sec:timing}. The timing results and new parameters are
presented in Section~\ref{sec:results}  and discussed in Section ~\ref{sec:discussions}.
 Finally, we summarize and present some prospects about the EPTA in Section~\ref{sec:conclusions}.

\section{Observations and data processing}
\label{sec:obs}

This paper presents the EPTA dataset, up to mid-2014, that was gathered
from the `historical' pulsar instrumentations at 
EFF, JBO, NRT and WSRT with, respectively, the EBPP (Effelsberg-Berkeley Pulsar Processor), 
DFB (Digital FilterBank), BON (Berkeley-Orl\'eans-Nan\c cay) and PuMa (Pulsar
Machine) backends.
The data recorded with the newest generation of instrumentations, e.g. PSRIX
 at EFF \citep{lkg+16} and PuMaII at WSRT \citep{kss08}, will be part of a future EPTA data release.

\begin{figure}
\includegraphics[height=80mm,angle=-90]{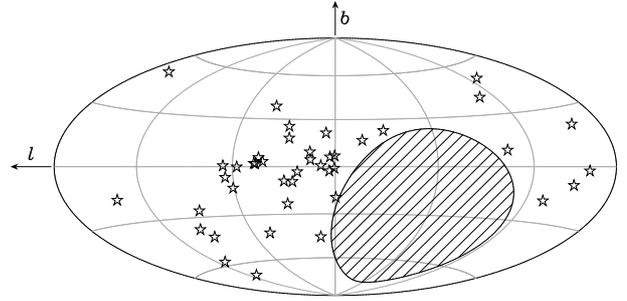}
\caption{Distribution of the 42 MSPs, represented with a star, in Galactic
coordinates (longitude $l$ and latitude $b$). The center of the plot is
oriented towards the Galactic Center. The hatched area is the part of the sky
(declination $\delta < -39^\circ$) that is not accessible to the EPTA.}
\label{fig:aitoff}
\end{figure}

 Compared to the dataset presented in
\citet{hlj+11}, in which timing of only five pulsars was presented, this
release includes 42 MSPs (listed in Table
\ref{tab:summary} with their distribution on the sky shown in
Fig.~\ref{fig:aitoff}). Among those 42 MSPs, 32 are members of binary systems. The  timing
solutions presented here span at least seven years, and for 16 of the MSPs the
baseline extends back $\sim 15$ years. For the five pulsars included in
\citet{hlj+11}, the baseline is extended by a factor 1.7-4.
When comparing our set of pulsars with the NANOGrav Nine-year Data Set
\citep{abb+15} (consisting of 37 MSPs) and the PPTA dataset \citep{mhb+13,rhc+16}
(consisting of 20 MSPs), we find
an overlap of 21 and 12 pulsars, respectively. However, we note that the
NANOGrav dataset contains data for  7 MSPs with a baseline less than two years.


 In this paper, we define an observing system  as a specific combination
of observatory, backend and frequency band. The radio telescopes and pulsar
backends used for the observations are described below.

\subsection{Effelsberg Radio Telescope}

The data from the 100-m Effelsberg Radio Telescope presented in this paper were
acquired using the  EBPP, an online
coherent dedispersion backend described in detail by \citet{bdz+97}. This
instrument can process a bandwidth (BW) up to 112 MHz depending on the DM value.
The signals from the two circular polarizations are split into 32 channels each
and sent to the dedisperser boards.  After the dedispersion takes place, the
output signals are folded (i.e. individual pulses are phase-aligned and summed)
using the topocentric pulse period.

EPTA timing observations at Effelsberg were made at a central frequency of 1410
MHz until April 2009 then moved to 1360 MHz  afterwards due to a change in the
receiver. Additional observations at S-Band (2639 MHz) began in November 2005
with observations at both frequencies taken during the same two-day observing
run. Typically, the observations occur on a monthly basis with an integration time per source
of about 30 minutes. The subintegration times range from 8 to 12 mins before
2009 and 2 mins thereafter. For 4 pulsars, namely PSRs J0030$+$0451,
J1024$-$0719, J1730$-$2304 and J2317$+$1439, there is a gap in the data from
1999 to 2005 as these sources were temporarily  removed from the observing list. Data reduction was
performed with the PSRCHIVE package \citep{hvm04}. The profiles were cleaned
of radio frequency interference (RFI) using the PSRCHIVE {\tt paz} tool but
also examined and excised manually with the {\tt pazi} tool. No standard
polarization calibration using a pulsed and linearly polarized noise diode  was
performed. However the EBPP automatically adjusts the power levels of both 
polarizations prior to each observation.  The TOAs were calculated by
cross-correlating the time-integrated,
frequency-scrunched, total intensity profile, with an analytic and noise free
template. This template was generated using the {\tt paas} tool to fit a set
of von Mises functions to a profile formed from high signal-to-noise ratio
(S/N) observations. In general, we used the standard `Fourier phase gradient'
algorithm \citep{tay92} implemented in PSRCHIVE to estimate the TOAs and their
uncertainties.  We used a different template for   each observing frequency,
including different templates for the 1410 and 1360 MHz observations. Local
time is kept by the on-site H-maser clock, which is corrected to Coordinated
Universal Time (UTC) using recorded offsets between the maser and the Global
Positioning System (GPS) satellites.

\subsection{Lovell Radio Telescope}

At Jodrell bank, the 76-m Lovell telescope is used in a regular monitoring
program to observe most of the pulsars presented in this paper. All TOAs used
here were generated by using the DFB, a clone of the Parkes Digital FilterBank.
 Each pulsar was observed with a typical cadence
of once every 10 days for 30 mins with a subintegration time of 10~s.  The DFB
came into operation in January 2009 observing at a central frequency of 1400
MHz with a BW of 128 MHz split into 512 channels.  From September 2009, the center frequency was
changed to 1520 MHz and the BW increased to 512 MHz (split into 1024 channels) of which
approximately 380 MHz was usable, depending on RFI conditions. As this is a
significant change, and to account for possible profile evolution with
observing frequency, both setups are considered as distinct observing systems
and different templates were used. 
Data cleaning and TOA generation  were done in a similar way to the
Effelsberg data. There is no standard polarization calibration
(through observations of a noise diode) applied
to the DFB data. However the power levels of both polarizations are
regularly and manually adjusted via a set of attenuators. Local time is kept by the on-site H-maser clock, which is
corrected to UTC using recorded offsets between
the maser and the GPS satellites.

\subsection{Nan\c cay Radio Telescope}
\label{sec:obs_nrt}
The Nan\c cay Radio Telescope is a meridian telescope with a collecting area
equivalent to a 94-m dish. The moving focal carriage that allows an observing
time of about one hour per source hosts the Low Frequency (LF) and High
Frequency (HF) receivers covering 1.1 to 1.8 GHz and 1.7 to 3.5 GHz, respectively.
 A large timing program of MSPs started in late 2004 with the commissioning of
the BON instrumentation, a member of the
ASP-GASP coherent dedispersion backend family \citep{d07}. A 128~MHz BW
is split into 32 channels by a CASPER\footnote{https://casper.berkeley.edu}
Serendip V board and then sent to servers to be coherently dedispersed and
folded to form 2-min subintegrations.

From 2004 to 2008 the BW was limited to 64 MHz and then extended to 128~MHz.
 At the same time, the NRT started to regularly observe a pulsed noise
diode prior to each observation in order  to properly correct for the
difference in gain and phase between the two polarizations. In August 2011, the
L-Band central frequency of the BON backend shifted from 1.4 GHz to 1.6 GHz to
accommodate the new wide-band NUPPI dedispersion backend \citep{ldc+14}.  Due to
known instrumental issues between November 2012 and April 2013
(i.e. loss of one of the polarization channels, mirroring of the spectrum),
these data have not been included in the analysis.

The flux density values at 1.4 GHz reported in Table \ref{tab:summary} are
derived from observations recorded  with the NUPPI instrument between MJD 55900
and 56700. The quasar 3C48 was chosen to be the reference source for the
absolute flux calibration. These flux density values have been corrected for
the declination-dependent illumination of the mirrors of the NRT. 
Although the NUPPI timing data are not included in this work, we used these
observations to estimate the median flux densities as no other EPTA data were
flux-calibrated. The NUPPI timing data will be part of a future EPTA
data release along with the data from other telescopes recorded with
new-generation instrumentations.

The data were reduced with the PSRCHIVE package  and automatically cleaned
for RFI. Except for pulsars with short orbital periods, all daily observations
are fully scrunched in time and frequency to form one single profile. For PSRs
J0610$-$2100, J0751$+$1807, J1738$+$0333, J1802$-$2124 the data were integrated
to form 6, 12, 16
and 8 min profiles respectively. The templates for the three observing
frequencies are constructed by phase-aligning the $\sim$10\% profiles with the
best S/N. The resulting integrated profiles are made noise free with the same
wavelet noise removal program as in \citet{dfg+13}. As stated above, we  used
the standard `Fourier phase gradient' from  PSRCHIVE to estimate the TOAs and
their uncertainties. However, we noticed that in the case of very low S/N
profiles, the reported uncertainties were underestimated. \citet{abb+15} also
observed that TOAs extracted from low S/N profiles deviate from a Gaussian
distribution and therefore excluded all TOAs where S/N <8 (see Appendix B of their paper
for more details). Here,
we made use of the Fourier domain Markov Chain Monte Carlo TOA estimator
(hereafter FDM) to properly estimate the error bars in this low S/N regime. We applied the FDM method
to PSRs J0034$-$0534, J0218$+$4232, J1455$-$3330, J2019$+$2425, J2033$+$1734.
All the BON data are time-stamped with a GPS-disciplined clock.

For PSR J1939$+$2134, archival data from 1990 to 1999 recorded with a swept-frequency
 local oscillator (hereafter referred to as DDS) at a frequency of 1410
MHz \citep{cbl+95} were added to the dataset. These data are time-stamped with
an on-site Rubidium clock, which is corrected to UTC using recorded offsets between
 the Rubidium clock and the Paris Observatory Universal Time.

\subsection{Westerbork Synthesis Radio Telescope}

The Westerbork Synthesis Radio Telescope is an East-West array
consisting of fourteen 25-m dishes, adding up to the equivalent size of a 94-m
dish when combined as a tied-array. From 1999 to 2010, an increasing
 number of MSPs were observed once a
month using the PuMa pulsar machine (a digital filterbank) at WSRT \citep{vkh+02}. In each observing
session, the pulsars were observed for 25 minutes each at one or more
frequencies centered at 350 MHz (10 MHz BW), 840 MHz (80 MHz BW) and 1380 MHz
(80 MHz spread across a total of 160 MHz BW). Up to 512 channels were used to
split the BW for the observations at 350 MHz. At 840 MHz and 1380 MHz, 64 channels were used per 10 MHz subband. For a more detailed description of
this instrumentation, see e.g. \citet{jsk+08}. Since 2007, the 840 MHz band was no longer used for
regular timing observations, however, an additional observing frequency
centered at 2273 MHz using 160 MHz BW was used for a selected set of the
observed pulsars. The data were dedispersed and folded offline using custom
software, and then
integrated over frequency and time to obtain a single profile for each
observation. 
Gain and phase difference between the two polarizations are adjusted during the phased-array calibration of the dishes.
To generate the TOAs, a high-S/N template based on the observations was used
for each observing frequency separately.
  Local time is kept by the on-site H-maser
clock, which is corrected to UTC using recorded
offsets between the maser and the GPS satellites.

\section{Data combination and timing}
\label{sec:timing}

The topocentric TOAs recorded at each observatory are first converted to the
Solar System barycenter (SSB) using the DE421 planetary ephemeris \citep{fwb09}  with
reference to the latest Terrestrial Time standard from the Bureau International
des Poids et Mesures (BIPM) \citep{pet10}. The DE421 model is a major
improvement on the DE200 ephemeris that was used for older
published ephemerides and later found to suffer from inaccurate values of planetary
masses \citep{sns+05,hbo06,vbs+08}.

We used TempoNest \citep{lah+14}, a Bayesian analysis software that uses the
Tempo2 pulsar timing package \citep{hem06, ehm06} and MULTINEST \citep{fhb09}, a
Bayesian inference tool, to evaluate and explore the parameter
space of the non-linear pulsar timing model. 
All pulsar timing parameters are sampled in
TempoNest with uniform priors. The timing model includes the
astrometric (right ascension, $\alpha$, declination, $\delta$, proper motion in
$\alpha$ and $\delta$, $\mu_{\alpha}$ and  $\mu_{\delta}$) and rotational parameters (period $P$
and period derivative $\dot{P}$).  If the pulsar is part of a binary system,
five additional parameters are incorporated to describe the Keplerian binary
motion: the orbital period $P_b$, the projected semi-major axis $x$ of the
pulsar orbit, the longitude of periastron $\omega$, the epoch  $T_0$ of the periastron
passage  and the eccentricity $e$. For some pulsars in our set, we require theory-independent PK parameters
\citep{dd85, dd86} to account for deviations from a Keplerian motion, or
parameters to describe changes in the viewing geometry of the systems. The
parameters we used include the precession of periastron
$\dot{\omega}$, the orbital period derivative $\dot{P_b}$, the Shapiro delay
(`range' $r$ and `shape' $s$; $s$ has a uniform prior in $\cos i$ space) and the apparent
derivative of the projected semi-major axis $\dot{x}$. These parameters are implemented
in Tempo2 under the `DD' binary model. In the case of quasi-circular orbits, the `ELL1'
model is preferred and replaces $\omega$, $T_0$ and $e$ with the two
Laplace-Lagrange parameters $\kappa$ and $\eta$ and the time of ascending node
$T_{\text{asc}}$ \citep{lcw+01}.  For the description of the Shapiro delay in PSRs
J0751$+$1807, J1600$-$3053 and J1918$-$0642 we adopted the orthometric
parametrization of the Shapiro delay introduced by \citet{fw10} with the
amplitude of the third harmonic of the Shapiro delay $h_3$ and the ratio of
successives harmonics $\varsigma$.
 
To combine the TOAs coming from the different observing systems  described in
Section~\ref{sec:obs}, we first corrected them for the phase difference between
the templates by cross-correlation of the reference template with the
other templates. We then fit for the arbitrary time offsets, known as JUMPs, between
the reference observing system and the remaining systems. These
JUMPs encompass, among other things: the difference in instrumental delays,
the use of different templates and the choice for the fiducial point on the
template. The JUMPs are analytically marginalized  over during the
TempoNest Bayesian analysis. In
order to properly weight the TOAs from each system, the
timing model includes a further two {\it ad~hoc} white noise parameters per observing
system. These parameters known as the error factor `EFAC', $E_f$, and the error
added in quadrature `EQUAD', $E_q$ (in units of seconds), relate to a TOA with uncertainty
$\sigma_p$ in seconds as:
\begin{equation}
\sigma = \sqrt{E_q^2 + E_f^2 {\sigma_p}^2}.
\end{equation}
Note that this definition of EFAC and EQUAD in TempoNest is different from the
definition employed in Tempo2 and the earlier timing software Tempo, where
$E_q$ was added in quadrature to $\sigma_p$ before applying $E_f$. The $E_f$
and $E_q$ parameters are  set with
uniform {priors in the 
logarithmic space (log-uniform priors)} in the $\log_{10}$-range $[-0.5,1.5], [-10, -3]$, respectively.
These prior ranges are chosen to be wide enough to include any value of EFAC
and EQUAD seen in our dataset.

Each pulsar timing model also includes two stochastic models to describe the
DM variations and an additional {achromatic} red noise process.
{Both processes are modeled as stationary, stochastic signals with 
a power-law spectrum of the form $S(f)\propto{}A^2f^{-\gamma}$,
where $S(f)$, $A$, and $\gamma$ are the power spectral density as function of
frequency $f$, the
amplitude and 
the spectral index, respectively. The power laws 
have a cutoff frequency at the lowest frequency, equal to the inverse of the data span, 
which is mathematically necessary for the subsequent calculation of the
covariance matrix \citep{hlm+09}.
It has been shown that this cutoff rises naturally for the achromatic red noise power law in pulsar timing data
because any low-frequency signal's power below the cutoff frequency is absorbed by the fitting of the pulsar's 
rotational frequency and frequency derivative  \citep{hlm+09,lbj+12}. 
It is possible to do the same for the DM variations model, by fitting a first
and a second DM derivative (parameters DM1 and DM2)
in the timing model \citep{lbj+14}. 
Implementation of the models is made using the time-frequency method of
\citet{lah+13}. 
Details on this process and applications can be found in \citet{ltm+15} and \citet{cll+15}.
In brief, denoting matrices with boldface letters, the red noise process 
time-domain signal, is expressed as a Fourier series,
${\rm \bf t}_{\rm TN}=\mathbfss{F}_{\rm TN}\textrm{{\bf a}}$, where $\mathbfss{F}_{\rm TN}$ 
is the sum of sines and cosines with coefficients given by the matrix $\textrm{{\bf a}}$. 
Fourier frequencies are sampled with integer multiples of the lowest frequency, 
and are sampled up to $1/14$\,days$^{-1}$. The 
Fourier coefficients are free parameters.} 

{The DM variations component 
is modeled similarly, with the only difference being that the time-domain signal 
is dependent on the observing frequency. According to the 
dispersion law from interstellar plasma, the delay in the arrival time 
of the pulse depends on the inverse square of the 
observing frequency, see e.g. \citet{lg12}. As such, the Fourier 
transform components are 
${\mathbfss{F}^{DM}_{ij}={\mathbfss{F}_{ij}}D_{i} D_{j}}$, 
where the i,j indices denote the residual index number, $D_i=1/(k\nu^2_i)$, 
and $k=2.41\times 10^{-16}$~Hz$^{-2}$cm$^{-3}$pc~s$^{-1}$, is the dispersion constant. 
This stochastic DM variations component is additional to the deterministic 
linear and quadratic components
implemented as part of the Tempo2 timing model.} 
In addition, we used the standard electron density model for the solar wind
included in Tempo2 with a value of 4~cm$^{-3}$ at 1 AU. This solar wind model
can be covariant with the measured astrometric parameters of the pulsar.

{The covariance matrix of each of these two components 
is then calculated with a function of the form \citep{ltm+15}:
\begin{equation}
\label{eq:BayesCovRed}
{\bf C} = {\bf C^{-1}_{\textrm{w}}} - {\bf 
C^{-1}_{\textrm{w}}}\mathbfss{F} \left[(\mathbfss{F})^{\textrm{{\bf T}}}{\bf 
C^{-1}_{\textrm{w}}}\mathbfss{F} + (\Psi)^{-1}\right]^{-1}  (\mathbfss{F})^{\textrm{{\bf T}}}{\bf C^{-1}_{\textrm{w}}} .
\end{equation}
The equation is valid for both the DM variations and achromatic 
red noise process, by using the corresponding Fourier transform 
$\mathbfss{F}$ and covariance matrix of the Fourier coefficients $\Psi=\langle
\textrm{a}_i\textrm{a}_j\rangle$. 
The ${\bf C_{\textrm{w}}}$ term is the white noise covariance matrix 
and is a diagonal matrix 
with the main diagonal formed by the residual uncertainties squared. 
The superscript ${\textrm{{\bf T}}}$ denotes the transpose of the matrix.}

{The power-law parameterization of the DM variations and red noise 
spectra means that the parameters we need to sample are 
the amplitudes and spectral indices of the power law. We do so by using 
uniform priors in the range $[0,7]$ for the spectral index and
log-uniform priors for the amplitudes, in the $\log_{10}$-range $[-20,-8]$. 
For discussion on the impact of our prior type selection, 
see \citet{lah+14} and \citet{cll+15}.
Here, we have used the least informative priors on the noise parameters. 
This means that the Bayesian inference will assign equal probability to 
these parameters if the data are insufficient to break the degeneracy between
them. This 
approach is adequate to derive a total noise covariance matrix (addition of 
white noise, red noise and DM variations covariance matrices) that allows 
robust estimation of the timing parameters. 
The prior ranges are set
to be wide enough to encompass any DM or red noise signal seen in the data. The
lower bound on the spectral index of the red noise process is set to zero as we
assume there is no blue process in the data. Together with the EFAC and EQUAD
values, the DM and red noise spectral indices and amplitudes are used by the timing 
software to form the timing residuals.}


\subsection{Criterion for Shapiro delay detectability}
\label{sec:criteria}
To assess the potential detectability of  Shapiro delay, we used the following
criterion. With the orthometric parametrization of Shapiro delay, we can compute the
 amplitude $h_3$ (in seconds) in the timing residuals \citep{fw10},
\begin{equation}
h_3 = \left( \frac{\sin i}{1+\cos i} \right)^3  m_c  T_\odot.
\end{equation}
Here, $c$ is the speed of light, $ T_\odot = 4.925~490~947$ $\mu$s is the mass
of the Sun in units of time. By assuming a median companion mass, $m_c$, given
by the mass function with $m_p=1.35$ M$_\odot$ and an inclination angle
$i=60^\circ$, we can predict an observable $h_{3o}$. We can then compare this
$h_{3o}$ value to the expected precision given by $\xi = \delta_{\text{TOAs}}
{\text{N}_{\text{TOAs}}}^{-1/2}$ where $ \delta_{\text{TOAs}}$ is the median
uncertainty of the TOAs and ${\text{N}_{\text{TOAs}}}$ the number of TOAs in
the dataset. The criterion $h_{3o} \gtrsim \xi$ associated with a non detection
of Shapiro delay would likely mean an unfavorable inclination angle, i.e. $i
\lesssim 60^\circ$.

\section{Timing results}
\label{sec:results}

In this section we summarize the timing results of the 42 MSPs obtained from
TempoNest. Among these sources, six pulsars, namely PSRs J0613$-$0200, J1012+5307,
J1600$-$3053, J1713+0747, J1744$-$1134 and J1909$-$3744, have been selected by
\citet{bps+15} to form the basis of the work presented by
\citet{ltm+15,tmg+15,bps+15}. The quoted uncertainties represent the $68.3\%$
Bayesian credible interval of the one-dimensional marginalized posterior
distribution of each parameter. The timing models are shown in 
Tables~\ref{tab:param1} to \ref{tab:param11}. These models, including the
stochastic parameters, are made publicly available on the
EPTA website\footnote{http://www.epta.eu.org/aom.html}. The reference
profiles at L-Band can be found in Fig.~\ref{plot:templates-1} and \ref{plot:templates-2}.
Throughout the paper, we refer to RMS as the weighted Root Mean Square timing
residuals. The details on the data sets used in this paper can be found in
Table~\ref{tab:data}.

\defcitealias{abb+15}{A15}
\defcitealias{rhc+16}{R16}
\defcitealias{mnf+16}{M16}
\defcitealias{vbc+09}{V09}

\begin{table*}
\begin{minipage}{180mm}
\caption{Summary of the 42-pulsar data set. The columns present the pulsar name
in the J2000 coordinate system, the observatories that contributed to the
dataset, the number of TOAs, the time span of the dataset, the median TOA
uncertainty ($\sigma_{\text{TOA}}$) taking into account the white noise
parameters `EFAC' and `EQUAD', the RMS timing
residual, the spin period, the orbital period and the median flux density of
the pulsar at 1400 MHz (see Section \ref{sec:obs_nrt} for more details about
the flux measurements). The last column gives the reference for the last
published timing solution where \citetalias{vbc+09}, \citetalias{abb+15}, \citetalias{rhc+16} relate to
\citet{vbc+09}, \citet{abb+15}, \citet{rhc+16}, respectively. The pulsars  indicated by $^\dag$ are also named
following the B1950 coordinate system, with the names  B1855$+$09, B1937$+$21 and
B1953$+$29 respectively.  The quoted RMS values are obtained from keeping the
noise parameters, DM and red noise models at the maximum likelihood value while
subtracting the DM signal from the residuals. Because of the degeneracy between
the DM and red noise models, especially where no multifrequency data are
available, the resulting RMS quoted here can be biased towards smaller values
(when the removed DM signal absorbed part of the red noise signal).   }
\label{tab:summary}
\begin{tabular}{lcrrrrrrrc}
\hline\hline
PSR JName & Observatories & $N_{\text{TOA}}$ & $T_{\text{span}}$  & $\sigma_{\text{TOA}}$ & RMS       & $P_{\rm Spin}$ &  $P_{\rm Orb}$  & S$_{1400}$ & References\\ 
          &               &           &     (yr)    & ($\mu s$) &  ($\mu s$) &  (ms)  &   (d)    &(mJy) &     \\ 
\hline\hline
\\ 
J0030$+$0451 & EFF, JBO, NRT & 907             & 15.1 &  3.79 & 4.1 & 4.9 & --- & 0.8  & \citet{aaa+09a}; \citetalias{abb+15} \\ 
J0034$-$0534 & NRT, WSRT & 276                 & 13.5 &  8.51 & 4.0 & 1.9 & 1.59 &  0.01  & \citet{hlk+04,aaa+10a}   \\ 
J0218$+$4232 & EFF, JBO, NRT, WSRT & 1196      & 17.6 & 10.51 & 7.4 & 2.3 & 2.03 & 0.6  &  \citet{hlk+04}  \\ 
J0610$-$2100 & JBO, NRT & 1034                 &  6.9 &  8.14 & 4.9 & 3.9 & 0.29 &  0.4 &  \citet{bjd+06}  \\ 
J0613$-$0200 & EFF, JBO, NRT, WSRT & 1369      & 16.1 &  2.57 & 1.8 & 3.1 & 1.20 & 1.7  &  \citetalias{vbc+09}; \citetalias{abb+15}; \citetalias{rhc+16} \\ 
\\ 
J0621$+$1002 & EFF, JBO, NRT, WSRT & 673       & 11.8 &  9.43 & 15.6 & 28.9 & 8.32 & 1.3  & \citet{sna+02,nsk08}   \\ 
J0751$+$1807 & EFF, JBO, NRT, WSRT & 1491      & 17.6 &  4.33 & 3.0 & 3.5 & 0.26 & 1.1  &  \citet{nss+05, nsk08}  \\ 
J0900$-$3144 & JBO, NRT & 875                  &  6.9 &  4.27 & 3.1 & 11.1 & 18.74 & 3.2  &  \citet{bjd+06}  \\ 
J1012$+$5307 & EFF, JBO, NRT, WSRT & 1459      & 16.8 &  2.73 & 1.6 & 5.3 & 0.60 & 3.0  &  \citet{lwj+09}; \citetalias{abb+15} \\ 
J1022$+$1001 & EFF, JBO, NRT, WSRT & 908       & 17.5 &  4.02 & 2.5 & 16.5 & 7.81 & 2.9  &  \citetalias{vbc+09}; \citetalias{rhc+16}  \\ 
\\ 
J1024$-$0719 & EFF, JBO, NRT, WSRT & 561       & 17.3 &  3.42 & 8.3 & 5.2 & --- & 1.3  &  \citetalias{vbc+09};\citet{egc+13}; \citetalias{abb+15}; \citetalias{rhc+16} \\
J1455$-$3330 & JBO, NRT & 524                  &  9.2 &  7.07 & 2.7 & 8.0 & 76.17 & 0.4  &  \citet{hlk+04}; \citetalias{abb+15} \\ 
J1600$-$3053 & JBO, NRT & 531                  &  7.7 &  0.55 & 0.46 & 3.6 & 14.35 & 2.0  &  \citetalias{vbc+09}; \citetalias{abb+15}; \citetalias{rhc+16} \\ 
J1640$+$2224 & EFF, JBO, NRT, WSRT & 595       & 17.3 &  4.48 & 1.8 & 3.2 & 175.46 & 0.4  &  \citet{llw+05}; \citetalias{abb+15} \\ 
J1643$-$1224 & EFF, JBO, NRT, WSRT & 759       & 17.3 &  2.53 & 1.7 & 4.6 & 147.02 & 3.9  &  \citetalias{vbc+09}; \citetalias{abb+15}; \citetalias{rhc+16} \\ 
\\ 
J1713$+$0747 & EFF, JBO, NRT, WSRT & 1188      & 17.7 &  0.59 & 0.68 & 4.6 & 67.83 &  4.9  & \citetalias{vbc+09};\citet{zsd+15}; \citetalias{abb+15}; \citetalias{rhc+16} \\ 
J1721$-$2457 & NRT, WSRT & 150                 & 12.8 & 24.28 & 11.7 & 3.5 & --- & 1.0  &  \citet{jsb+10}  \\ 
J1730$-$2304 & EFF, JBO, NRT & 285             & 16.7 &  4.17 & 1.6 & 8.1 & --- & 2.7  &  \citetalias{vbc+09}; \citetalias{rhc+16}  \\ 
J1738$+$0333 & JBO, NRT & 318                  &  7.3 &  5.95 & 3.0 & 5.9 & 0.35 & 0.3  &  \citet{fwe+12}; \citetalias{abb+15} \\ 
J1744$-$1134 & EFF, JBO, NRT, WSRT & 536       & 17.3 &  1.21 & 0.86 & 4.1 & --- & 1.6 &  \citetalias{vbc+09}; \citetalias{abb+15}; \citetalias{rhc+16} \\ 
\\ 
J1751$-$2857 & JBO, NRT & 144                  &  8.3 &  3.52 & 3.0 & 3.9 & 110.75 & 0.4  &  \citet{sfl+05}  \\ 
J1801$-$1417 & JBO, NRT & 126                  &  7.1 &  3.81 & 2.6 & 3.6 & --- & 1.1  &   \citet{lfl+06} \\ 
J1802$-$2124 & JBO, NRT & 522                  &  7.2 &  3.38 & 2.7 & 12.6 & 0.70 & 0.9  &  \citet{fsk+10}  \\ 
J1804$-$2717 & JBO, NRT & 116                  &  8.4 &  7.23 & 3.1 & 9.3 & 11.13 & 1.0  &  \citet{hlk+04}  \\ 
J1843$-$1113 & JBO, NRT, WSRT & 224            & 10.1 &  2.48 & 0.71 & 1.8 & --- & 0.5  &  \citet{hfs+04}  \\ 
\\ 
J1853$+$1303 & JBO, NRT & 101                  &  8.4 &  3.58 & 1.6 & 4.1 & 115.65 & 0.5  &  \citet{gsf+11}; \citetalias{abb+15}\\ 
J1857$+$0943$^\dag$ & EFF, JBO, NRT, WSRT & 444 & 17.3 &  2.57 & 1.7 & 5.4 & 12.33 & 3.3  &  \citetalias{vbc+09}; \citetalias{abb+15}; \citetalias{rhc+16}  \\ 
J1909$-$3744 & NRT & 425                       &  9.4 &  0.26 & 0.13 & 2.9 & 1.53 & 1.1  &  \citetalias{vbc+09}; \citetalias{abb+15}; \citetalias{rhc+16}  \\ 
J1910$+$1256 & JBO, NRT & 112                  &  8.5 &  3.39 & 1.9 & 5.0 & 58.47 & 0.5  &  \citet{gsf+11}; \citetalias{abb+15}  \\ 
J1911$+$1347 & JBO, NRT & 140                  &  7.5 &  1.78 & 1.4 & 4.6 & --- & 0.6  &  \citet{lfl+06}  \\ 
\\ 
J1911$-$1114 & JBO, NRT & 130                  &  8.8 &  8.82 & 4.8 & 3.6 & 2.72 & 0.5  &  \citet{tsb+99}  \\ 
J1918$-$0642 & JBO, NRT, WSRT & 278            & 12.8 &  3.18 & 3.0 & 7.6 & 10.91 & 1.2  &  \citet{jsb+10}; \citetalias{abb+15}  \\ 
J1939$+$2134$^\dag$ & EFF, JBO, NRT, WSRT & 3174 & 24.1 & 0.49 & 34.5 & 1.6 & --- & 8.3  &  \citetalias{vbc+09}; \citetalias{abb+15}; \citetalias{rhc+16}  \\ 
J1955$+$2908$^\dag$ & JBO, NRT & 157           &  8.1 & 14.92 & 6.5 & 6.1 & 117.35 & 0.5  &  \citet{gsf+11}; \citetalias{abb+15}  \\ 
J2010$-$1323 & JBO, NRT & 390                  &  7.4 &  2.89 & 1.9 & 5.2 & --- & 0.5  &  \citet{jbo+07}; \citetalias{abb+15}  \\ 
\\ 
J2019$+$2425 & JBO, NRT & 130                  &  9.1 & 26.86 & 9.6 & 3.9 & 76.51 & 0.1  &  \citet{nss01}  \\ 
J2033$+$1734 & JBO, NRT & 194                  &  7.9 & 18.24 & 12.7 & 5.9 & 56.31 & 0.1  &  \citet{spl04}  \\ 
J2124$-$3358 & JBO, NRT & 544                  &  9.4 &  5.57 & 3.2 & 4.9 & --- & 2.7  &  \citetalias{vbc+09}; \citetalias{rhc+16}  \\ 
J2145$-$0750 & EFF, JBO, NRT, WSRT & 800       & 17.5 &  2.64 & 1.8 & 16.1 & 6.84 & 4.0  &  \citetalias{vbc+09}; \citetalias{abb+15}; \citetalias{rhc+16}  \\ 
J2229$+$2643 & EFF, JBO, NRT & 316             &  8.2 & 11.18 & 4.2 & 3.0 & 93.02 & 0.1  &  \citet{wdk+00}  \\ 
\\ 
J2317$+$1439 & EFF, JBO, NRT, WSRT & 555       & 17.3 &  7.78 & 2.4 & 3.4 & 2.46 & 0.3  &  \citet{cnt96}; \citetalias{abb+15}  \\ 
J2322$+$2057 & JBO, NRT & 229                  &  7.9 & 12.47 & 5.9 & 4.8 & --- & 0.03  &  \citet{nt95}  \\ 
\hline
\end{tabular}
\end{minipage}
\end{table*}

\subsection{PSR J0030$+$0451}
A timing ephemeris for this isolated pulsar has been published by
\citet{aaa+09a} with a joint analysis of gamma-ray data from the \textit{Fermi} Gamma-ray
Space Telescope. Because the authors used the older DE200 version of the Solar
System ephemeris model, we report here updated astrometric measurements.
While our measured proper motion is consistent with the \citet{aaa+09a} value,
we get a significantly lower parallax value $\pi=2.79\pm0.23 $ mas that we
attribute partly to the errors in the DE200 ephemeris. Indeed reverting back to
the DE200 in our analysis yields an increased value of the parallax by 0.3 mas
but still below the parallax $\pi=4.1\pm0.3$ mas determined by \citet{aaa+09a}.

\subsection{PSR J0034$-$0534}
PSR J0034$-$0534 is a very faint MSP when observed at L-Band with a flux
density $S_{1400}=0.01$ mJy leading to profiles with very low S/N compared to
most other MSPs considered here. Helped by the better timing precision at
 350 MHz, we were able to improve on the previously
published composite proper motion $\mu=31\pm9$
mas yr$^{-1}$  by \citet{hll+05}  to $\mu=12.1\pm0.5$  mas yr$^{-1}$. We also
measure the eccentricity  $e=(4.3\pm0.7) \times 10^{-6}$ of this system for the
first time. Even with our improved timing precision characterized by a timing
residuals RMS of 4 \us, the detection of the parallax signature (at most 2.4
\us~according to \citet{aaa+10a}) is still out of reach.
 
 \begin{table*}
\caption{Timing model parameters for PSRs J0030$+$0451, J0034$-$0534,
J0218$+$4232 and J0610$-$2100. Figures in parentheses represent the
$68.3\%$ confidence uncertainties in the last digit quoted and come from the
one-dimensional marginalized posterior distribution of each parameter. The
measured timing parameters are introduced in Section~\ref{sec:timing}. The
derived parameters show the Galactic longitude ($l$) and latitude ($b$), the
parallax distance corrected from the Lutz-Kelker bias ($d$),
the composite proper motion ($\mu$). The position, spin period and DM are given for the reference epoch of MJD 55000.
 The three kinematic contributions
($\dot{P}_{\text{shk}}$, $\dot{P}_{\text{kz}}$ and $\dot{P}_{\text{dgr}}$ ) to the intrinsic period
derivative ($\dot{P}_{\text{int}}$) are introduced in
Section~\ref{sec:dis_shk}. For binary pulsars, the minimum companion mass,
assuming a pulsar mass of 1.2 M$_\odot$, is also indicated on the last line.}

\begin{minipage}{180mm}
\label{tab:param1}
\begin{tabular}{lllll}
\hline\hline
PSR Name & J0030$+$0451 & J0034$-$0534 & J0218$+$4232 & J0610$-$2100 \\ 
\hline
MJD range & 51275 --- 56779 & 51770 --- 56705 & 50370 --- 56786 & 54270 --- 56791 \\ 
 Number of TOAs & 907 & 276 & 1196 & 1034 \\ 
 RMS timing residual ($\mu s$) & 4.1 & 4.0 & 7.4 & 4.9 \\ 
 Reference epoch (MJD) & 55000 & 55000 & 55000 & 55000 \\ 
\\ 
Measured parameters\\ 
\\ 
Right ascension, $\alpha$ & 00:30:27.42836(6) & 00:34:21.83422(8) & 02:18:06.357299(19) & 06:10:13.595462(17) \\ 
Declination, $\delta$ & 04:51:39.707(3) & $-$05:34:36.722(3) & 42:32:17.3821(4) & $-$21:00:27.9313(4) \\ 
Proper motion in $\alpha$ (mas\,yr$^{-1}$) & $-$5.9(5) & 7.9(3) & 5.31(7) & 9.0(1) \\ 
Proper motion in $\delta$ (mas\,yr$^{-1}$) & $-$0.2(11) & $-$9.2(6) & $-$3.15(13) & 16.78(12) \\ 
Period, $P$ (ms) & 4.86545328635201(19) & 1.87718188583171(10) & 2.32309053151224(8) & 3.861324766195(3) \\ 
Period derivative, $\dot{P}$ ($\times 10^{-20}$) & 1.0172(3) & 0.49784(13) & 7.73955(7) & 1.2298(19) \\ 
Parallax, $\pi$ (mas) & 2.79(23) & --- & --- & --- \\ 
DM (cm$^{-3}$pc) & 4.329(6) & 13.7658(19) & 61.2488(17) & 60.67(3) \\ 
DM1 (cm$^{-3}$pc yr$^{-1}$) & 0.0007(5) & $-$0.0001(1) & $-$0.0003(2) & $-$0.014(8)\\
DM2 (cm$^{-3}$pc yr$^{-2}$) & 0.0001(1) & $-$0.000030(17) & 0.000056(20)&  0.002(1) \\
\\ 
Orbital period, $P_b$ (d) & --- & 1.58928182532(14) & 2.02884611561(9) & 0.2860160068(6) \\ 
Epoch of periastron, $T_0$ (MJD) & --- & 48766.98(4) & 49150.883(16) & 52814.303(13) \\ 
Projected semi-major axis, $x$ (lt-s) & --- & 1.4377662(5) & 1.9844344(4) & 0.0734891(4) \\ 
Longitude of periastron, $\omega_0$ (deg) & --- & 313(9) & 49(3) & 67(16) \\ 
Orbital eccentricity, $e$ & --- & 4.3(7)$\times 10^{-6}$ & 6.8(4)$\times 10^{-6}$ & 2.9(8)$\times 10^{-5}$ \\ 
$\kappa = e \times \sin \omega_0$ & --- & $-$3.1(7)$\times 10^{-6}$ & 5.1(4)$\times 10^{-6}$ & 2.7(8)$\times 10^{-5}$ \\ 
$\eta = e \times \cos \omega_0$ & --- & 3.0(6)$\times 10^{-6}$ & 4.5(4)$\times 10^{-6}$ & 1.2(8)$\times 10^{-5}$ \\ 
Time of asc. node (MJD) & --- & 48765.5995019(5) & 49150.6089170(3) & 52814.249581(3) \\ 
\\ 
Derived parameters\\ 
\\ 
Gal. longitude, $l$ (deg) & 113.1 & 111.5 & 139.5 & 227.7 \\ 
 Gal. latitude, $b$ (deg) & $-$57.6 & $-$68.1 & $-$17.5 & $-$18.2 \\ 
LK Px Distance, $d$ (pc) & $354_{-27}^{+31}$& --- & --- & --- \\ 
 Composite PM, $\mu$ (mas\,yr$^{-1}$) &5.9(5) & 12.1(5) & 6.18(9) & 19.05(11) \\ 
 $\dot{P}_{\text{shk}} (\times 10^{-20}) $	&0.015(3)	&0.036	&0.057	&1.2	\\
 $\dot{P}_{\text{kz}}	(\times 10^{-20}) $	&$-$0.078(7) 	&$-$0.056 &$-$0.034	&$-$0.082	\\
 $\dot{P}_{\text{dgr}} (\times 10^{-20}) $	&$-$0.0030(3)	&$-$0.00086 & 0.013	&0.011	\\
 $\dot{P}_{\text{int}} (\times 10^{-20}) $	&1.084(6)	&0.518	&7.7	&0.0955	\\
 Characteristic age, $\tau_c$ (Gyr) & 7.1 & 5.7 & 0.48 & 64.0 \\ 
 Surface magnetic field, $B$ ($\times 10^8$ G) & 2.3 & 1.0 & 4.3 & 0.6 \\ 
Min. companion mass (M$_{\odot}$) & --- & 0.13 & 0.16 & 0.02 \\
\hline
 \end{tabular}
\end{minipage}
\end{table*}

\subsection{PSR J0218$+$4232}
The broad shape of the pulse profile of this pulsar (with a duty cycle of about
50\%, see Figure~\ref{plot:templates-1}) and its low flux density limit
our timing precision to about 7~$\mu$s  and, therefore, its use for GWB
detection.  \citet{dyc+14} recently published the pulsar composite proper
motion $\mu = 6.53\pm0.08$ mas~yr$^{-1}$ from very long baseline interferometry
(VLBI). With EPTA data, we find  $\mu=6.14\pm0.09$ mas~yr$^{-1}$. This value is
in disagreement with the VLBI result. A possible explanation for this
discrepancy is
that \citet{dyc+14} overfitted their model with five parameters for five
observing epochs.  \citet{dyc+14} also
reported a distance  $d=6.3_{-2.3}^{+8.0}$ kpc from VLBI parallax measurement.
\citet{vl14} later argued that the \citet{dyc+14}  parallax suffers from the Lutz-Kelker bias
and corrected the distance to be $d=3.2_{-0.6}^{+0.9}$ kpc. This
distance is consistent with the 2.5 to 4 kpc range estimated from the
properties of the white dwarf companion to PSR J0218+4232 \citep{bkk02}. Even with the
\citet{vl14} $3\sigma$ lowest distance estimate, the parallax would induce a
signature on the timing residuals of less than 800 ns \citep{lk04}, which is far from
our current timing precision. We therefore cannot further constrain the
distance with our current dataset. Our measurement of the system's eccentricity
$e=(6.8\pm0.4)\times10 ^{-6} $ is significantly lower than the previously
reported value $e = (22 \pm 2) \times 10^{-6}$ by \citet{hlk+04}.

\subsection{PSR J0610$-$2100}
With a very low-mass companion ($0.02~\text{M}_{\odot} <
\text{M}_c<0.05~\text{M}_{\odot}$), PSR J0610$-$2100 is a member of the
`black widow' family, which are a group of (often) eclipsing binary MSPs believed
to be ablating their companions. Here we report on a newly measured
eccentricity, $e=(2.9\pm0.8) \times 10^{-5}$, and an improved proper motion
($\mu_{\alpha}=9.0\pm0.1$ mas~yr$^{-1}$ and $\mu_{\delta}=16.78\pm0.12$
mas~yr$^{-1}$) compared to the previous values ($\mu_{\alpha}=7\pm3$
mas~yr$^{-1}$ and $\mu_{\delta}=11\pm3$ mas~yr$^{-1}$) from \citet{bjd+06}
derived with slightly more than two years of data. It is interesting to note
that, in contrast to another well studied black widow pulsar, PSR J2051$-$0827 \citep{lvt+11}, no secular
variations of the orbital parameters are detected in this system.
There is also no evidence for eclipses of the radio signal in our data.

We checked our data for possible orbital-phase dependent  DM-variation that could
account for the new measurement of the eccentricity. We found no evidence for
this within our DM precision. We also obtained consistent results for the eccentricity and longitude of
periastron after removing TOAs for given orbital phase ranges.

\subsection{PSR J0613$-$0200}
For PSR J0613$-$0200, we measure a parallax $\pi=1.25\pm 0.13$ mas that is
 consistent with the value published in \citet{vbc+09} ($\pi=0.8\pm
0.35$ mas). In addition,
we report on the first detection of the  orbital period derivative $\dot{P_b} =
(4.8 \pm 1.1) \times
10^{-14}$ thanks to our 16-yr baseline. This result will be discussed further
in Section \ref{sec:dis_shk}. Finally, we improve on the precision of the
proper motion with $\mu_{\alpha}=-1.822\pm0.008$ mas~yr$^{-1}$ and
$\mu_{\delta}=-10.355\pm0.017$ mas~yr$^{-1}$.

\subsection{PSR J0621$+$1002}
Despite being the slowest rotating MSP of this dataset with a period of almost 30 ms,
PSR J0621$+$1002 has a profile with a narrow peak feature of width $\sim500$
\us. We are able to measure the precession of the periastron
$\dot{\omega}=0.0113\pm0.0006$ deg yr$^{-1}$  and find it
to be within 1 $\sigma$ of the value reported by \citet{nsk08}  using
 Arecibo data. We also find a similar value of the proper motion
to \citet{sna+02}.

\subsection{PSR J0751$+$1807}
PSR J0751$+$1807 is a 3.5-ms pulsar in an approximately 6-h orbit. \citet{nss+05}
originally reported a parallax $\pi=1.6\pm0.8$~mas and a measurement of the
orbital period derivative $\dot{P_b}=(-6.4 \pm 0.9) \times 10^{-14}$. Together
with their detection of the Shapiro delay, they initially derived a large
pulsar mass $m_p=2.1\pm0.2 \text{M}_{\odot}$.  \citet{nsk08} later corrected
the orbital period derivative measurement to $\dot{P}_b=(-3.1\pm 0.5) \times 10^{-14}$,
giving a much lower pulsar mass $m_p=1.26 \pm 0.14 \text{M}_{\odot}$. Here we
report on a  parallax $\pi=0.82 \pm 0.17$ mas and $\dot{P}_{b} = (-3.5 \pm
0.25) \times 10^{-14}$ that is similar to the value in \citet{nsk08}. 
However, we measured a precise composite proper motion of
$13.7\pm0.3$ mas yr$^{-1}$, inconsistent with the result ($6\pm2$ mas yr$^{-1}$) from \citet{nss+05}.
\citet{nsk08} explained the issue found with the timing solution presented
in \citet{nss+05} but did not provide an update of the proper motion for
comparison with our value.
We are also able to measure an apparent change in the semi-major axis
$\dot{x}=(-4.9\pm0.9) \times 10^{-15}$. Finally, we applied the orthometric
parametrization of the Shapiro delay to get $h_3=(3.0 \pm 0.6) \times 10^{-7}$ and $\varsigma=0.81\pm0.17$.
 The interpretation of these results will be discussed in Section \ref{sec:dis_mass}.

\subsection{PSR J0900$-$3144}
With about seven years of timing data available for PSR J0900$-$3144 (discovered
by \citep{bjd+06}) we  detect the proper motion for the first time,
revealing it to be one of the lowest composite proper-motion objects among our
data set with $\mu=2.26\pm0.07$ mas yr$^{-1}$. We also uncover a marginal
signature of the parallax $\pi=0.77\pm 0.44$ mas. However, we do not detect the
signature of the Shapiro delay despite the improvement in timing precision
compared to \citet{bjd+06}.
Following the criterion introduced in Section~\ref{sec:criteria}, we
 get $h_{3o}=0.4 \mu$s. With $\delta_{\text{TOAs}} = 4.27\mu$s and N$_\text{TOAs} =
875$, we find $\xi =0.14$~$\mu$s. Hence, given $ \xi < h_{3o}$, we argue for $i \lesssim 60^\circ$ to explain
the lack of Shapiro delay detection in this system.

\begin{table*}
\caption{Timing model parameters for PSRs J0613$-$0200, J0621$+$1002,
J0751$+$1807 and J0900$-$3144. See caption of Table~\ref{tab:param1} for a
description of this table.}
\begin{minipage}{180mm}
\label{tab:param2}
\begin{tabular}{lllll}
\hline\hline
PSR Name & J0613$-$0200 & J0621$+$1002 & J0751$+$1807 & J0900$-$3144 \\ 
\hline
MJD range & 50931 --- 56795 & 52482 --- 56780 & 50363 --- 56793 & 54286 --- 56793 \\ 
 Number of TOAs & 1369 & 673 & 1491 & 875 \\ 
 RMS timing residual ($\mu s$) & 1.8 & 15.6 & 3.0 & 3.1 \\ 
 Reference epoch (MJD) & 55000 & 55000 & 55000 & 55000 \\ 
\\ 
Measured parameters\\ 
\\ 
Right ascension, $\alpha$ & 06:13:43.975672(2) & 06:21:22.11436(3) & 07:51:09.155331(13) & 09:00:43.953088(8) \\ 
Declination, $\delta$ & $-$02:00:47.22533(7) & 10:02:38.7352(15) & 18:07:38.4864(10) & $-$31:44:30.89520(13) \\ 
Proper motion in $\alpha$ (mas\,yr$^{-1}$) & 1.822(8) & 3.23(12) & $-$2.73(5) & $-$1.01(5) \\ 
Proper motion in $\delta$ (mas\,yr$^{-1}$) & $-$10.355(17) & $-$0.5(5) & $-$13.4(3) & 2.02(7) \\ 
Period, $P$ (ms) & 3.061844088094608(15) & 28.8538611940574(16) & 3.47877083927942(4) & 11.1096493380938(6) \\ 
Period derivative, $\dot{P}$ ($\times 10^{-20}$)& 0.959013(14) & 4.730(5) & 0.77874(5) & 4.8880(11) \\ 
Parallax, $\pi$ (mas) & 1.25(13) & --- & 0.82(17) & 0.77(44) \\ 
DM (cm$^{-3}$pc) & 38.7746(14) & 36.47(3) & 30.246(6) & 75.707(8) \\ 
DM1 (cm$^{-3}$pc yr$^{-1}$) & 0.00002(7)  & $-$0.0094(3) & 0.0000(2) &  0.0009(7)\\
DM2 (cm$^{-3}$pc yr$^{-2}$) & $-$0.000002(7) & 0.0011(2) & 0.00004(4) & $-$0.0002(3) \\
\\ 
Orbital period, $P_b$ (d) & 1.198512575184(13) & 8.3186812(3) & 0.263144270792(7) & 18.7376360594(9) \\ 
Epoch of periastron, $T_0$ (MJD) & 53113.953(4) & 49746.86675(19) & 51800.283(7) & 52682.295(5) \\ 
Projected semi-major axis, $x$ (lt-s) & 1.09144409(6) & 12.0320732(4) & 0.3966158(3) & 17.24881126(15) \\ 
Longitude of periastron, $\omega_0$ (deg) & 47.2(11) & 188.774(9) & 92(9) & 70.41(10) \\ 
Orbital eccentricity, $e$ & 5.40(10)$\times 10^{-6}$ & 0.00245724(7) & 3.3(5)$\times 10^{-6}$ & 1.0490(17)$\times 10^{-5}$ \\ 
$\kappa = e \times \sin \omega_0$ & 3.96(10)$\times 10^{-6}$ & --- & 3.3(5)$\times 10^{-6}$ & 9.883(17)$\times 10^{-6}$ \\ 
$\eta = e \times \cos \omega_0$ & 3.67(11)$\times 10^{-6}$ & --- & 3.8(50)$\times 10^{-7}$ & 3.517(17)$\times 10^{-6}$ \\ 
Time of asc. node (MJD) & 53113.796354200(16) & --- & 51800.21586826(4) & 52678.63028819(13) \\ 
\\ 
Orbital period derivative, $\dot{P_b}$ & 4.8(11)$\times 10^{-14}$ & --- & $-$3.50(25)$\times 10^{-14}$ & --- \\ 
First derivative of $x$, $\dot{x}$ & --- & --- & $-$4.9(9)$\times 10^{-15}$ & --- \\ 
Periastron advance, $\dot{\omega}$ (deg/yr) & --- & 0.0113(6) & --- & --- \\ 
Third harmonic of Shapiro, $h_3$ ($\mu s$) & --- & --- & 0.30(6) & --- \\
Ratio of harmonics amplitude, $\varsigma$ & --- & --- & 0.81(17) & --- \\

\\ 
Derived parameters\\ 
\\ 
Gal. longitude, $l$ (deg) & 210.4 & 200.6 & 202.7 & 256.2 \\ 
 Gal. latitude, $b$ (deg) & $-$9.3 & $-$2.0 & 21.1 & 9.5 \\ 
LK Px Distance, $d$ (pc) & $777_{-70}^{+84}$& --- & $999_{-146}^{+202}$& $815_{-211}^{+378}$ \\ 
 Composite PM, $\mu$ (mas\,yr$^{-1}$) &10.514(17) & 3.27(14) & 13.7(3) & 2.26(7) \\ 
$\dot{P}_{\text{shk}} (\times 10^{-20}) $     &0.064(7)    &0.1       &0.16(3)       &0.011(5)       \\
$\dot{P}_{\text{kz}}  (\times 10^{-20}) $     &$-$0.0039(4)   &$-$0.0016 &$-$0.015(2)       &$-$0.012(4)       \\
$\dot{P}_{\text{dgr}} (\times 10^{-20}) $     &0.010(1)     &0.24 &0.018(4)       &$-$0.06(3)       \\
$\dot{P}_{\text{int}} (\times 10^{-20}) $     &0.889(8)    &4.39       &0.62(3)       &4.95(3)       \\
 Characteristic age, $\tau_c$ (Gyr) & 5.5 & 10.4 & 8.9 & 3.6 \\ 
 Surface magnetic field, $B$ ($\times 10^8$ G) & 1.7 & 11.4 & 1.5 & 7.5 \\ 
Min. companion mass (M$_{\odot}$) & 0.12 & 0.41 & 0.12 & 0.33 \\
\hline
 \end{tabular}
\end{minipage}
\end{table*}

\subsection{PSR J1012$+$5307}
\citet{lwj+09} previously presented a timing solution using a subset of these
EPTA data to perform a test on gravitational dipole radiation and variation of the
gravitational constant, $\dot{G}$. The  $\dot{x}$ and $\dot{P_b}$ parameters we
present here are consistent with the values from  \citet{lwj+09} but we
improve on the uncertainties of these parameters by factors of two and three,
respectively. Nonetheless, we note that our value for the parallax $\pi=0.71\pm 0.17$
mas differs by less than 2$\sigma$ from the value measured by \citet{lwj+09} using
the DE405 ephemeris. 

\subsection{PSR J1022$+$1001}
As recently pointed out by \citet{van13}, this source requires a high level of
polarimetric calibration in order to reach the best timing precision. Indeed,
by carefully calibrating their data, \citet{van13} greatly improved on the
timing model of \citet{vbc+09}  and successfully unveiled the precession of the
periastron $\dot{\omega }=0.0097\pm0.0023$ deg yr$^{-1}$, the presence of 
Shapiro delay and the secular variation of $\dot{x}$. Here we find similar
results with $\dot{\omega }=0.010 \pm 0.002$ deg yr$^{-1}$ and a 2-$\sigma$ 
consistent $\dot{x}$ with a completely  independent dataset.
 Nonetheless, we can not confirm the measurement of Shapiro delay
with our dataset. For this pulsar, we get $h_{3o}=0.62 \mu$s. With $\xi =
0.14$ $\mu$s, our constraint implies that the inclination angle $i \lesssim 60^\circ$, in
agreement with the result presented by \citet{van13}.

\subsection{PSR J1024$-$0719}
\citet{hbo06} were the first to announce a parallax $\pi=1.9\pm 0.4$ mas for
this nearby and isolated MSP that shows a large amount of red noise
\citep{cll+15}.  More recently, \citet{egc+13} used a subset of this
EPTA dataset to produce an
ephemeris and detected gamma-ray pulsations from this pulsar.  The authors assumed
the LK bias corrected distance \citep{vwc+12} from the \citet{hbo06} parallax
value to estimate its gamma-ray efficiency. However, it should be noted that
\citet{vbc+09} did not report on the measurement of the parallax using an
extended version of the \citet{hbo06} dataset. 
With this independent dataset we detect a parallax $\pi=0.80\pm0.17$ mas, a
value inconsistent with the early measurement reported by \citet{hbo06}. A
possible explanation for this discrepancy could be that \citet{hbo06} did not
include a red noise model in their analysis.

\subsection{PSR J1455$-$3330}
The last timing solution for this pulsar was published by \citet{hlk+04} and
characterized by an RMS of 67 \us. Thanks to our 9 years of data
with an RMS of less than 3 \us, we successfully detect the signature of
the  proper motion $\mu_{\alpha}=7.88\pm0.08$ mas~yr$^{-1}$ and
$\mu_{\delta}=-2.23\pm0.19$ mas~yr$^{-1}$, the parallax $\pi=1.04\pm 0.35$ mas
and the secular variation of the semi-major axis,
$\dot{x}=(-1.7\pm0.4)\times10^{-14}$ for the first time.

\subsection{PSR J1600$-$3053}
This 3.6-ms pulsar can be timed at very high precision thanks to the $\sim 45$
\us~wide peak on the right edge of its profile (see
Fig.~\ref{plot:templates-1}). We present here a precise measurement of the
parallax $\pi = 0.64 \pm 0.07$ mas, a value  marginally consistent with the
$\pi = 0.2 \pm 0.15$ mas from \citet{vbc+09}. We also show a large improvement
on the Shapiro delay detection through the use of the orthometric
parametrization \citep{fw10} with $h_3= (3.3\pm0.2) \times 10^{-7}$ and
$\varsigma=0.68\pm0.05$. The resulting mass measurement of this system
is discussed in Section~\ref{sec:dis_mass}.

\begin{table*}
\caption{Timing model parameters for PSRs J1012$+$5307, J1022$+$1001,
J1024$-$0719 and J1455$-$3330. See caption of Table~\ref{tab:param1} for a
description of this table.
$^{\dag}$For the observer, we report here the values from the
analysis in the ecliptic coordinate system, longitude $\lambda=153.865866885(16)^{\circ}$, latitude $\beta =-0.063930(14)^{\circ}$, proper motion in $\lambda$, $\mu_{\lambda}=-15.93(2)$ mas yr$^{-1}$ and proper motion in $\beta$, $\mu_{\beta}=-10(15)$ mas yr$^{-1}$. 
$^{\ddag}$ The reason for the  negative intrinsic period derivative
$\dot{P}_{\text{int}}$ of PSR J1024$-$0719 is explained in
Section~\ref{sec:dis_shk}.}
\begin{minipage}{180mm}
\begin{tabular}{lllll}
\hline\hline
PSR Name & J1012$+$5307 & J1022$+$1001 & J1024$-$0719 & J1455$-$3330 \\ 
\hline
MJD range & 50647 --- 56794 & 50361 --- 56767 & 50460 --- 56764 & 53375 --- 56752 \\ 
 Number of TOAs & 1459 & 908 & 561 & 524 \\ 
 RMS timing residual ($\mu s$) & 1.6 & 2.5 & 8.3 & 2.7 \\ 
 Reference epoch (MJD) & 55000 & 55000 & 55000 & 55000 \\ 
\\ 
Measured parameters\\ 
\\ 
Right ascension, $\alpha$ & 10:12:33.437521(5) & 10:22:57.9992(15)$^{\dag}$ & 10:24:38.675378(5) & 14:55:47.969869(14) \\ 
Declination, $\delta$ & 53:07:02.29999(6) & 10:01:52.78(6)$^{\dag}$ & $-$07:19:19.43395(15) & $-$33:30:46.3801(4) \\ 
Proper motion in $\alpha$ (mas\,yr$^{-1}$) & 2.609(8) & $-$18.2(64)$^{\dag}$ & $-$35.28(3) & 7.88(8) \\ 
Proper motion in $\delta$ (mas\,yr$^{-1}$) & $-$25.482(11) & $-$3(16)$^{\dag}$ & $-$48.18(7) & $-$2.23(19) \\ 
Period, ${P}$ (ms) & 5.255749101970103(19) & 16.45292995606771(11) & 5.1622046403157(3) & 7.987204929333(3) \\ 
Period derivative, $\dot{P}$ ($\times 10^{-20}$) & 1.712730(17) & 4.3322(4) & 1.8553(4) & 2.428(4) \\ 
Parallax, $\pi$ (mas) & 0.71(17) & 0.72(20) & 0.80(17) & 1.04(35) \\ 
DM (cm$^{-3}$pc) & 9.0172(14) & 10.250(4) & 6.485(10) & 13.563(7) \\ 
DM1 (cm$^{-3}$pc yr$^{-1}$) & 0.00016(2) & 0.0004(1) & 0.0025(8) & $-$0.002(4)\\
DM2 (cm$^{-3}$pc yr$^{-2}$) & 0.000016(2) & 0.00026(5) & $-$0.0007(2) & 0.001(1) \\
\\ 
Orbital period, $P_b$ (d) & 0.604672722901(13) & 7.8051348(11) & --- & 76.174568631(9) \\ 
Epoch of periastron, $T_0$ (MJD) & 50700.229(13) & 50246.7166(7) & --- & 48980.1330(10) \\ 
Projected semi-major axis, $x$ (lt-s) & 0.58181703(12) & 16.7654104(5) & --- & 32.362222(3) \\ 
Longitude of periastron, $\omega_0$ (deg) & 88(8) & 97.68(3) & --- & 223.460(5) \\ 
Orbital eccentricity, $e$ & 1.30(16)$\times 10^{-6}$ & 9.7229(14)$\times 10^{-5}$ & --- & 1.69636(12)$\times 10^{-4}$ \\ 
$\kappa = e \times \sin \omega_0$ & 1.30(16)$\times 10^{-6}$ & --- & --- & --- \\ 
$\eta = e \times \cos \omega_0$ & 5.1(173)$\times 10^{-8}$ & --- & --- & --- \\ 
Time of asc. node (MJD) & 50700.08174604(3) & --- & --- & --- \\ 
\\ 
Orbital period derivative, $\dot{P_b}$ & 6.1(4)$\times 10^{-14}$ & --- & --- & --- \\ 
First derivative of $x$, $\dot{x}$ & 2.0(4)$\times 10^{-15}$ & 1.79(12)$\times 10^{-14}$ & --- & $-$1.7(4)$\times 10^{-14}$ \\ 
Periastron advance, $\dot{\omega}$ (deg/yr) & --- & 0.0097(23) & --- & --- \\ 
\\ 
Derived parameters\\ 
\\ 
Gal. longitude, $l$ (deg) & 160.3 & 242.4 & 251.7 & 330.7 \\ 
 Gal. latitude, $b$ (deg) & 50.9 & 43.7 & 40.5 & 22.6 \\ 
LK Px Distance, $d$ (pc) & $1148_{-175}^{+241}$& $1092_{-182}^{+258}$& $1083_{-163}^{+226}$& $797_{-179}^{+304}$ \\ 
 Composite PM, $\mu$ (mas\,yr$^{-1}$) &25.615(11) & 19(9) & 59.72(6) & 8.19(9) \\ 
$\dot{P}_{\text{shk}} (\times 10^{-20}) $     &1.0(2)       &1.6(1.4)       & 4.8(10)	&0.10(4)       \\
$\dot{P}_{\text{kz}}  (\times 10^{-20}) $     &$-$0.077(7)      &$-$0.24(2) & $-$0.057(5)      &$-$0.035(7)       \\
$\dot{P}_{\text{dgr}} (\times 10^{-20}) $     &0.016(3)       &$-$0.010(2)   & $-$0.021(4)     &0.03(1)       \\
$\dot{P}_{\text{int}} (\times 10^{-20}) $     &0.8(2)       &3.0(1.4)     & -2.9(10)$^{\ddag}$      &2.33(4)       \\
 Characteristic age, $\tau_c$ (Gyr) & 10.3 & 8.7 & --- & 5.4 \\ 
 Surface magnetic field, $B$ ($\times 10^8$ G) & 2.1 & 7.1 & --- & 4.4 \\ 
Min. companion mass (M$_{\odot}$) & 0.10 & 0.66 & --- & 0.23\\
\hline
 \end{tabular}
\end{minipage}
\end{table*}

\subsection{PSR J1640$+$2224}
\citet{llw+05} used early Arecibo and Effelsberg data to report on the tentative detection of Shapiro delay for
this wide binary system in a 6-month orbit. From this measurement they deduced
the orientation of the system to be nearly edge-on ($78^{\circ} < i <
88^{\circ}$) and a companion mass for the white dwarf $m_p=0.15_{-0.05}^{+0.08}
\text{M}_{\odot}$. We cannot constrain the Shapiro delay with the
current EPTA data, even though our data comprise almost twice the number of
TOAs with a similar overall timing precision. The parallax signature in the
residuals also remains  undetected (based on Bayesian evidence\footnote{A
difference of 3 in the log evidence between two models is usually required to
justify the introduction of an additional parameter \citep{kr95}.}) but we find a significant
$\dot{x}=(1.07\pm0.16)\times 10^{-14}$, consistent with the upper limit set by \citet{llw+05}.

\subsection{PSR J1643$-$1224}
Using PPTA data, \citet{vbc+09} previously announced a parallax value $\pi =
2.2 \pm 0.4$ mas  that is marginally consistent with our value of $\pi = 1.17
\pm 0.26$ mas. We get a similar proper motion and  $\dot{x}=(-4.79\pm0.15) \times
10^{-14}$, albeit measured with a greater precision.

\subsection{PSR J1713$+$0747}
\label{sec:1713}
PSR J1713$+$0747 is one of the most precisely timed pulsars over two decades
\citep{vbc+09, zsd+15}.
Our proper motion and parallax values are consistent with the ones from \citet{vbc+09} and
\citet{zsd+15}. Nonetheless we can not detect any hint of the orbital period derivative $\dot{P_{b}}$.
The measurement of the Shapiro delay yields the following masses of the system,
$m_p=1.33_{-0.08}^{+0.09} \text{M}_{\odot}$ and $m_c=0.289_{-0.011}^{+0.013}
\text{M}_{\odot}$, in very good agreement with \citet{zsd+15}.

\begin{figure}
\centering
\includegraphics[height=80mm,angle=-90]{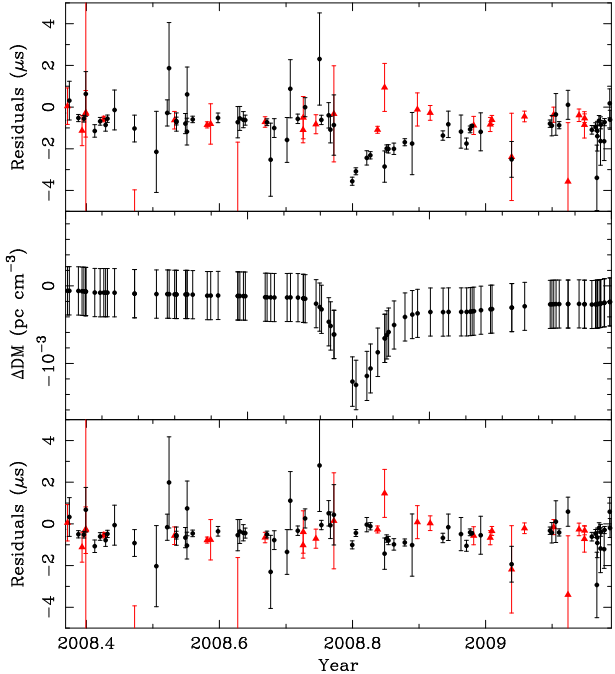}
\caption{Top panel: zoom-in on the PSR J1713$+$0747 residuals (black dots and red
triangles are L-Band and S-Band data respectively). Middle panel: DM signal
from the maximum likelihood DM model incorporating the shapelet basis functions
(see Section~\ref{sec:1713} for details). The bottom panel shows the residuals after
subtraction of the DM signal. The uncertainties on the DM signal come directly
from the 1-$\sigma$ uncertainties on the shapelet amplitudes used to model the
event, obtained from the full Bayesian analysis.}
\label{plot:1713_event}
\end{figure}

When inspecting the residuals of PSR J1713$+$0747 we noticed successive TOAs
towards the end of 2008 that arrived significantly earlier ($\sim3$ \us) than
predicted by our ephemeris (see top panel of Figure~\ref{plot:1713_event}).
After inspection of the original archives and comparison with other high
precision datasets like those on PSRs J1744$-$1134 and J1909$-$3744, we ruled out any
instrumental or clock issue as an explanation for this shift. We therefore
attribute this effect to a deficiency of the electron content towards the line
of sight of the pulsar. This  event has also been observed by the other PTAs
\citep{zsd+15,cks+15} and interpreted as possibly a kinetic shell propagating
through the interstellar medium \citep{cks+15} followed by a rarefaction of the electron
content.

To model this DM event we used shapelet basis functions.  A thorough
description of the shapelet formalism can be found in \cite{ref03}, with
astronomical uses being described in e.g., \cite{rb03,km04,lah15}.  Shapelets
are a complete ortho-normal set of basis functions that allow us to recreate
the effect of non-time-stationary DM variations in a statistically robust
manner, simultaneously with the rest of the analysis.  We used the Bayesian
evidence to determine the number of shapelet coefficients to include in the
model (only one coefficient was necessary in this study, i.e. the shapelet is
given by a Gaussian).  Our priors on the
location of the event span the entire dataset, while we assume an event width
of between five days and one year. The maximum likelihood results indicate
an event centered around MJD 54761 with a width of 10 days. The resulting DM signal
(including the shapelet functions) and the residuals corrected from it are
plotted in the middle and bottom panels of Fig.~\ref{plot:1713_event}
respectively. The DM model hence predicts a drop of $(1.3\pm0.4) \times 10^{-3}$
pc~cm$^{-3}$.

\begin{table*}
\caption{Timing model parameters for PSRs J1600$-$3053, J1640$+$2224,
J1643$-$1224 and J1713+0747. See caption of Table~\ref{tab:param1} for a
description of this table.}
\begin{minipage}{180mm}
\begin{tabular}{lllll}
\hline\hline
PSR Name & J1600$-$3053 & J1640$+$2224 & J1643$-$1224 & J1713+0747 \\ 
\hline
MJD range & 53998 --- 56795 & 50459 --- 56761 & 50459 --- 56778 & 50360 --- 56810 \\ 
 Number of TOAs & 531 & 595 & 759 & 1188 \\ 
 RMS timing residual ($\mu s$) & 0.46 & 1.8 & 1.7 & 0.68 \\ 
 Reference epoch (MJD) & 55000 & 55000 & 55000 & 55000 \\ 
\\ 
Measured parameters\\ 
\\ 
Right ascension, $\alpha$ & 16:00:51.903338(4) & 16:40:16.744834(7) & 16:43:38.161498(8) & 17:13:49.5331754(5) \\ 
Declination, $\delta$ & $-$30:53:49.37542(18) & 22:24:08.84121(13) & $-$12:24:58.6735(6) & 07:47:37.492536(16) \\ 
Proper motion in $\alpha$ (mas\,yr$^{-1}$) & $-$0.940(19) & 2.087(20) & 6.04(4) & 4.923(3) \\ 
Proper motion in $\delta$ (mas\,yr$^{-1}$) & $-$6.94(7) & $-$11.29(4) & 4.07(15) & $-$3.909(5) \\ 
Period, $P$ (ms) & 3.59792851006493(3) & 3.16331586776034(5) & 4.62164152573380(10) & 4.570136598154477(12) \\ 
Period derivative, $\dot{P}$ ($\times 10^{-20}$) & 0.95014(6) & 0.28161(11) & 1.8461(3) & 0.852919(13) \\ 
Parallax, $\pi$ (mas) & 0.64(7) & --- & 1.17(26) & 0.90(3) \\ 
DM (cm$^{-3}$pc) & 52.3245(16) & 18.422(10) & 62.411(5) & 15.9930(3) \\ 
DM1 (cm$^{-3}$pc yr$^{-1}$) & $-$0.0003(1) & $-$0.0000(2) & $-$0.0013(3) & 0.00006(3)\\
DM2 (cm$^{-3}$pc yr$^{-2}$) & 0.000012(47) & 0.00006(8) & 0.0000(1) & 0.000006(5)\\
\\ 
Orbital period, $P_b$ (d) & 14.34845777290(15) & 175.460664603(11) & 147.017397756(17) & 67.8251309745(14) \\ 
Epoch of periastron, $T_0$ (MJD) & 52506.3739(4) & 51626.1804(3) & 49283.9337(5) & 48741.9737(3) \\ 
Projected semi-major axis, $x$ (lt-s) & 8.8016546(5) & 55.3297223(5) & 25.0726144(7) & 32.34241956(15) \\ 
Longitude of periastron, $\omega_0$ (deg) & 181.835(9) & 50.7343(5) & 321.8488(10) & 176.1989(15) \\ 
Orbital eccentricity, $e$ & 1.73723(8)$\times 10^{-4}$ & 7.97299(8)$\times 10^{-4}$ & 5.05746(9)$\times 10^{-4}$ & 7.49421(7)$\times 10^{-5}$ \\ 
\\ 
First derivative of $x$, $\dot{x}$  & $-$2.8(5)$\times 10^{-15}$ & 1.07(16)$\times 10^{-14}$ & $-$4.79(15)$\times 10^{-14}$ & --- \\ 
Inclination angle, $i$ (deg) & $68.6_{-3.8}^{+3.4}$ & --- & --- & 71.8(6) \\ 
Longitude of ascending node, $\Omega$ (deg) &  --- & --- & --- & 89.9(17) \\
Companion mass, $m_c$ ($\text{M}_\odot$) & $0.208_{-0.043}^{+0.059}$ & --- & --- & 0.290(12) \\ 
Third harmonic of Shapiro, $h_3$ ($\mu s$) & 0.33(2) & --- & --- & --- \\
Ratio of harmonics amplitude, $\varsigma$ & 0.68(5) & --- & --- & --- \\
\\ 
Derived parameters\\ 
\\ 
Gal. longitude, $l$ (deg) & 344.1 & 41.1 & 5.7 & 28.8 \\ 
 Gal. latitude, $b$ (deg) & 16.5 & 38.3 & 21.2 & 25.2 \\ 
LK Px Distance, $d$ (pc) & $1492_{-150}^{+187}$& --- & $758_{-127}^{+185}$& $1108_{-33}^{+35}$ \\ 
 Composite PM, $\mu$ (mas\,yr$^{-1}$) &7.00(7) & 11.49(4) & 7.28(9) & 6.286(4) \\ 
$\dot{P}_{\text{shk}} (\times 10^{-20}) $     &0.064(8)       &0.12       &0.05(1)      &0.049(2)       \\
$\dot{P}_{\text{kz}}  (\times 10^{-20}) $     &$-$0.0137(9)      &$-$0.033 &$-$0.018(3)     &$-$0.0277(4)       \\
$\dot{P}_{\text{dgr}} (\times 10^{-20}) $     &0.043(7)       &0.0013     &0.029(8)      &0.020(2)       \\
$\dot{P}_{\text{int}} (\times 10^{-20}) $     &0.86(1)       &0.196      &1.79(2)       &0.812(2)       \\
 Characteristic age, $\tau_c$ (Gyr) & 6.7 & 25.6 & 4.1 & 8.9 \\ 
 Surface magnetic field, $B$ ($\times 10^8$ G) & 1.8 & 0.8 & 2.9 & 1.9 \\ 
Min. companion mass (M$_{\odot}$) & 0.19 & 0.23 & 0.11 & 0.26 \\
\hline
 \end{tabular}
\end{minipage}
\end{table*}

\subsection{PSR J1721$-$2457}
Thanks to an additional five years of data compared to  \citet{jsb+10}, the
proper motion of this isolated MSP is now better constrained. Our current
timing precision is most likely limited by the pulsar's large duty cycle (see
Fig.~\ref{plot:templates-1}) and the apparent absence of sharp features in the
profile. The flux density of this pulsar is also quite low with a value of 1 mJy at 1400MHz.

\subsection{PSR J1730$-$2304}
This low-DM and isolated MSP has a profile with multiple pulse components (see
Fig.~\ref{plot:templates-1}). As this pulsar lies very near to the ecliptic
plane ($\beta= 0.19^{\circ}$), we are unable to constrain its proper motion in
declination, similar to the previous study \citep{vbc+09}. Assuming the NE2001 distance, the
expected parallax timing signature would be as large as 2.3 \us. We report here on a
tentative detection of the parallax, $\pi = 0.86\pm0.32$ mas.

\subsection{PSR J1738$+$0333}
After the determination of the masses in this system from optical observations
\citep{akk+12}, \citet{fwe+12} used the precise measurements of the proper motion, parallax
 and $\dot{P_b}$ in this binary system to put constraints on
scalar-tensor theories of gravity. Our measured proper motion remains
consistent with their measurements. With a longer baseline and  more
observations recorded with the sensitive Arecibo Telescope, \citet{fwe+12}
were able to detect the parallax and the orbital period derivative of the
system. However, we do not yet reach the sensitivity to detect these two
parameters with our dataset.

\subsection{PSR J1744$-$1134}
This isolated MSP was thought to show long-term timing noise by \citet{hbo06}
even with a dataset shorter than 3 years.
In our data set we detect a (red) timing noise component \citep[see][]{cll+15}.
The RMS of the time-domain noise signal is
$\sim\,0.4\mu$s, but has a peak-to-peak variation of $\sim\,2\mu$s. The
higher latter value, however, is due to a bump which appears localized
in time (MJD $\sim$ 54000 to 56000). As discussed in \citet{cll+15},
non-stationary noise
from instrumental instabilities may cause such effects, but data with
better multi-telescope coverage are necessary to verify such a
possibility. This is further investigated in Lentati et al. (submitted)
using a more extended dataset from the International Pulsar Timing Array (IPTA) 
\citep{vlh+16}.

\begin{table*}
\caption{Timing model parameters for PSRs J1721$-$2457, J1730$-$2304,
J1738$+$0333 and J1744$-$1134. See caption of Table~\ref{tab:param1} for a
description of this table.}
\begin{minipage}{180mm}
\begin{tabular}{lllll}
\hline\hline
PSR Name & J1721$-$2457 & J1730$-$2304 & J1738$+$0333 & J1744$-$1134 \\ 
\hline
MJD range & 52076 --- 56737 & 50734 --- 56830 & 54103 --- 56780 & 50460 --- 56761 \\ 
 Number of TOAs & 150 & 268 & 318 & 536 \\ 
 RMS timing residual ($\mu s$) & 11.7 & 1.6 & 3.0 & 0.86 \\ 
 Reference epoch (MJD) & 55000 & 55000 & 55000 & 55000 \\ 
\\ 
Measured parameters\\ 
\\ 
Right ascension, $\alpha$ & 17:21:05.4979(3) & 17:30:21.66835(13) & 17:38:53.966375(11) & 17:44:29.4075373(14) \\ 
Declination, $\delta$ & $-$24:57:06.17(5) & $-$23:04:31.16(4) & 03:33:10.8720(4) & $-$11:34:54.69437(11) \\ 
Proper motion in $\alpha$ (mas\,yr$^{-1}$) & 1.9(12) & 20.7(7) & 7.08(6) & 18.810(6) \\ 
Proper motion in $\delta$ (mas\,yr$^{-1}$) & $-$25(16) & 9(12) & 4.97(19) & $-$9.36(3) \\ 
Period, $P$ (ms) & 3.496633783466(6) & 8.1227980469486(7) & 5.850095860612(5) & 4.074545941825154(15) \\ 
Period derivative, $\dot{P}$ ($\times 10^{-20}$) & 0.556(7) & 2.0196(11) & 2.410(4) & 0.89347(4) \\ 
Parallax, $\pi$ (mas) & --- & 0.86(32) & --- & 2.38(8) \\ 
DM (cm$^{-3}$pc) & 48.33(15) & 9.622(9) & 33.798(18) & 3.1312(17) \\ 
DM1 (cm$^{-3}$pc yr$^{-1}$) & $-$0.00(2) & 0.001(1) & $-$0.01(1) & $-$0.01(1)\\
DM2 (cm$^{-3}$pc yr$^{-2}$) & $-$0.002(4) & $-$0.0004(3) & 0.000(2) & 0.000(2)\\
\\ 
Orbital period, $P_b$ (d) & --- & --- & 0.35479073990(3) & --- \\ 
Epoch of periastron, $T_0$ (MJD) & --- & --- & 52500.25(3) & --- \\ 
Projected semi-major axis, $x$ (lt-s) & --- & --- & 0.3434304(4) & --- \\ 
Longitude of periastron, $\omega_0$ (deg) & --- & --- & 52(27) & --- \\ 
Orbital eccentricity, $e$ & --- & --- & 3.6(18)$\times 10^{-6}$ & --- \\ 
$\kappa = e \times \sin \omega_0$ & --- & --- & 2.9(20)$\times 10^{-6}$ & --- \\ 
$\eta = e \times \cos \omega_0$ & --- & --- & 2.2(16)$\times 10^{-6}$ & --- \\ 
Time of asc. node (MJD) & --- & --- & 52500.1940106(3) & --- \\ 
\\ 
Derived parameters\\ 
\\ 
Gal. longitude, $l$ (deg) & 0.4 & 3.1 & 27.7 & 14.8 \\ 
 Gal. latitude, $b$ (deg) & 6.8 & 6.0 & 17.7 & 9.2 \\ 
LK Px Distance, $d$ (pc) & --- & $904_{-216}^{+382}$& --- & $419_{-13}^{+14}$ \\ 
 Composite PM, $\mu$ (mas\,yr$^{-1}$) &26(16) & 23(5) & 8.65(12) & 21.009(15) \\ 
$\dot{P}_{\text{shk}} (\times 10^{-20}) $     & 0.7(9)      &0.9(5)       &0.15      &0.183(6)       \\
$\dot{P}_{\text{kz}}  (\times 10^{-20}) $     & $-$0.00298(5) &$-$0.004(1) &$-$0.024    &$-$0.00248(6)       \\
$\dot{P}_{\text{dgr}} (\times 10^{-20}) $     & 0.047(4)    &0.07(3)      &0.039     &0.013(1)       \\
$\dot{P}_{\text{int}} (\times 10^{-20}) $     & 0.0(7)      &1.0(6)       &2.24      &0.699(7)       \\
 Characteristic age, $\tau_c$ (Gyr) & $>7.9$ & 12.3 & 4.1 & 9.2 \\ 
 Surface magnetic field, $B$ ($\times 10^8$ G) & $<1.6$ & 2.9 & 3.7 & 1.7 \\ 
Min. companion mass (M$_{\odot}$) & --- & --- & 0.08 & --- \\
\hline
 \end{tabular}
\end{minipage}
\end{table*}

\subsection{PSR J1751$-$2857}
\citet{sfl+05} announced this wide ($P_b = 111$~days) binary MSP after
timing it for 4 years with an RMS of 28 $\mu s$ without a detection of the proper
motion. With 6 years of data at a much lower RMS, we are able to constrain its
proper motion ($\mu_{\alpha}=-7.4\pm0.1$ mas~yr$^{-1}$ and
$\mu_{\delta}=-4.3\pm1.2$ mas~yr$^{-1}$) and detect $\dot{x}=(4.6\pm0.8) \times
10^{-14}$.

\subsection{PSR J1801$-$1417}
This isolated MSP was discovered by  \citet{lfl+06}. With increased timing
precision, we measure a new composite proper motion $ \mu = 11.3\pm0.3$ mas
yr$^{-1}$. As our dataset for this pulsar does not include multifrequency
information; we can not rule out DM variations.

\subsection{PSR J1802$-$2124}
\citet{fsk+10} recently reported on the mass measurement of this system by
combining TOAs from the Green Bank, Parkes and Nan\c cay radio telescopes.
Therefore, our dataset shows no improvement in the determination of the system
parameters but gives consistent results to \citet{fsk+10}.

\subsection{PSR J1804$-$2717}
With an RMS timing residual improved by a factor 25 compared to the last
results published by \citet{hlk+04}, we obtain a reliable measurement of the
proper motion of this system. Assuming the distance based on the NE2001 model
$d_{\text{NE2001}} = 780$ pc, the parallax timing signature can amount to 1.5
\us, still below our current timing precision.

\begin{table*}
\caption{Timing model parameters for PSRs J1751$-$2857, J1801$-$1417,
J1802$-$2124 and J1804$-$2717. See caption of Table~\ref{tab:param1} for a
description of this table.}
\begin{minipage}{180mm}
\begin{tabular}{lllll}
\hline\hline
PSR Name & J1751$-$2857 & J1801$-$1417 & J1802$-$2124 & J1804$-$2717 \\ 
\hline
MJD range & 53746 --- 56782 & 54206 --- 56782 & 54188 --- 56831 & 53766 --- 56827 \\ 
 Number of TOAs & 144 & 126 & 522 & 116 \\ 
 RMS timing residual ($\mu s$) & 3.0 & 2.6 & 2.7 & 3.1 \\ 
 Reference epoch (MJD) & 55000 & 55000 & 55000 & 55000 \\ 
\\ 
Measured parameters\\ 
\\ 
Right ascension, $\alpha$ & 17:51:32.693197(17) & 18:01:51.073331(19) & 18:02:05.33522(2) & 18:04:21.133087(19) \\ 
Declination, $\delta$ & $-$28:57:46.520(3) & $-$14:17:34.526(2) & $-$21:24:03.653(8) & $-$27:17:31.335(4) \\ 
Proper motion in $\alpha$ (mas\,yr$^{-1}$) & $-$7.4(1) & $-$10.89(12) & $-$1.13(12) & 2.56(15) \\ 
Proper motion in $\delta$ (mas\,yr$^{-1}$) & $-$4.3(12) & $-$3.0(10) & $-$3(4) & $-$17(3) \\ 
Period, $P$ (ms) & 3.914873259435(3) & 3.6250967171671(17) & 12.6475937923794(16) & 9.343030844543(4) \\ 
Period derivative, $\dot{P}$ ($\times 10^{-20}$) & 1.121(3) & 0.530(3) & 7.291(3) & 4.085(5) \\ 
Parallax, $\pi$ (mas) & --- & --- & 1.24(57) & --- \\ 
DM (cm$^{-3}$pc) & 42.84(3) & 57.26(4) & 149.614(9) & 24.74(4) \\ 
DM1 (cm$^{-3}$pc yr$^{-1}$) & $-$0.01(1) & 0.004(7) & $-$0.002(2) & $-$0.005(6)\\
DM2 (cm$^{-3}$pc yr$^{-2}$) & 0.001(2) & 0.000(2) & 0.0005(6) & 0.000(1)\\
\\ 
Orbital period, $P_b$ (d) & 110.74646080(4) & --- & 0.698889254216(9) & 11.128711967(3) \\ 
Epoch of periastron, $T_0$ (MJD) & 52491.574(4) & --- & 52595.851(14) & 49615.080(9) \\ 
Projected semi-major axis, $x$ (lt-s) & 32.5282215(20) & --- & 3.718853(3) & 7.2814525(7) \\ 
Longitude of periastron, $\omega_0$ (deg) & 45.508(11) & --- & 29(7) & 158.7(3) \\ 
Orbital eccentricity, $e$ & 1.2795(3)$\times 10^{-4}$ & --- & 2.9(3)$\times 10^{-6}$ & 3.406(16)$\times 10^{-5}$ \\ 
$\kappa = e \times \sin \omega_0$ & --- & --- & 1.4(4)$\times 10^{-6}$ & --- \\ 
$\eta = e \times \cos \omega_0$ & --- & --- & 2.59(17)$\times 10^{-6}$ & --- \\ 
Time of asc. node (MJD) & --- & --- & 52595.79522502(4) & --- \\ 
\\ 
First derivative of $x$, $\dot{x}$  & 4.6(8)$\times 10^{-14}$ & --- & -3(5)$\times 10^{-15}$ & --- \\ 
Sine of inclination angle, $\sin{i}$ & --- & --- & 0.971(13) & --- \\ 
Companion mass, $m_c$ ($\text{M}_\odot$) & --- & --- & 0.83(19) & --- \\ 
\\ 
Derived parameters\\ 
\\ 
Gal. longitude, $l$ (deg) & 0.6 & 14.5 & 8.4 & 3.5 \\ 
 Gal. latitude, $b$ (deg) & $-$1.1 & 4.2 & 0.6 & $-$2.7 \\ 
LK Px Distance, $d$ (pc) & --- & --- & $640_{-195}^{+436}$& --- \\ 
 Composite PM, $\mu$ (mas\,yr$^{-1}$) &8.5(6) & 11.3(3) & 3(4) & 17(3) \\ 
$\dot{P}_{\text{shk}} (\times 10^{-20}) $     &0.077       &0.17       &0.02(5)     &0.53       \\
$\dot{P}_{\text{kz}}  (\times 10^{-20}) $     &$-$0.0002   &$-$0.0012  &$-$0.000083(4) &$-$0.0014       \\
$\dot{P}_{\text{dgr}} (\times 10^{-20}) $     &0.045       &0.05       &0.07(6)     &0.071       \\
$\dot{P}_{\text{int}} (\times 10^{-20}) $     &0.999       &0.31       &7.19(8)      &3.49       \\
 Characteristic age, $\tau_c$ (Gyr) & 6.2 & 18.5 & 2.8 & 4.2 \\ 
 Surface magnetic field, $B$ ($\times 10^8$ G) & 2. & 1.1 & 9.7 & 5.8 \\ 
Min. companion mass (M$_{\odot}$) & 0.18 & --- & 0.76 & 0.19 \\
\hline
 \end{tabular}
\end{minipage}
\end{table*}

\subsection{PSR J1843$-$1113}
This isolated pulsar discovered by \citet{hfs+04} is the second fastest-spinning
MSP in our dataset. Its mean flux density (S$_{1400} = 0.6$ mJy)
is among the lowest, limiting our current timing precision to $\sim$ 1 \us. For
the first time, we report the detection of the proper motion
$\mu_{\alpha}=-1.91\pm0.07$ mas~yr$^{-1}$ and $\mu_{\delta}=-3.2\pm0.3$
mas~yr$^{-1}$ and still low-precision parallax $\pi = 0.69 \pm 0.33$ mas.

\subsection{PSR J1853$+$1303}
Our values of proper motion and semi-major axis change are consistent with the
recent work by \citet{gsf+11} using high-sensitivity Arecibo and Parkes data,
though there is no evidence for the signature of the parallax in our data
 , most likely due to our less precise dataset.

\subsection{PSR J1857$+$0943 (B1855+09)}
Our measured parallax $\pi=0.7\pm0.26$ mas is lower than, but still compatible
with, the value reported by \citet{vbc+09}. We also report a marginal detection
of $\dot{x} = (-2.7 \pm 1.1) \times 10^{-15}$. Our measurement of the Shapiro
delay is also similar to the  previous result from \citet{vbc+09}.

\subsection{PSR J1909$-$3744}
PSR J1909$-$3744 \citep{jbv+03} is the most precisely timed source with a RMS
timing residual of about 100 ns. As these authors pointed out, this pulsar's
profile has a narrow peak with a pulse duty cycle of 1.5\% (43$\mu$s) at FWHM (see
Fig.~\ref{plot:templates-2}). Unfortunately its declination makes it only
visible with the NRT but it will be part of the SRT timing campaign. We improved
the precision of the measurement of the orbital period derivative $\dot{P_b}$ by a factor
of six compared to \citet{vbc+09} and our constraint on $\dot{x}$ is
consistent with their tentative detection.

\subsection{PSR J1910$+$1256}
We get similar results as recently published by \citet{gsf+11} with Arecibo and
Parkes data. In addition, we uncover a marginal signature of the parallax
$\pi = 1.44 \pm 0.74$ mas, consistent with the upper limit set by  \citet{gsf+11}.

\begin{table*}
\caption{Timing model parameters for PSRs J1843$-$1113, J1853$+$1303,
J1857$+$0943 and J1909$-$3744. See caption of Table~\ref{tab:param1} for a
description of this table.}
\begin{minipage}{180mm}
\begin{tabular}{lllll}
\hline\hline
PSR Name & J1843$-$1113 & J1853$+$1303 & J1857$+$0943 & J1909$-$3744 \\ 
\hline
MJD range & 53156 --- 56829 & 53763 --- 56829 & 50458 --- 56781 & 53368 --- 56794 \\ 
 Number of TOAs & 224 & 101 & 444 & 425 \\ 
 RMS timing residual ($\mu s$) & 0.71 & 1.6 & 1.7 & 0.13 \\ 
 Reference epoch (MJD) & 55000 & 55000 & 55000 & 55000 \\ 
\\ 
Measured parameters\\ 
\\ 
Right ascension, $\alpha$ & 18:43:41.261917(12) & 18:53:57.318765(12) & 18:57:36.390605(4) & 19:09:47.4335737(7) \\ 
Declination, $\delta$ & $-$11:13:31.0686(7) & 13:03:44.0693(4) & 09:43:17.20714(10) & $-$37:44:14.51561(3) \\ 
Proper motion in $\alpha$ (mas\,yr$^{-1}$) & $-$1.91(7) & $-$1.61(9) & $-$2.649(17) & $-$9.519(3) \\ 
Proper motion in $\delta$ (mas\,yr$^{-1}$) & $-$3.2(3) & $-$2.79(17) & $-$5.41(3) & $-$35.775(10) \\ 
Period, $P$ (ms) & 1.8456663232093(6) & 4.0917974456530(10) & 5.36210054870034(9) & 2.947108069766629(7) \\ 
Period derivative, $\dot{P}$ ($\times 10^{-20}$) & 0.9554(7) & 0.8724(14) & 1.78447(17) & 1.402518(14) \\ 
Parallax, $\pi$ (mas) & 0.69(33) & --- & 0.70(26) & 0.87(2) \\ 
DM (cm$^{-3}$pc) & 59.964(8) & 30.576(20) & 13.303(4) & 10.3925(4) \\ 
DM1 (cm$^{-3}$pc yr$^{-1}$) & 0.002(4) & 0.002(4) & 0.0017(2) & $-$0.00032(3)\\
DM2 (cm$^{-3}$pc yr$^{-2}$) & 0.0005(9) & $-$0.0005(8) & $-$0.00018(8) & 0.00004(1)\\
\\ 
Orbital period, $P_b$ (d) & --- & 115.65378824(3) & 12.3271713831(3) & 1.533449474329(13) \\ 
Epoch of periastron, $T_0$ (MJD) & --- & 52890.256(18) & 46432.781(3) & 53114.72(4) \\ 
Projected semi-major axis, $x$ (lt-s) & --- & 40.7695169(14) & 9.2307819(9) & 1.89799099(6) \\ 
Longitude of periastron, $\omega_0$ (deg) & --- & 346.65(6) & 276.47(7) & 180(9) \\ 
Orbital eccentricity, $e$ & --- & 2.368(3)$\times 10^{-5}$ & 2.170(4)$\times 10^{-5}$ & 1.22(11)$\times 10^{-7}$ \\ 
$\kappa = e \times \sin \omega_0$ & --- & --- & --- & $-$2.3(1900)$\times 10^{-10}$ \\ 
$\eta = e \times \cos \omega_0$ & --- & --- & --- & $-$1.22(11)$\times 10^{-7}$ \\ 
Time of asc. node (MJD) & --- & --- & --- & 53113.950741990(10) \\ 
\\ 
Orbital period derivative, $\dot{P_b}$ & --- & --- & --- & 5.03(5)$\times 10^{-13}$ \\ 
First derivative of $x$, $\dot{x}$  & --- & 2.4(7)$\times 10^{-14}$ & $-$2.7(11)$\times 10^{-15}$ & $0.6(17)\times 10^{-16}$ \\ 
Sine of inclination angle, $\sin{i}$ & --- & --- & 0.9987(6) & 0.99771(13) \\ 
Companion mass, $m_c$ ($\text{M}_{\odot}$) & --- & --- & 0.27(3) & 0.213(3) \\ 
\\ 
Derived parameters\\ 
\\ 
Gal. longitude, $l$ (deg) & 22.1 & 44.9 & 42.3 & 359.7 \\ 
 Gal. latitude, $b$ (deg) & $-$3.4 & 5.4 & 3.1 & $-$19.6 \\ 
LK Px Distance, $d$ (pc) & $1092_{-318}^{666}$& --- & $1098_{-254}^{+439}$& $1146_{-28}^{+30}$ \\ 
 Composite PM, $\mu$ (mas\,yr$^{-1}$) &3.8(3) & 3.22(15) & 6.03(3) & 37.020(10) \\ 
$\dot{P}_{\text{shk}} (\times 10^{-20}) $     &0.007(4)      &0.0091     &0.05(2)     &1.12(3)       \\
$\dot{P}_{\text{kz}}  (\times 10^{-20}) $     &$-$0.0004(3)  &$-$0.0016  &$-$0.0010(3)  &$-$0.0242(6)       \\
$\dot{P}_{\text{dgr}} (\times 10^{-20}) $     &0.014(9)      &$-$0.0028  &$-$0.0002(30)  &0.031(3)       \\
$\dot{P}_{\text{int}} (\times 10^{-20}) $     &0.94(1)       &0.868      &1.73(2)      &0.27(3)       \\
 Characteristic age, $\tau_c$ (Gyr) & 3.1 & 7.5 & 4.9 & 17.4 \\ 
 Surface magnetic field, $B$ ($\times 10^8$ G) & 1.3 & 1.9 & 3.1 & 0.9 \\ 
Min. companion mass (M$_{\odot}$) & --- & 0.22 & 0.22 & 0.18 \\
\hline
 \end{tabular}
\end{minipage}
\end{table*}

\subsection{PSR J1911$+$1347}
 With a pulse width at 50\% of the main peak amplitude (see
Fig.~\ref{plot:templates-2}), $W_{50} = 89 $ $\mu$s (only twice the width of
J1909$-$3744), this isolated MSP is potentially a good candidate for PTAs.
Unfortunately it has so far been observed at the JBO and NRT observatories only and
no multifrequency observations are available. Based on this work, this pulsar
has now been included in the observing list at the other EPTA telescopes.
Despite the good timing precision we did not detect the parallax  but we did
measure the proper motion for the first time with $\mu_{\alpha}=-2.90\pm0.04$
mas~yr$^{-1}$ and $\mu_{\delta}=-3.74\pm0.06$ mas~yr$^{-1}$.

\subsection{PSR J1911$-$1114}
The last ephemeris for this pulsar was published by \citet{tsb+99} 16 years
ago using the DE200 planetary ephemeris. Our EPTA dataset spans three times
longer than the one from \citet{tsb+99}. We hence report here on a greatly
improved position, proper motion ($\mu_{\alpha}=-13.75\pm0.16$ mas~yr$^{-1}$
and $\mu_{\delta}=-9.1\pm1.0$ mas~yr$^{-1}$) and a new eccentricity
$e=(1.6\pm1.0) \times 10^{-6}$, lower by a factor of 10 than the previous
measurement.

\subsection{PSR J1918$-$0642}
PSR J1918$-$0642 is another MSP studied by \citet{jsb+10} with EPTA data.
Compared to \citet{jsb+10} we extended the baseline with an additional five
years of data. We unveil the signature of Shapiro delay in this system
with $h_3=(8.6\pm1.2)\times 10^{-7}$ and $\varsigma =0.91\pm0.04$.
The masses of the system are discussed in Section~\ref{sec:dis_mass}.

\begin{table*}
\caption{Timing model parameters for PSRs J1910$+$1256, J1911$+$1347,
J1911$-$1114 and J1918$-$0642. See caption of Table~\ref{tab:param1} for a
description of this table.}
\begin{minipage}{180mm}
\begin{tabular}{lllll}
\hline\hline
PSR Name & J1910$+$1256 & J1911$+$1347 & J1911$-$1114 & J1918$-$0642 \\ 
\hline
MJD range & 53725 --- 56828 & 54095 --- 56827 & 53815 --- 57027 & 52095 --- 56769 \\ 
 Number of TOAs & 112 & 140 & 130 & 278 \\ 
 RMS timing residual ($\mu s$) & 1.9 & 1.4 & 4.8 & 3.0 \\ 
 Reference epoch (MJD) & 55000 & 55000 & 55000 & 55000 \\ 
\\ 
Measured parameters\\ 
\\ 
Right ascension, $\alpha$ & 19:10:09.701439(12) & 19:11:55.204679(5) & 19:11:49.28233(3) & 19:18:48.033114(7) \\ 
Declination, $\delta$ & 12:56:25.4869(4) & 13:47:34.38398(15) & $-$11:14:22.481(3) & $-$06:42:34.8896(4) \\ 
Proper motion in $\alpha$ (mas\,yr$^{-1}$) & 0.28(9) & $-$2.90(4) & $-$13.75(16) & $-$7.16(4) \\ 
Proper motion in $\delta$ (mas\,yr$^{-1}$) & $-$7.37(15) & $-$3.74(6) & $-$9.1(10) & $-$5.95(11) \\ 
Period, $P$ (ms) & 4.983584018674(3) & 4.6259625397749(6) & 3.625745633114(5) & 7.6458728874589(14) \\ 
Period derivative, $\dot{P}$ ($\times 10^{-20}$) & 0.9675(17) & 1.6927(9) & 1.395(4) & 2.5686(17) \\ 
Parallax, $\pi$ (mas) & 1.44(74) & --- & --- & --- \\ 
DM (cm$^{-3}$pc) & 38.094(11) & 30.987(6) & 31.02(11) & 26.610(11) \\ 
DM1 (cm$^{-3}$pc yr$^{-1}$) & $-$0.003(6) & 0.000(2) & $-$0.02(2) & 0.003(3)\\
DM2 (cm$^{-3}$pc yr$^{-2}$) & 0.0000(8) & $-$0.0002(5) & 0.003(3) & 0.0003(5)\\
\\ 
Orbital period, $P_b$ (d) & 58.466742964(14) & --- & 2.7165576619(7) & 10.9131777490(4) \\ 
Epoch of periastron, $T_0$ (MJD) & 54079.3152(14) & --- & 50456.5(3) & 51575.775(7) \\ 
Projected semi-major axis, $x$ (lt-s) & 21.1291036(7) & --- & 1.7628746(9) & 8.3504665(10) \\ 
Longitude of periastron, $\omega_0$ (deg) & 105.998(9) & --- & 121(34) & 219.60(20) \\ 
Orbital eccentricity, $e$ & 2.3023(4)$\times 10^{-4}$ & --- & 1.6(10)$\times 10^{-6}$ & 2.039(8)$\times 10^{-5}$ \\ 
$\kappa = e \times \sin \omega_0$ & --- & --- & 1.4(11)$\times 10^{-6}$ & --- \\ 
$\eta = e \times \cos \omega_0$ & --- & --- & $-$8.4(91)$\times 10^{-7}$ & --- \\ 
Time of asc. node (MJD) & --- & --- & 50455.6117845(13) & --- \\ 
\\ 
First derivative of $x$, $\dot{x}$  & $-$2.0(6)$\times 10^{-14}$ & --- & --- & $0.9(1.8)\times 10^{-15}$ \\ 
Third harmonic of Shapiro, $h_3$ ($\mu s$) & --- & --- & --- & 0.86(12) \\
Ratio of harmonics amplitude, $\varsigma$ & --- & --- & --- & 0.91(4) \\
\\ 
Derived parameters\\ 
\\ 
Gal. longitude, $l$ (deg) & 46.6 & 47.5 & 25.1 & 30.0 \\ 
 Gal. latitude, $b$ (deg) & 1.8 & 1.8 & $-$9.6 & $-$9.1 \\ 
LK Px Distance, $d$ (pc) & $554_{-186}^{+461}$& --- & --- & --- \\ 
 Composite PM, $\mu$ (mas\,yr$^{-1}$) &7.37(15) & 4.73(6) & 16.5(6) & 9.31(7) \\ 
$\dot{P}_{\text{shk}} (\times 10^{-20}) $     &0.04(3)      &0.052     &0.29     &0.2       \\
$\dot{P}_{\text{kz}}  (\times 10^{-20}) $     &$-$0.0002(1) &$-$0.00041 &$-$0.0084  &$-$0.016       \\
$\dot{P}_{\text{dgr}} (\times 10^{-20}) $     &$-$0.003(3)  &$-$0.031   &0.025    &0.039       \\
$\dot{P}_{\text{int}} (\times 10^{-20}) $     &0.93(3)      &1.67      &1.09     &2.35       \\
 Characteristic age, $\tau_c$ (Gyr) & 8.5 & 4.4 & 5.3 & 5.2 \\ 
 Surface magnetic field, $B$ ($\times 10^8$ G) & 2.2 & 2.8 & 2.0 & 4.3 \\ 
Min. companion mass (M$_{\odot}$) & 0.18 & --- & 0.11 & 0.22 \\
\hline
 \end{tabular}
\end{minipage}
\end{table*}

\subsection{PSR J1939$+$2134 (B1937+21)}
Thanks to the addition of early Nan\c cay DDS TOAs, our dataset span over 24
years for this pulsar. This pulsar has been long known to show significant DM
variations as well as a high level of timing noise  \citep{ktr94}; see
residuals in Fig.~\ref{plot:residuals-2}. A
possible interpretation of this red noise is the presence of an asteroid belt
around the pulsar \citep{scm+13}. Despite this red noise, the timing signature
of the parallax has successfully been extracted to get $\pi=0.22 \pm 0.08$ mas,
a value consistent with \citet{ktr94} and \citet{vbc+09}.

\subsection{PSR J1955$+$2908 (B1953+29)}
PSR J1955$+$2908 is another MSP recently analyzed by \citet{gsf+11}.  With an
independent dataset, we get similar results to \citet{gsf+11}. We report here on
the tentative detection of $\dot{x}=(4.0\pm1.4) \times 10^{-14}$.

\subsection{PSR J2010$-$1323}
This isolated MSP was discovered a decade ago \citep{jbo+07} and no update on
the pulsar's parameters has been published since then. Hence we announce here
the detection of the proper motion  $\mu_{\alpha}=-2.53\pm0.09$ mas~yr$^{-1}$
and $\mu_{\delta}=-5.7\pm0.4$ mas~yr$^{-1}$. Assuming the NE2001 distance of 1
kpc, the parallactic timing signature would amount to 1.17 $\mu s$ but was not
detected in our data.

\subsection{PSR J2019$+$2425}
Compared to the Arecibo 430-MHz dataset used by \citet{nss01}, the EPTA timing
precision for this pulsar is limited due to its low flux density at 1400 MHz. Because
of this we are not able to measure the secular change of the projected
semi-major axis $\dot{x}$.

\begin{table*}
\caption{Timing model parameters for PSRs J1939$+$2134, J1955$+$2908,
J2010$-$1323 and J2019+2425. See caption of Table~\ref{tab:param1} for a
description of this table.}
\begin{minipage}{180mm}
\begin{tabular}{lllll}
\hline\hline
PSR Name & J1939$+$2134 & J1955$+$2908 & J2010$-$1323 & J2019+2425 \\ 
\hline
MJD range & 47958 --- 56778 & 53813 --- 56781 & 54089 --- 56785 & 53451 --- 56788 \\ 
 Number of TOAs & 3174 & 157 & 390 & 130 \\ 
 RMS timing residual ($\mu s$) & 34.5 & 6.5 & 1.9 & 9.6 \\ 
 Reference epoch (MJD) & 55000 & 55000 & 55000 & 55000 \\ 
\\ 
Measured parameters\\ 
\\ 
Right ascension, $\alpha$ & 19:39:38.561224(2) & 19:55:27.87574(3) & 20:10:45.920637(11) & 20:19:31.94082(8) \\ 
Declination, $\delta$ & 21:34:59.12570(4) & 29:08:43.4599(6) & $-$13:23:56.0668(7) & 24:25:15.0130(19) \\ 
Proper motion in $\alpha$ (mas\,yr$^{-1}$) & 0.070(4) & $-$0.77(19) & 2.53(9) & $-$8.8(6) \\ 
Proper motion in $\delta$ (mas\,yr$^{-1}$) & $-$0.401(5) & $-$4.7(3) & $-$5.7(4) & $-$19.9(7) \\ 
Period, $P$ (ms) & 1.55780656108493(5) & 6.133166606620(5) & 5.2232710972195(3) & 3.934524144385(9) \\ 
Period derivative, $\dot{P}$ ($\times 10^{-20}$) & 10.51065(3) & 2.979(5) & 0.4832(6) & 0.695(7) \\ 
Parallax, $\pi$ (mas) & 0.22(8) & --- & --- & --- \\ 
DM (cm$^{-3}$pc) & 71.0237(13) & 104.54(6) & 22.174(11) & 17.17(12) \\ 
DM1 (cm$^{-3}$pc yr$^{-1}$) & 0.0000(4) & $-$0.00(1) & 0.0009(6) & $-$0.04(3)\\
DM2 (cm$^{-3}$pc yr$^{-2}$) & 0.00003(4) & $-$0.002(2) & $-$0.0004(3) & 0.004(4)\\
\\ 
Orbital period, $P_b$ (d) & --- & 117.34909924(8) & --- & 76.51163605(8) \\ 
Epoch of periastron, $T_0$ (MJD) & --- & 46112.470(4) & --- & 50054.652(12) \\ 
Projected semi-major axis, $x$ (lt-s) & --- & 31.412661(11) & --- & 38.767653(3) \\ 
Longitude of periastron, $\omega_0$ (deg) & --- & 29.452(10) & --- & 159.07(6) \\ 
Orbital eccentricity, $e$ & --- & 3.3021(7)$\times 10^{-4}$ & --- & 1.1113(11)$\times 10^{-4}$ \\ 
\\ 
First derivative of $x$, $\dot{x}$  & --- & 4.0(14)$\times 10^{-14}$ & --- & --- \\ 
\\ 
Derived parameters\\ 
\\ 
Gal. longitude, $l$ (deg) & 57.5 & 65.8 & 29.4 & 64.7 \\ 
 Gal. latitude, $b$ (deg) & $-$0.3 & 0.4 & $-$23.5 & $-$6.6 \\ 
LK Px Distance, $d$ (pc) & $3266_{-658}^{+1020}$& --- & --- & --- \\ 
 Composite PM, $\mu$ (mas\,yr$^{-1}$) &0.407(5) & 4.8(3) & 6.2(4) & 21.7(7) \\ 
$\dot{P}_{\text{shk}} (\times 10^{-20}) $     &0.00020(6)   &0.16     &0.051     &0.67       \\
$\dot{P}_{\text{kz}}  (\times 10^{-20}) $     &$-$0.0000069(26)  &$-$3e-05 &$-$0.056    &$-$0.0053       \\
$\dot{P}_{\text{dgr}} (\times 10^{-20}) $     &$-$0.04(2)     &$-$0.27    &0.02 &$-$0.042       \\
$\dot{P}_{\text{int}} (\times 10^{-20}) $     &10.55(2)      &3.09     &0.469     &0.0717       \\
 Characteristic age, $\tau_c$ (Gyr) & 0.2 & 3.1 & 17.7 & 87.0 \\ 
 Surface magnetic field, $B$ ($\times 10^8$ G) & 4.1 & 4.4 & 1.6 & 0.5 \\ 
Min. companion mass (M$_{\odot}$) & --- & 0.17 & --- & 0.29 \\
\hline
 \end{tabular}
\end{minipage}
\end{table*}

\subsection{PSR J2033$+$1734}
In spite of a narrow peak of width $\sim 160$ $\mu$s this MSP has a very large
timing RMS of 14 $\mu$s. With the absence of obvious systematics in the
residuals, we attribute the poor timing precision to the extremely low flux
density of this pulsar at 1400 MHz, $S_{1400} = 0.1$ mJy where all of our
observations were performed. Indeed this pulsar was discovered by
\citet{rtj+96} with the Arecibo telescope at 430 MHz and later followed up by
\citet{spl04} still at 430 and 820 MHz with the Green Bank 140-ft telescope.
Here we report with an independent dataset at 1400 MHz a similar proper motion
result to \citet{spl04}. 

\subsection{PSR J2124$-$3358}
For the isolated PSR J2124$-$3358, our measured  proper motion is consistent
with the already precise value published by \citet{vbc+09}. Our parallax
$\pi=2.50\pm0.36$ mas is also consistent with their results but with a better
precision.

\subsection{PSR J2145$-$0750}
Despite its rotational period of 16 ms PSR J2145$-$0750 is characterized by a
timing RMS of 1.8 $\mu$s  thanks to its narrow leading peak and large average
flux density, $S_{1400}=7.2$~mJy. The EPTA dataset does not show any
evidence for a variation of the orbital period  of PSR J2145$-$0750 or a
precession of periastron, even though \citet{vbc+09} reported a marginal
detection  with a slightly shorter data span characterized by a higher RMS
timing residual. On the other hand, we detect a significant
$\dot{x}=(8.2\pm0.7)\times10^{-15}$, which is not consistent with the
marginal detection, $\dot{x}=(-3\pm1.5)\times10^{-15}$, reported by \citet{vbc+09}. 

\subsection{PSR J2229$+$2643}
With eight years of data on PSR J2229$+$2643, we measure
$\mu_{\alpha}=-1.73\pm0.12$ mas~yr$^{-1}$ and $\mu_{\delta}=-5.82\pm0.15$
mas~yr$^{-1}$. Our measured $\mu_{\delta}$ is inconsistent with the last timing solution by
\citet{wdk+00} using the DE200 ephemeris ($\mu_{\alpha}=1\pm4$ mas~yr$^{-1}$ and $\mu_{\delta}=-17\pm4$
mas~yr$^{-1}$). Given our much smaller timing residual RMS, our use of the
superior DE421 model and longer baseline, we are confident our value is
more reliable. The expected timing signature of the parallax (0.7 \us) is too small
to be detected with the current dataset. Note that the early Effelsberg data
recorded with the EPOS backend included in \citet{wdk+00} are not part of this
dataset. 

\begin{table*}
\caption{Timing model parameters for PSRs J2033$+$1734, J2124$-$3358,
J2145$-$0750 and J2229+2643. See caption of Table~\ref{tab:param1} for a
description of this table.}
\begin{minipage}{180mm}
\begin{tabular}{lllll}
\hline\hline
PSR Name & J2033$+$1734 & J2124$-$3358 & J2145$-$0750 & J2229+2643 \\ 
\hline
MJD range & 53898 --- 56789 & 53365 --- 56795 & 50360 --- 56761 & 53790 --- 56796 \\ 
 Number of TOAs & 194 & 544 & 800 & 316 \\ 
 RMS timing residual ($\mu s$) & 12.7 & 3.2 & 1.8 & 4.2 \\ 
 Reference epoch (MJD) & 55000 & 55000 & 55000 & 55000 \\ 
\\ 
Measured parameters\\ 
\\ 
Right ascension, $\alpha$ & 20:33:27.51418(7) & 21:24:43.847820(11) & 21:45:50.460593(9) & 22:29:50.885423(18) \\ 
Declination, $\delta$ & 17:34:58.5249(17) & $-$33:58:44.9190(3) & $-$07:50:18.4876(4) & 26:43:57.6812(4) \\ 
Proper motion in $\alpha$ (mas\,yr$^{-1}$) & $-$5.9(5) & $-$14.04(8) & $-$9.58(4) & $-$1.73(12) \\ 
Proper motion in $\delta$ (mas\,yr$^{-1}$) & $-$9.1(8) & $-$50.14(14) & $-$8.86(10) & $-$5.82(15) \\ 
Period, $P$ (ms) & 5.948957630705(7) & 4.9311149439851(3) & 16.05242391938130(15) & 2.97781934162567(11) \\ 
Period derivative, $\dot{P}$ ($\times 10^{-20}$) & 1.108(9) & 2.0569(5) & 2.9788(3) & 0.1522(4) \\ 
Parallax, $\pi$ (mas) & --- & 2.50(36) & 1.53(11) & --- \\ 
DM (cm$^{-3}$pc) & 25.00(13) & 4.585(9) & 8.983(3) & 22.72(3) \\ 
DM1 (cm$^{-3}$pc yr$^{-1}$) & $-$0.03(2) & 0.0005(7) & 0.00019(5) & 0.0008(5)\\
DM2 (cm$^{-3}$pc yr$^{-2}$) & 0.002(4) & 0.0000(3) & 0.000006(26) & 0.0001(3)\\
\\ 
Orbital period, $P_b$ (d) & 56.30779617(7) & --- & 6.83890261532(19) & 93.01589390(5) \\ 
Epoch of periastron, $T_0$ (MJD) & 49878.125(11) & --- & 50313.7121(7) & 49419.709(3) \\ 
Projected semi-major axis, $x$ (lt-s) & 20.1631167(16) & --- & 10.1641056(3) & 18.9125228(5) \\ 
Longitude of periastron, $\omega_0$ (deg) & 78.09(7) & --- & 200.81(4) & 14.337(11) \\ 
Orbital eccentricity, $e$ & 1.2861(14)$\times 10^{-4}$ & --- & 1.9323(12)$\times 10^{-5}$ & 2.5525(5)$\times 10^{-4}$ \\ 
$\kappa = e \times \sin \omega_0$ & --- & --- & $-$6.866(12)$\times 10^{-6}$ & --- \\ 
$\eta = e \times \cos \omega_0$ & --- & --- & $-$1.8062(12)$\times 10^{-5}$ & --- \\ 
Time of asc. node (MJD) & --- & --- & 50309.89724107(6) & --- \\ 
\\ 
First derivative of $x$, $\dot{x}$  & --- & --- & 8.2(7)$\times 10^{-15}$ & --- \\ 
\\ 
Derived parameters\\ 
\\ 
Gal. longitude, $l$ (deg) & 60.9 & 10.9 & 47.8 & 87.7 \\ 
 Gal. latitude, $b$ (deg) & $-$13.2 & $-$45.4 & $-$42.1 & $-$26.3 \\ 
LK  Px Distance, $d$ (pc) & --- & $382_{-47}^{+61}$& $645_{-41}^{+47}$& --- \\ 
 Composite PM, $\mu$ (mas\,yr$^{-1}$) &10.8(7) & 52.07(14) & 13.05(8) & 6.07(15) \\ 
$\dot{P}_{\text{shk}} (\times 10^{-20}) $     &0.34     &1.2(2)     &0.43(3)      &0.038       \\
$\dot{P}_{\text{kz}}  (\times 10^{-20}) $     &$-$0.041 &$-$0.06(1) &$-$0.30(2)      &$-$0.053       \\
$\dot{P}_{\text{dgr}} (\times 10^{-20}) $     &$-$0.079 &0.008(1) &$-$0.007(1)    &$-$0.03       \\
$\dot{P}_{\text{int}} (\times 10^{-20}) $     &0.891    &0.9(2)   &2.85(1)      &0.198       \\
 Characteristic age, $\tau_c$ (Gyr) & 10.6 & 8.9 & 8.9 & 23.9 \\ 
 Surface magnetic field, $B$ ($\times 10^8$ G) & 2.3 & 2.1 & 6.9 & 0.8 \\ 
Min. companion mass (M$_{\odot}$) & 0.17 & --- & 0.39 & 0.11 \\
\hline
 \end{tabular}
\end{minipage}
\end{table*}

\subsection{PSR J2317$+$1439}
Compared to \citet{cnt96} we are able to constrain the proper motion
($\mu_{\alpha}=-1.19\pm0.07$ mas~yr$^{-1}$ and $\mu_{\delta}=3.33\pm0.13$
mas~yr$^{-1}$) and eccentricity $e=(5.7\pm1.6) \times 10^{-7}$ of the system
through the use of the ELL1 parametrization.  We also detect a marginal
signature of the parallax $\pi=0.7\pm0.3$ mas. 

\subsection{PSR J2322$+$2057}
PSR J2322$+$2057 is an isolated MSP with a pulse profile consisting of two
peaks separated by $\simeq$ 200\dg (see Fig.~\ref{plot:templates-2}).
\citet{nt95} were the last to publish a timing solution for this last source in
our dataset. We measure a proper motion consistent with their results albeit
with much greater precision, $\mu=24.0\pm0.4$  mas yr$^{-1}$.

\begin{table*}
\caption{Timing model parameters for PSRs J2317$+$1439 and J2322+2057. See
caption of Table~\ref{tab:param1} for a description of this table.}
\begin{minipage}{180mm}
\label{tab:param11}
\begin{tabular}{lll}
\hline\hline
PSR Name & J2317$+$1439 & J2322+2057 \\ 
\hline
MJD range & 50458 --- 56794 & 53905 --- 56788 \\ 
 Number of TOAs & 555 & 229 \\ 
 RMS timing residual ($\mu s$) & 2.4 & 5.9 \\ 
 Reference epoch (MJD) & 55000 & 55000 \\ 
\\ 
Measured parameters\\ 
\\ 
Right ascension, $\alpha$ & 23:17:09.236614(11) & 23:22:22.33516(7) \\ 
Declination, $\delta$ & 14:39:31.2563(4) & 20:57:02.6772(14) \\ 
Proper motion in $\alpha$ (mas\,yr$^{-1}$) & $-$1.19(7) & $-$18.4(4) \\ 
Proper motion in $\delta$ (mas\,yr$^{-1}$) & 3.33(13) & $-$15.4(5) \\ 
Period,  $P$ (ms) & 3.44525112564488(18) & 4.8084282894641(17) \\ 
Period derivative, $\dot{P}$ ($\times 10^{-20}$) & 0.2433(3) & 0.9661(20) \\ 
Parallax, $\pi$ (mas) & 0.7(3) & --- \\ 
DM (cm$^{-3}$pc) & 21.902(6) & 13.36(4) \\ 
DM1 (cm$^{-3}$pc yr$^{-1}$) & $-$0.0007(8) & $-$0.003(5) \\
DM2 (cm$^{-3}$pc yr$^{-2}$) & $-$0.0002(2) & $-$0.000(1) \\
\\ 
Orbital period, $P_b$ (d) & 2.45933150327(12) & --- \\ 
Epoch of periastron, $T_0$ (MJD) & 49300.92(11) & --- \\ 
Projected semi-major axis, $x$ (lt-s) & 2.31394874(18) & --- \\ 
Longitude of periastron, $\omega_0$ (deg) & 66(16) & --- \\ 
Orbital eccentricity, $e$ & 5.7(16)$\times 10^{-7}$ & --- \\ 
$\kappa = e \sin \omega_0$ & 5.2(16)$\times 10^{-7}$ & --- \\ 
$\eta = e \cos \omega_0$ & 2.3(16)$\times 10^{-7}$ & --- \\ 
Time of asc. node (MJD) & 49300.4724327(3) & --- \\ 
\\ 
Derived parameters\\ 
\\ 
Gal. longitude, $l$ (deg) & 91.4 & 96.5 \\ 
 Gal. latitude, $b$ (deg) & $-$42.4 & $-$37.3 \\ 
LK Px Distance, $d$ (pc) & $1011_{-220}^{+348}$& --- \\ 
 Composite PM, $\mu$ (mas\,yr$^{-1}$) &3.53(13) & 24.0(4) \\ 
$\dot{P}_{\text{shk}} (\times 10^{-20}) $     &0.011(4)      &0.54             \\
$\dot{P}_{\text{kz}}  (\times 10^{-20}) $     &$-$0.10(3)    &$-$0.09              \\
$\dot{P}_{\text{dgr}} (\times 10^{-20}) $     &$-$0.017(6)   &$-$0.02              \\
$\dot{P}_{\text{int}} (\times 10^{-20}) $     &0.35(4)       &0.538              \\
Characteristic age, $\tau_c$ (Gyr) & 15.6 & 14.2 \\ 
 Surface magnetic field, $B$ ($\times 10^8$ G) & 1.1 & 1.6 \\ 
Min. companion mass (M$_{\odot}$) & 0.16 & --- \\
\hline
 \end{tabular}
\end{minipage}
\end{table*}

\section{Discussion}
\label{sec:discussions}

\subsection{Distances}
\label{sec:dis_distances}

In Table~\ref{tab:distances}, we present the parallaxes measured from our
data, based on the distance-dependent curvature of the wave-front
coming from the pulsar. This curvature causes an arrival-time delay
$\tau$ (in seconds) with a periodicity of six months and a maximal amplitude of
\citep{lk04}:
\begin{equation}\label{eq:px}
\tau = \frac{d_{\odot}^2 \cos^2 \beta}{2 c d}
\end{equation}
where $d_{\odot}$ is the distance of the Earth to the Sun, $d$ is the
distance of the SSB to the pulsar, $c$ is the
speed of light and $\beta$ is the ecliptic latitude of the pulsar.

Because of the asymmetric error-volume, parallax measurements with
significance less than $\sim 4\sigma$, are unreliable as the
Lutz-Kelker bias dominates the measurement \citep{lk73,vlm10}. The
Lutz-Kelker-corrected parallax values as well as the derived
distances\footnote{We remind the reader that the most likely distance
is not necessarily equal to the inverse of the most likely parallax,
given the non-linearity of the inversion.} are also given in
Table~\ref{tab:distances}, based on the analytical corrections proposed by
\citet{vwc+12} and the flux density values shown in Table~\ref{tab:summary}.

In total, we present 22 new parallax measurements. Seven of these new
measurements are for MSPs that had no previous distance measurement,
but all of these are still strongly biased since their significance is
at best $3\sigma$. For five pulsars (specifically for PSRs J0030+0451,
J1012+5307, J1022+1001, J1643$-$1224 and J1857+0943) our parallax
measurement is of comparable significance than the previously
published value, but with the exception of PSR J1857+0943, our
measurement precision is better than those published previously; and
the lower significance is a consequence of the smaller parallax value
measured (as predicted by the bias-correction). Our measurement for
PSR J1857+0943 is slightly less precise than the value published by
\citet{vbc+09}, but consistent within $1\sigma$.

Finally, we present improved parallax measurements for ten pulsars:
PSRs J0613$-$0200, J0751+1807, J1024$-$0719, J1600$-$3053, J1713+0747,
J1744$-$1134, J1909$-$3744, J1939+2134, J2124$-$3358 and J2145$-$0750.
For seven of these the previous measurement was already free of bias,
for the remaining three (PSRs J0613$-$0200, J0751+1807 and
J2124$-$3358) our update reduces the bias to below the $1\sigma$
uncertainty level (with two out of three moving in the direction predicted
by the bias-correction code). For three pulsars with previously
published parallax measurements we only derive upper limits, but two
of these previous measurements (for PSRs J0218+4232 and J1853+1303)
were of low significance and highly biased. Only PSR J1738+0333's
parallax was reliably measured with GBT and Arecibo data
\citep{fwe+12} and not confirmed by us. Four pulsars had a known
parallax before the creation of the NE2001 model, namely PSRs J1713$+$0747
\citep{cfw94}, J1744$-$1134 \citep{tbm+99}, J1857$+$0943 and J1939$+$2134
\citep{ktr94}. These pulsars are therefore not included in our analysis of the
NE2001 distance (see below), leaving us with a total of 21 parallaxes.

\subsubsection{Distance comparison with NE2001 predictions}
\begin{figure}
\includegraphics[height=85mm,angle=-90]{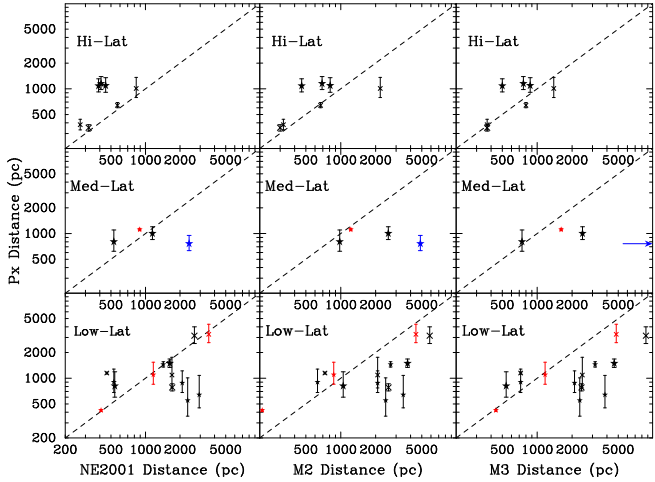}
\caption{Comparison between the Lutz-Kelker bias corrected parallax distances
(in ordinates) and the DM distances (in abscissa) for different Galactic
latitudes $b$ on logarithmic scales. The DM distances in the left, middle and
right panels are derived from the NE2001, M2 and M3 models respectively. Top
panels: the stars show pulsars with $b>40^{\circ}$ and the crosses pulsars with
$b<-40^{\circ}$. Middle panels: the stars show pulsars with
$40^{\circ}>b>20^{\circ}$ and the crosses pulsars with
$-40^{\circ}<b<-20^{\circ}$. Bottom panels: the stars show pulsars with
$20^{\circ}>b>0^{\circ}$ and the crosses pulsars with
$-20^{\circ}<b<0^{\circ}$. The red symbols indicate pulsars with a known
parallax before NE2001 was created, namely PSRs J1713$+$0747, J1744$-$1134, J1857$+$0943 and J1939$+$2134. The
blue symbol indicates PSR J1643$-$1224 where its corresponding M3 distance is
infinite and represented by an arrow. }
\label{fig:dist1}
\end{figure}

When comparing the bias-corrected distances presented in
Table~\ref{tab:distances} with those predicted by the widely used NE2001
 electron-density model for the Milky Way \citep{cl02},
 we find that the model performs reasonably well overall.
However, significant offsets exist, primarily at high positive
latitudes and large distance ($d>2$ kpc) into the Galactic plane.
In Fig~\ref{fig:dist1}, we plot this comparison for three ranges of Galactic
 latitude $b$ (defined as low: $\left|b\right| < 20^\circ$, intermediate:
 $20^\circ < \left|b\right| < 40^\circ$ and high: $\left|b\right| > 40^\circ$)
 highlighting the weakness of NE2001 at high latitude.
We find a mean uncertainty of 64\%, 55\% and 117\% respectively for the NE2001
distances to be consistent with our measurement.
 On average, the NE2001 distances would require an uncertainty of 80\%. This
value is significantly higher than the 25\% uncertainty typically assumed in
the literature for  this model; or than the fractional uncertainties displayed
in Figure~12 of  \citet{cl02}.

\subsubsection{Distance comparison with M2 and M3 predictions}

To improve on the shortcomings of NE2001, \citet[hereafter S12]{sch12} recently introduced 
two new models of the Galactic electron density based on \citet[][hereafter
TC93]{tc93} and NE2001, referred to as M2 and M3 in S12. In these two models, the author selected a set 
of 45 lines-of-sight to update the original  TC93 and NE2001 thick disk and 
fit for an exponential scale height of 1.59 and 1.31 kpc.
In the selection process of these 45 lines-of-sight, S12 excluded pulsars lying
in the Galactic plane, i.e. $|b|<5^{\circ}$; see Section 4.2 of S12 for
additional details.

 The distance estimates given by M2 and M3 are reported in the fourth and fifth
columns of Table~\ref{tab:distances}. Except for seven and five pulsars, respectively, the new M2 and M3 distances are
systematically higher than the NE2001 distances. In the case of PSR J1643$-$1224,
M3 even  predicts an infinite distance as it is unable to account for enough  
free electrons in the Galactic model towards this line-of-sight.

In Fig.~\ref{fig:dist1} we show the comparison between the parallax distances
and the NE2001, M2 and M3 distances as a function of the three Galactic latitude 
ranges defined in the previous section. As can be seen, the M2 and M3 predictions for
high latitude pulsars are a slightly better  match to the parallax distances than
NE2001. However, for low latitude, the M2 and M3 distances are significantly higher
than the parallax distances. To be consistent with the parallax distances, M2 
requires uncertainties of 95\%, 200\% and 53\% while M3 requires  113\%, 202\%  and
41\% for low, intermediate and high latitude respectively.  This result is  not surprising as low latitude
pulsars have been excluded in the S12 analysis. On average, M2 and M3 require an
uncertainty of 96\% and 102\%, significantly higher than our estimated uncertainty for
NE2001.

\begin{figure}
\includegraphics[width=90mm]{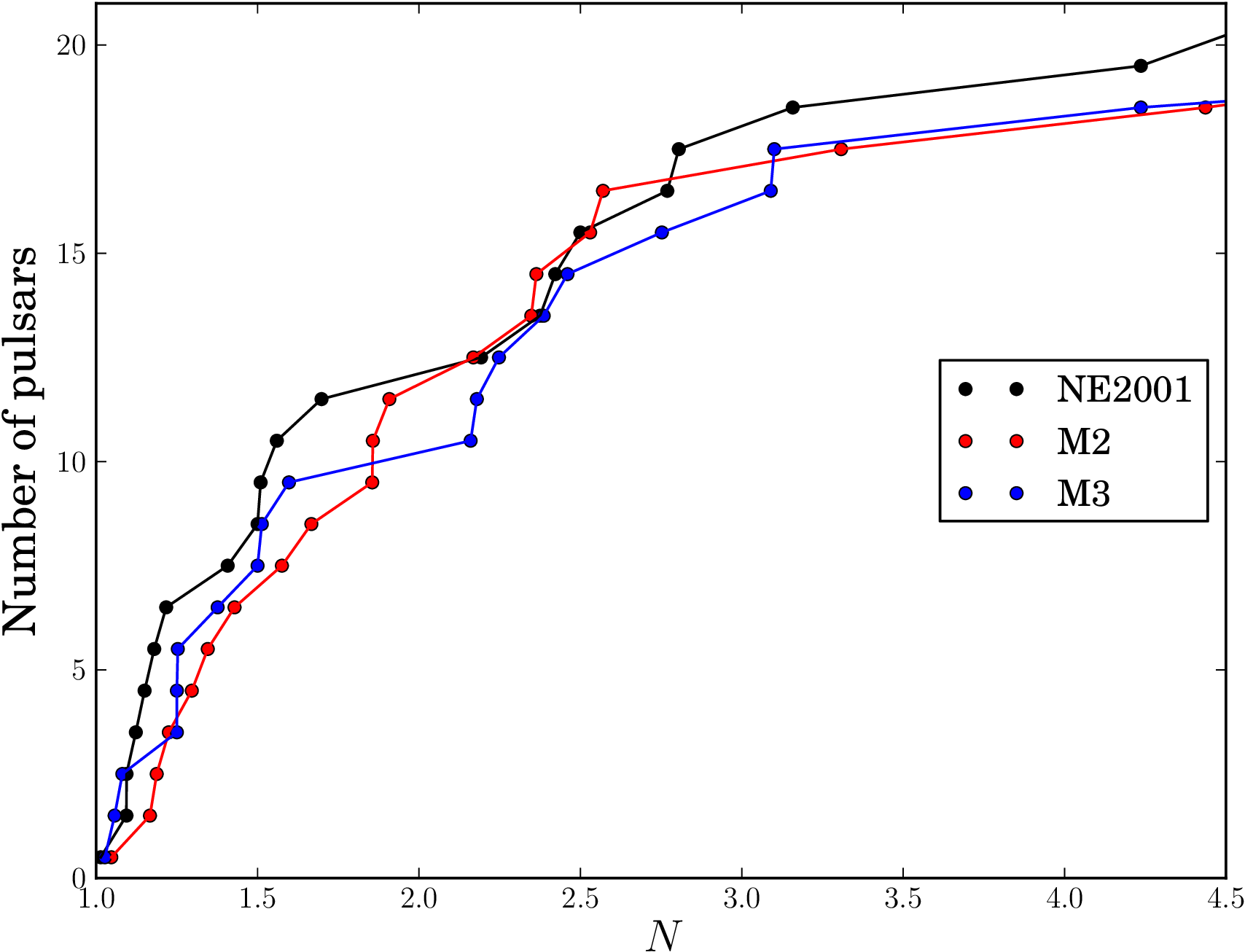}
\caption{Cumulative distribution of the $N$ factor between  
the DM distance and the parallax distance (see Eq.~\ref{eq:Nfactor}). These distributions include
the 21 pulsars with  measured parallaxes in Table~\ref{tab:distances}. The DM distances are derived 
from the NE2001, M2 and M3 models and represented in black, red and blue respectively.}
\label{fig:dist2}
\end{figure}

In Fig.~\ref{fig:dist2} we follow the method introduced by S12 to further compare the
quality of the DM models and plot 
the cumulative distribution of the $N$ factor:
\begin{equation}
\label{eq:Nfactor}
N = 
\begin{cases}
  {D_\text{model}}/{D_\pi}, & \text{if}\ D_\text{model} > D_\pi \\
  {D_\pi}/{D_\text{model}}, & \text{otherwise}
\end{cases}
\end{equation}
with $D_\pi$  and $D_\text{model}$ being the parallax distance and distance
from a given Galactic electron density model (NE2001, M2 or M3), respectively.
 As can be seen, the NE2001 model 
provides on average slightly better distance estimates (lower $N$)
than the M2 or M3 models. M3 gives more accurate distance than M2 for the first
half of lines-of-sight (when the prediction of both models is the best) but
gets superseded by M2 when N increases.


\begin{table*}
\begin{minipage}{180mm}
\caption{Summary of pulsar parallaxes and distance estimates. The columns give
the pulsar name, the DM, the distance based on the NE2001 electron density
model $D_{\text{NE2001}}$ \citep{cl02}, the distance based on the M2 and M3 models, 
$D_{\text{M2}}$ and $D_{\text{M3}}$ \citep{sch12}, an upper limit on the distance $D_{\dot{P}}$
(only indicated when this limit is $<15$ kpc; see text), the previously
published parallax value $\pi_{hist}$, our new measurement of the parallax
$\pi$ and the LK-bias corrected parallax $\pi_{corr}$ with the corresponding
distance  $D_{\pi}$.  For clarity, the values in bold show the updated or new
parallax measurements as part of this work.  The references for $\pi_{hist}$
can be found in Table \ref{tab:summary}.}
\label{tab:distances}
\begin{tabular}{crcccccccc}
\hline
PSR JName    &   DM   & $D_{\text{NE2001}}$ & $D_{\text{M2}}$ & $D_{\text{M3}}$ & $D_{\dot{P}}$ &  $\pi_{hist}$ & $\pi$ &    $\pi_{corr}$           & $D_{\pi}$ \\
	     &  (cm$^{-3}$pc) &  (kpc) &  (kpc)   &  (kpc)  &   (kpc) &  (mas)  &  (mas)   &(mas)         & (kpc)\\
\hline
J0030$+$0451 &   4.33 &  0.32 &	0.30 & 0.37 &  --- & $4.1\pm0.3$ &     $\mathbf{2.79\pm0.23}$ & $\mathbf{2.71_{-0.23}^{+0.23}}$ & $\mathbf{0.35_{-0.03}^{+0.03}}$ \\
J0034$-$0534 &  13.76 &  0.54 & 1.27 & 1.23 &  --- &--- & --- & --- & ---\\
J0218$+$4232 &  61.25 &  2.67 & 5.85 & 8.67 & --- & $0.16 \pm 0.09$ & --- & $0.22_{-0.05}^{+0.07}$ &
$3.15^{+0.85}_{-0.60}$\footnote{For PSR J0218$+$4232 the parallax was obtained
through VLBI observations \citep{dyc+14} but the inferred large distance was
later corrected by \citet{vl14} for the Lutz-Kelker bias.}  \\
J0610$-$2100 &  60.66 &  3.54 & 5.64 & 8.94 & $<3.85$ &  --- & --- & --- & ---\\
J0613$-$0200 &  38.78 &  1.71 & 2.58 & 2.41 & $<11.19$ &  $0.8 \pm 0.35$  & $\mathbf{1.25\pm0.13}$ & $\mathbf{1.21_{-0.13}^{+0.13}}$ & $\mathbf{0.78_{-0.07}^{+0.08}}$ \\
J0621$+$1002 &  36.45 &  1.36 & 2.02 & 1.90 & --- &  --- & --- & --- & ---\\
J0751$+$1807 &  30.25 &  1.15 & 2.57 & 2.46 & $<4.71$ &  $1.6\pm0.8$ &     $\mathbf{0.82\pm0.17}$ & $\mathbf{0.74_{-0.17}^{+0.17}}$ & $\mathbf{1.07_{-0.17}^{+0.24}}$ \\
J0900$-$3144 &  75.70 &  0.54 & 1.05 & 0.54 & --- &  --- &             $\mathbf{0.77\pm0.44}$ & $\mathbf{0.35_{-0.16}^{+0.32}}$ & $\mathbf{0.81_{-0.21}^{+0.38}}$ \\
J1012$+$5307 &   9.02 &  0.41 & 0.69 & 0.76 & $<2.14$ &  $1.22 \pm 0.26$&  $\mathbf{0.71\pm0.17}$ & $\mathbf{0.70_{-0.15}^{+0.15}}$ & $\mathbf{1.15_{-0.17}^{+0.24}}$ \\
J1022$+$1001 &  10.25 &  0.45 & 0.81 & 0.87 & $<3.3$ &  $1.8 \pm 0.3$ &  $\mathbf{0.72\pm0.20}$ & $\mathbf{0.70_{-0.17}^{+0.18}}$ & $\mathbf{1.09_{-0.18}^{+0.26}}$ \\
J1024$-$0719 &   6.49 &  0.39 & 0.46 & 0.50 & $<0.43$&  $1.9 \pm 0.4$ &  $\mathbf{0.80\pm0.17}$ & $\mathbf{0.75_{-0.16}^{+0.16}}$ & $\mathbf{1.08_{-0.16}^{+0.23}}$ \\
J1455$-$3330 &  13.56 &  0.53 & 0.98 & 0.74 & --- &  --- &             $\mathbf{1.04\pm0.35}$ & $\mathbf{0.49_{-0.24}^{+0.35}}$ & $\mathbf{0.80_{-0.18}^{+0.30}}$ \\
J1600$-$3053 &  52.32 &  1.63 & 3.77 & 4.62 & --- &  $0.2 \pm 0.15$ &  $\mathbf{0.64\pm0.07}$ & $\mathbf{0.62_{-0.07}^{+0.07}}$ & $\mathbf{1.49_{-0.15}^{+0.19}}$ \\
J1640$+$2224 &  18.42 &  1.16 & 1.61 & 2.63 & $<3.43$&  --- &             --- & --- & ---\\
J1643$-$1224 &  62.41 &  2.40 & 4.86 & $>50$& --- &  $2.2 \pm 0.4$ &   $\mathbf{1.17\pm0.26}$ & $\mathbf{0.99_{-0.27}^{+0.26}}$ & $\mathbf{0.76_{-0.13}^{+0.19}}$ \\
J1713$+$0747 &  15.99 &  0.89 & 1.22 & 1.61 & --- &  $0.94 \pm 0.05$&  $\mathbf{0.90\pm0.03}$ & $\mathbf{0.90_{-0.03}^{+0.03}}$ & $\mathbf{1.11_{-0.03}^{+0.04}}$ \\
J1721$-$2457 &  48.68 &  1.30 & 1.67 & 1.84 & $<0.96$&  --- &             --- & --- & ---\\
J1730$-$2304 &   9.61 &  0.53 & 0.63 & 0.72 & $<1.85$&  --- &             $\mathbf{0.86\pm0.32}$ & $\mathbf{0.21_{-0.07}^{+0.38}}$ & $\mathbf{0.90_{-0.22}^{+0.38}}$ \\
J1738$+$0333 &  33.80 &  1.43 & 2.60 & 3.16 & --- &  $0.68\pm 0.05$&   --- & $0.67_{-0.05}^{+0.05}$ & $1.45_{-0.10}^{+0.11}$\\
J1744$-$1134 &   3.13 &  0.41 & 0.21 & 0.44 & $<1.9$&  $2.4 \pm 0.1$ &   $\mathbf{2.38\pm0.08}$ & $\mathbf{2.37_{-0.08}^{+0.08}}$ & $\mathbf{0.42_{-0.01}^{+0.01}}$ \\ 
J1751$-$2857 &  42.90 &  1.10 & 1.51 & 1.73 & $<5.92$&  --- &             --- & --- & ---\\
J1801$-$1417 &  57.19 &  1.52 & 1.90 & 2.17 & $<3.47$&  --- &             --- & --- & ---\\
J1802$-$2124 & 149.63 &  2.94 & 3.46 & 3.84 & --- &  --- &             $\mathbf{1.24\pm0.57}$ & $\mathbf{0.08_{-0.03}^{+0.13}}$ & $\mathbf{0.64_{-0.19}^{+0.44}}$ \\
J1804$-$2717 &  24.57 &  0.78 & 1.29 & 1.13 & $<4.73$&  --- &             --- & --- & ---\\
J1843$-$1113 &  59.95 &  1.70 & 2.08 & 2.45 & --- &  --- &             $\mathbf{0.69\pm0.33}$ & $\mathbf{0.11_{-0.04}^{+0.15}}$ & $\mathbf{1.09_{-0.32}^{+0.67}}$ \\
J1853$+$1303 &  30.65 &  2.08 & 1.08 & 2.10 & --- &  $1.0\pm 0.3$ &    --- & $0.19_{-0.08}^{+0.42}$ & $0.88_{-0.20}^{+0.34}$ \\
J1857$+$0943 &  13.30 &  1.17 & 0.87 & 1.17 & --- &  $1.1\pm 0.2$ &    $\mathbf{0.70\pm0.26}$ & $\mathbf{0.20_{-0.10}^{+0.31}}$ & $\mathbf{1.10_{-0.25}^{+0.44}}$ \\
J1909$-$3744 &  10.39 &  0.46 & 0.73 & 0.72 & $<1.42$& $0.79\pm 0.02$ &  $\mathbf{0.87\pm0.02}$ & $\mathbf{0.87_{-0.02}^{+0.02}}$ & $\mathbf{1.15_{-0.03}^{+0.03}}$ \\
J1910$+$1256 &  38.10 &  2.33 & 2.44 & 2.33 & ---  & --- &            $\mathbf{1.44\pm0.74}$ & $\mathbf{0.11_{-0.04}^{+0.11}}$ & $\mathbf{0.55_{-0.19}^{+0.46}}$ \\
J1911$+$1347 &  30.98 &  2.07 & 1.88 & 2.07 & ---  &             --- & --- & ---\\
J1911$-$1114 &  30.97 &  1.23 & 2.01 & 1.86 & $<6.01$&  --- &             --- & --- & ---\\
J1918$-$0642 &  26.54 &  1.24 & 1.79 & 1.75 & --- &  --- &             --- & --- & ---\\
J1939$+$2134 &  71.02 &  3.56 & 4.45 & 4.81 & --- &  $0.13 \pm 0.07$&  $\mathbf{0.22\pm0.08}$ & $\mathbf{0.19_{-0.06}^{+0.07}}$ & $\mathbf{3.27_{-0.66}^{+1.02}}$ \\
J1955$+$2908 & 104.55 &  4.64 & 6.73 & 6.75 & --- &  --- &             --- & --- & ---\\
J2010$-$1323 &  22.18 &  1.02 & 1.78 & 1.95 & --- &  --- &             --- & --- & ---\\
J2019$+$2425 &  17.15 &  1.50 & 1.16 & 1.50 & $<1.67$&  --- &             --- & --- & ---\\
J2033$+$1734 &  25.01 &  2.00 & 1.85 & 2.17 & $<10.51$ &  --- &             --- & --- & ---\\
J2124$-$3358 &   4.58 &  0.27 & 0.32 & 0.37 & $<0.67$ &  $3.1\pm0.6$ &     $\mathbf{2.50\pm0.36}$ & $\mathbf{2.31_{-0.36}^{+0.36}}$ & $\mathbf{0.38_{-0.05}^{+0.06}}$ \\ 
J2145$-$0750 &   8.98 &  0.57 & 0.67 & 0.79 & --- &  $1.6\pm0.3$ &     $\mathbf{1.53\pm0.11}$ & $\mathbf{1.51_{-0.11}^{+0.11}}$ & $\mathbf{0.64_{-0.04}^{+0.05}}$ \\
J2229$+$2643 &  22.66 &  1.43 & 1.94 & 2.23 & --- &  --- & --- & --- & ---\\
J2317$+$1439 &  21.90 &  0.83 & 2.19 & 1.39 & --- &  --- &             $\mathbf{0.7\pm0.3}$ & $\mathbf{0.55_{-0.18}^{+0.24}}$ & $\mathbf{1.01_{-0.22}^{+0.35}}$ \\
J2322$+$2057 &  13.55 &  0.80 & 1.06 & 1.12 & $<1.81$&  --- & --- & --- & ---\\
\hline
\end{tabular}
\end{minipage}
\end{table*}


\subsection{Proper motions and 2-D spatial velocities}
\label{sec:dis_vel}

Stellar evolution modeling by \citet{tb96} and \citet{cc97} predicted that the
recycled MSP population would have a smaller spatial velocity than the normal
pulsar population. A study by \citet{tsb+99} found a mean transverse velocity
\vt for MSPs of $85\pm13$~km~s$^{-1}$ based on a sample of 23 objects. They
noted that this  value is four times lower than the ordinary young pulsar
velocity. The authors also observed isolated MSPs to have a velocity two-thirds
smaller than the binary MSPs. 
With an ever increasing number of MSPs, further studies by \citet{hll+05} and
\citet{gsf+11} found no statistical evidence for a difference in the velocity
distribution of isolated and binary MSPs. \citet{hll+05} reported on \vt
$=76\pm16$ and \vt $=89\pm15$   km~s$^{-1}$ for isolated and binary MSPs
respectively while \citet{gsf+11} found \vt $=68\pm16$  and \vt $=96\pm15$
km~s$^{-1}$ for isolated and binary MSPs. 
All these results are in agreement with other  work by \citet{lkn+06}.

Within our sample of 42 MSPs, we measured seven new proper motions, of which
three are for 
isolated MSPs (PSRs J1843$-$1113, J1911$+$1347 and J2010$-$1323) and 4 are
for binary MSPs (PSRs J0034$-$0534, J0900$-$3144, J1751$-$2857 and J1804$-$2717).
In addition, we improved the precision of the proper-motion measurement by a
factor of ten for seven other MSPs (PSRs J0610$-$2100, J0613$-$0200, J1455$-$3330,
J1801$-$1417, J1911$-$1114, J2229$+$2643 and J2317$+$1439). 

These improvements in the proper motion as well as the distance estimates
presented in Section~\ref{sec:dis_distances} and recent discoveries of MSPs
published elsewhere led us to re-examine the distribution of $V_T$, the transverse
velocity of MSPs in km~s$^{-1}$, where

\begin{equation}
  V_T =\text{ 4.74 km s}^{-1} \times \mu \times d.
\end{equation}
Again, $\mu$ is the proper motion in mas yr$^{-1}$ and $d$ the distance to the
pulsar in kpc. In this
analysis we considered all known MSPs listed in the ATNF pulsar catalogue, but
discarding pulsars in globular clusters, double neutron stars or pulsars with
$P>20$ ms. This represents  19 isolated and 57 binary pulsars for a total of 76
MSPs. In comparison, the last published MSP velocity study by \citet{gsf+11}
made use of 10 isolated and 27 binary MSPs with $P$ below 10 ms. If we choose
to restrict our sample to pulsars with $P$ below 10 ms,  only 6 binary pulsars
would not pass our criteria. The selected isolated and binary pulsars are
listed in Tables \ref{tab:iso_velocities} and \ref{tab:bin_velocities}
respectively. The distances used in the calculation of $V_T$ and reported in
the third column of Tables \ref{tab:iso_velocities} and
\ref{tab:bin_velocities} are the best distance estimates available, either
coming from the Lutz-Kelker-corrected parallax or the NE2001 model.

We find an average velocity of $88\pm17$ km~s$^{-1}$ and $93\pm13$ km~s$^{-1}$
for the isolated and binary MSPs,  respectively. For the entire MSP dataset, we
get an average velocity of $92\pm10$ km~s$^{-1}$. Our results are consistent
with the work by \citet{hll+05} and \citet{gsf+11}.

When we keep only the pulsars with a more reliable distance estimate (i.e.
pulsars with a parallax measurement), 8 isolated and 20
binary  MSPs are left in our sample. In this case, we find an average velocity
of $75\pm10$ km~s$^{-1}$ and 
$56\pm3$ km~s$^{-1}$ for the isolated and binary MSPs respectively. 
Conversely, we get an average velocity of $98\pm29$ km~s$^{-1}$ and $113\pm20$
km~s$^{-1}$ for the pulsars with a distance coming from the Galactic electron
density models. The explanations for this discrepancy are twofold: the
NE2001 model is overestimating the distances for low Galactic latitude as shown
in Fig.~\ref{fig:dist1} and our sample of 2-D velocities is biased against
distant low-velocity MSPs. Nearby
pulsars are likely to have a parallax and a proper-motion measurement whereas
distant pulsars would most likely have a distance estimate from the NE2001 model and
a proper-motion measurement for the high-velocity pulsars only.

\begin{figure}
\includegraphics[height=80mm,angle=-90]{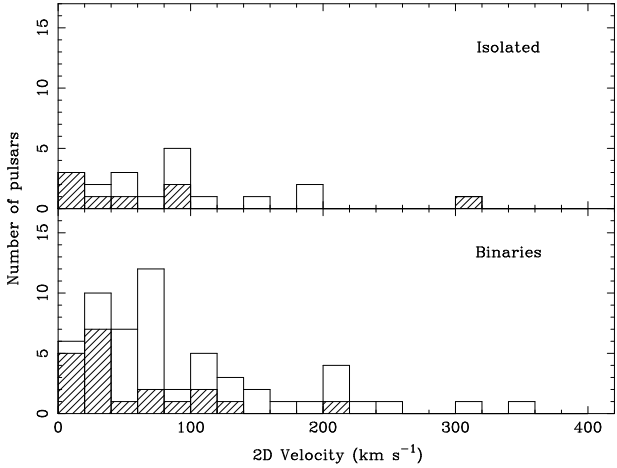}
\caption{Histogram of the 2-D velocity distribution for a sample of 19 isolated
MSPs (top panel) and 57 binary MSPs (bottom panel). The respective average
velocities are $88\pm17$ km s$^{-1}$ and $96\pm12$ km s$^{-1}$. The hatched
part of the histogram shows the pulsars with a distance estimate from the
parallax measurement (8 isolated and 21 binary MSPs).}
\label{fig:2D_hist}
\end{figure}

Fig.~\ref{fig:2D_hist} shows the histogram of the velocities for both the
isolated and binary MSPs populations. A two-sample Kolgomorov-Smirnov (KS) test
between the full isolated and binary MSPs velocity distributions results in a
KS-statistic of 0.14 and a p-value of 0.92. If we perform the same test on the
pulsars with a parallax distance, we get a KS-statistic of 0.25 and a p-value
of 0.81. For both cases, we therefore cannot reject the null hypothesis and we argue that there is
no statistical evidence for the measurements to be drawn from different
distributions. This supports the scenario that both isolated and binary MSPs
evolve from the same population of binary pulsars.



\begin{table*}
\caption{Summary of the transverse motion of the isolated MSPs. The columns
indicate the pulsar name, the composite proper motion, the distance and the
corresponding transverse velocity. The last column shows the last reference
with published proper motion and distance measurements. The distances refer to
the best distance estimates available, either the parallax when uncertainties
are given or the NE2001 distance (indicated by $^\dag$) where a 80\% error is
assumed.
Values in bold face indicate the new proper-motion measurements.}
\label{tab:iso_velocities}
\begin{tabular}{ccrrr}
\hline
PSR JName & $\mu$        & Distance  &  2D Velocity       & Reference\\ 
          & (mas yr$^{-1})$ & (pc)  &  (km s$^{-1})$      & \\ 
\hline
J0030$+$0451	&	 $5.9 \pm 0.5$ 		&	 $ 350 \pm   30$ 	&	 $ 9.8 \pm  1.2$ 	&	This work\\
J0645$+$5158	&	 $7.60 \pm 0.20$ 	&	 $ 700 \pm  200$ 	&	 $  25 \pm    7$ 	&	\citet{slr+14}\\
J0711$-$6830	&	 $21.08 \pm 0.08$ 	&	 $860^\dag$ 	&	$  86 \pm   69$ 	&	\citet{vbc+09}\\
J1024$-$0719	&	 $59.72 \pm 0.06$ 	&	 $1080\pm230$ 	&	 $ 306 \pm   65$ 	&	This work\\
J1453$+$1902	&	 $7.5 \pm 2.2$ 		&	 $1150^\dag$ 	&	 $  41 \pm   35$ 	&	\citet{lmcs07}\\
J1721$-$2457	&	 $25.5 \pm 15.3$ 	&	 $1300^\dag$ 	&	 $ 157 \pm  157$ 	&	This work\\
J1730$-$2304	&	 $22.6 \pm 4.8$ 	&	 $ 900 \pm  300$ 	&	 $  96 \pm   38$ 	&	This work\\
J1744$-$1134	&	 $21.009 \pm 0.014$ 	&	 $ 420 \pm   10$ 	&	 $41.8 \pm  1.0$ 	&	This work\\
J1801$-$1417	&	 $11.30 \pm 0.27$ 	&	 $1520^\dag$ 	&	 $  81 \pm   65$ 	&	This work\\
J1843$-$1113	&	 $\mathbf{3.76 \pm 0.22}$ 	&	 $1090 \pm  670$ 	&	 $  19 \pm   12$ 	&	This work\\
J1905$+$0400	&	 $8.2 \pm 0.4$ 	&		 $1700^\dag$ 	&	 $  66 \pm   53$ 	&	\citet{gsf+11}\\
J1911$+$1347	&	 $\mathbf{4.73 \pm 0.05}$ 	&	 $2070^\dag$ 	&	 $  46 \pm   37$ 	&	This work\\
J1923$+$2515	&	 $24.3 \pm 6.8$ 	&	 $1630^\dag$ 	&	 $ 188 \pm  159$ 	&	\citet{lbr+13}\\
J1939$+$2134	&	 $0.407 \pm 0.005$ 	&	 $3270 \pm 1020$ 	&	 $ 6.3 \pm  2.0$ 	&	This work\\
J1944$+$0907	&	 $21.6 \pm 2.5$ 	&	 $1790^\dag$ 	&	 $ 183 \pm   148$ 	&	\citet{clm+05}\\
J1955$+$2527	&	 $3.1 \pm 0.7$ 		&	 $7510^\dag$ 	&	 $ 110 \pm   92$ 	&	\citet{dfc+12}\\
J2010$-$1323	&	 $\mathbf{6.24 \pm 0.33}$ 	&	 $1030^\dag$ 	&	 $  30 \pm   24$ 	&	This work\\
J2124$-$3358	&	 $52.07 \pm 0.13$ 	&	 $ 380 \pm   60$ 	&	 $  94 \pm   15$ 	&	This work\\
J2322$+$2057	&	 $24.0 \pm 0.4$ 	&	 $800^\dag$ 	&	 $  91 \pm   73$ 	&	This work\\
\hline
\end{tabular}
\end{table*}

\begin{table*}
\caption{Summary of the transverse motion of the binary MSPs. The columns
indicate the pulsar name, the composite proper motion, the distance and the
corresponding transverse velocity. The last column shows the last reference
with published proper motion and distance measurements. The distances refer to
the best distance estimates available, either the parallax when uncertainties
are given or the NE2001 distance (indicated by $^\dag$) where a 80\% error is
assumed.
Values in bold face indicate the new proper-motion measurements.}
\label{tab:bin_velocities}
\begin{tabular}{ccrrr}
\hline
PSR JName & $\mu$        & Distance  &  2D Velocity       & Reference\\ 
          & (mas yr$^{-1})$ & (pc)  &  (km s$^{-1})$      & \\ 
\hline
J0034$-$0534	&	 $12.1 \pm 0.5$ 	&	 $540^\dag$ 	&	 $  31 \pm   25$ 	&	This work\\
J0101$-$6422	&	 $15.6 \pm 1.7$ 	&	 $560^\dag$ 	&	 $  41 \pm   33$ 	&	\citet{kcj+12}\\
J0218$+$4232	&	 $6.18 \pm 0.09$ 	&	 $3150^\dag$ 	&	 $  92 \pm   25$ 	&	\citet{vl14}\\
J0437$-$4715	&	 $141.29 \pm 0.06$ 	&	 $156.0 \pm  1.0$ 	&	 $104.5 \pm  0.7$ 	&	\citet{dvtb08}\\
J0610$-$2100	&	 $19.05 \pm 0.11$ 	&	 $3540^\dag$ 	&	 $ 320 \pm  256$ 	&	This work\\
J0613$-$0200	&	 $10.514 \pm 0.016$ 	&	 $ 780 \pm   80$ 	&	 $  39 \pm    4$ 	&	This work\\
J0636$+$5129	&	 $4.7 \pm 0.9$ 		&	 $490^\dag$ 	&	 $  11 \pm    9$ 	&	\citet{slr+14}\\
J0751$+$1807	&	 $13.66 \pm 0.23$ 	&	 $1070 \pm  240$ &	 $  69 \pm   16$ 	&	This work\\
J0900$-$3144	&	 $\mathbf{2.26 \pm 0.06}$ 	&	 $ 810 \pm  380$ 	&	 $   9 \pm    4$ 	&	This work\\
J1012$+$5307	&	 $25.615 \pm 0.010$ 	&	 $1150 \pm  240$ 	&	 $ 140 \pm   29$ 	&	This work\\
J1017$-$7156	&	 $9.96 \pm 0.06$ 	&	 $2980^\dag$ 	&	 $ 141 \pm   113$ 	&	\citet{nbb+14}\\
J1023$+$0038	&	 $17.98 \pm 0.04$ 	&	 $1370 \pm   40$ 	&	 $116.8 \pm  3.4$ 	&	\citet{dab+12}\\
J1045$-$4509	&	 $8.01 \pm 0.20$ 	&	 $1960^\dag$ 	&	 $  74 \pm   60$ 	&	\citet{vbc+09}\\
J1125$-$5825	&	 $10.28 \pm 0.30$ 	&	 $2620^\dag$ 	&	 $ 128 \pm   102$ 	&	\citet{nbb+14}\\
J1231$-$1411	&	 $104.4 \pm 2.2$ 	&	 $440^\dag$ 	&	 $ 218 \pm  100$ 	&	\citet{rcc+11}\\
J1300$+$1240	&	 $96.15 \pm 0.07$ 	&	 $450^\dag$ 	&	$ 205 \pm   164$ 	&	\citet{kw03}\\
J1337$-$6423	&	 $9.2 \pm 5.4$ 		&	 $5080^\dag$ 	&	 $ 222 \pm  220$ 	&	\citet{nbb+14}\\
J1405$-$4656	&	 $48.3 \pm 6.9$ 	&	 $580^\dag$ 	&	 $ 133 \pm   108$ 	&	\citet{btb+15}\\
J1431$-$4715	&	 $10.6 \pm 3.6$ 	&	 $1560^\dag$ 	&	 $  78 \pm   68$ 	&	\citet{btb+15}\\
J1446$-$4701	&	 $4.47 \pm 0.22$ 	&	 $1460^\dag$ 	&	 $  31 \pm   25$ 	&	\citet{nbb+14}\\
J1455$-$3330	&	 $\mathbf{8.19 \pm 0.09}$ 	&	 $ 800 \pm  300$ 	&	 $  31 \pm   12$ 	&	This work\\
J1543$-$5149	&	 $5.9 \pm 1.7$ 		&	 $2420^\dag$ 	&	 $  68 \pm   58$ 	&	\citet{nbb+14}\\
J1600$-$3053	&	 $7.00 \pm 0.07$ 	&	 $1490 \pm  190$ 	&	 $  49 \pm    6$ 	&	This work\\
J1603$-$7202	&	 $7.84 \pm 0.09$ 	&	 $1170^\dag$ 	&	 $  43 \pm   35$ 	&	\citet{vbc+09}\\
J1640$+$2224	&	 $11.485 \pm 0.030$ 	&	 $1160^\dag$ 	&	 $  63 \pm   51$ 	&	This work\\
J1643$-$1224	&	 $7.28 \pm 0.08$ 	&	 $ 760 \pm  190$ 	&	 $  26 \pm    7$ 	&	This work\\
J1708$-$3506	&	 $5.7 \pm 1.3$ 		&	 $2790^\dag$ 	&	 $  75 \pm   63$ 	&	\citet{nbb+14}\\
J1709$+$2313	&	 $10.2 \pm 0.9$ 	&	 $1410^\dag$ 	&	 $  68 \pm   55$ 	&	\citet{lwf+04}\\
J1713$+$0747	&	 $6.2865 \pm 0.0032$ 	&	 $1110 \pm   40$ 	&	 $33.1 \pm  1.2$ 	&	This work\\
J1719$-$1438	&	 $11.2 \pm 2.0$ 	&	 $1210^\dag$ 	&	 $  64 \pm   53$ 	&	\citet{nbb+14}\\
J1731$-$1847	&	 $6.2 \pm 2.9$ 		&	 $2550^\dag$ 	&	 $  75 \pm   69$ 	&	\citet{nbb+14}\\
J1738$+$0333	&	 $8.65 \pm 0.12$ 	&	 $1470 \pm  100$ 	&	 $  60 \pm    4$ 	&	\citet{fwe+12}\\
J1745$-$0952	&	 $23.9 \pm 2.5$ 	&	 $1830^\dag$ 	&	 $ 207 \pm  167$ 	&	\citet{jsb+10}\\
J1745$+$1017	&	 $7.8 \pm 1.0$ 		&	 $1260^\dag$ 	&	 $  47 \pm   38$ 	&	\citet{bgc+13}\\
J1751$-$2857	&	 $\mathbf{8.5 \pm 0.6}$ &	 $1110^\dag$ 	&	 $  45 \pm   36$ 	&	This work\\
J1801$-$3210	&	 $13.6 \pm 8.2$ 	&	 $4030^\dag$ 	&	 $ 260 \pm  260$ 	&	\citet{nbb+14}\\
J1802$-$2124	&	 $3.5 \pm 3.2$ 		&	 $ 640 \pm  440$ 	&	 $  11 \pm   12$ 	&	This work\\
J1804$-$2717	&	 $\mathbf{17.3 \pm 2.4}$&	 $780^\dag$ 	&	 $  64 \pm   52$ 	&	This work\\
J1816$+$4510	&	 $6.1 \pm 0.9$ 		&	 $2410^\dag$ 	&	 $  70 \pm   57$ 	&	\citet{slr+14}\\
J1853$+$1303	&	 $3.22 \pm 0.15$ 	&	 $ 880 \pm  340$ 	&	 $  13 \pm    5$ 	&	\citet{gsf+11}\\
J1857$+$0943	&	 $6.028 \pm 0.022$ 	&	 $1100 \pm  440$ 	&	 $  31 \pm   13$ 	&	This work\\
J1903$+$0327	&	 $5.60 \pm 0.11$ 	&	 $6360^\dag$ 	&	 $ 169 \pm   135$ 	&	\citet{fbw+11}\\
J1909$-$3744	&	 $37.020 \pm 0.009$ 	&	 $1150 \pm   30$ 	&	 $ 202 \pm    5$ 	&	This work\\
J1910$+$1256	&	 $7.37 \pm 0.15$ 	&	 $ 550 \pm  460$ 	&	 $  19 \pm   16$ 	&	This work\\
J1911$-$1114	&	 $16.5 \pm 0.5$ 	&	 $1220^\dag$ 	&	 $  95 \pm   76$ 	&	This work\\
J1918$-$0642	&	 $9.31 \pm 0.07$ 	&	 $1240^\dag$ 	&	 $  55 \pm   44$ 	&	This work\\
J1949$+$3106	&	 $5.95 \pm 0.08$ 	&	 $6520^\dag$ 	&	 $ 184 \pm   147$ 	&	\citet{dfc+12}\\
J1955$+$2908	&	 $4.75 \pm 0.26$ 	&	 $4640^\dag$ 	&	 $ 104 \pm   84$ 	&	This work\\
J1959$+$2048	&	 $30.4 \pm 0.6$ 	&	 $2490^\dag$ 	&	 $ 359 \pm  287$ 	&	\citet{aft94}\\
J2019$+$2425	&	 $21.7 \pm 0.7$ 	&	 $1490^\dag$ 	&	 $ 153 \pm   123$ 	&	This work\\
J2033$+$1734	&	 $10.8 \pm 0.7$ 	&	 $2000^\dag$ 	&	 $ 102 \pm   82$ 	&	This work\\
J2043$+$1711	&	 $13.0 \pm 2.0$ 	&	 $1760^\dag$ 	&	 $ 108 \pm   88$ 	&	\citet{gfc+12}\\
J2051$-$0827	&	 $5.3 \pm 1.0$ 		&	 $1040^\dag$ 	&	 $  26 \pm   21$ 	&	\citet{dlk+01}\\
J2129$-$5721	&	 $13.31 \pm 0.10$ 	&	 $ 420 \pm  200$ 	&	 $  26 \pm   13$ 	&	\citet{vbc+09}\\
J2145$-$0750	&	 $13.05 \pm 0.07$ 	&	 $ 640 \pm   50$ &	 $40 \pm  3$ 	&	This work\\
J2229$+$2643	&	 $6.07 \pm 0.14$ 	&	 $1430^\dag$ 	&	 $  41 \pm   33$ 	&	This work\\
J2317$+$1439	&	 $3.53 \pm 0.12$ 	&	 $1010 \pm  350$ 	&	 $  17 \pm    6$ 	&	This work\\
\hline
\end{tabular}
\end{table*}


\subsection{Shklovskii and Galactic acceleration contributions}
\label{sec:dis_shk}

The observed pulse period derivatives, $\dot{P}$, reported in Tables \ref{tab:param1}
 to \ref{tab:param11} are different from their intrinsic values
$\dot{P}_{\text{int}}$. This is because it  includes the 'Shklovskii' contribution due
to the transverse velocity of the pulsar \citep[$\dot{P}_{\text{shk}}$,][]{shk70}, the
acceleration from the differential Galactic rotation ($\dot{P}_{\text{dgr}}$) and the
acceleration towards the Galactic disk ($\dot{P}_{\text{kz}}$) \citep{dt91,nt95}.
Hence $\dot{P}_{\text{int}}$ can be written as

\begin{equation}
\label{eq:pdot}
  \dot{P}_{\text{int}}=\dot{P}-\dot{P}_{\text{shk}}-\dot{P}_{\text{dgr}}-\dot{P}_{\text{kz}},
\end{equation}
where the Shklovskii contribution $\dot{P}_{\text{shk}}$ is given by
\begin{equation}
  \frac{\dot{P}_{\text{shk}}}{P} = \frac{\mu^2 d}{c}.
\end{equation}
Again $d$ is our best distance estimate for the pulsar and $\mu$ our measured
composite proper motion. The equation for $\dot{P}_{\text{dgr}}$
is taken from \citet{nt95} with updated values for the distance to the Galactic center
$R_0=8.34\pm0.16$ kpc and the Galactic rotation speed at the Sun
$\Theta=240\pm8$~km~s$^{-1}$ \citep{rmb+14}. $\dot{P}_{\text{kz}}$ is taken from the
linear interpolation of the $K_z$ model in \citet[see Fig.~8]{hf04}.

To compute these contributions with full error propagation we use  the distances from
Table~\ref{tab:distances} and the proper motions shown in
Tables~\ref{tab:iso_velocities} and \ref{tab:bin_velocities}. These values are
reported for each pulsar at the bottom of Tables~\ref{tab:param1} to
\ref{tab:param11}. The magnitudes of all three corrective
terms to $\dot{P}$ depend on the distance $d$ to the pulsar. 
Alternatively, as the pulsar braking torque causes the spin period to increase
(i.e. $\dot{P}$ to be positive) in systems where no mass transfer is taking
place, we used this constraint to set an upper limit, $D_{\dot{P}}$, on the
distance to the pulsar by assuming all the observed $\dot{P}$ is a result of
kinematic and Galactic acceleration effects. This upper limit  $D_{\dot{P}}$ is
shown in column 5 of Table~\ref{tab:distances} for 19 pulsars, where this upper
limit is below 15 kpc.

For all pulsars except PSRs J0610$-$2100, J1024$-$0719 and J1721$-$2457, the
upper limits $D_{\dot{P}}$ are consistent with both the NE2001 and M3
distances, $D_{\text{NE2001}}$ and $D_{\text{M3}}$ respectively. For PSR
J0610$-$2100, $D_{\text{M3}}=8.94$  kpc is ruled out by
$D_{\dot{P}} < 3.89$ kpc. We note that for this pulsar, $D_{\text{M3}}$ is 2.5 times higher than
$D_{\text{NE2001}}$. For PSR J1721$-$2457, both $D_{\text{NE2001}}$ and
$D_{\text{M3}}$ are ruled out by $D_{\dot{P}} < 0.96$ kpc. The case of PSR J1024$-$0719 is discussed below.


\begin{table*}
\begin{minipage}{180mm}
\centering
\caption{Summary of the kinematic and relativistic contributions to the
observed orbital period derivative $\dot{P_b}$. The columns indicate the pulsar
name, the Lutz-Kelker-corrected parallax distance (we made the errors symmetric by always taking the highest of the two error estimates given in Table~\ref{tab:distances}), the observed orbital period
derivative $\dot{P_b}$,  the contributions to $\dot{P_b}$ from the Shklovskii
effect, Galactic potential, differential Galactic rotation and gravitational
wave radiation assuming GR. The last column shows the estimated distance assuming all
$\dot{P_b}$ arises from these contributions.
$^{\dag}$ Assuming $m_p=1.4$ M$_\odot$ and $i=60^\circ$.
$^{\ddag}$ For PSR J1012$+$5307, we take $m_c=0.16\pm0.02$ M$_{\odot}$  and $i=52\pm4^{\circ}$ from
\citet{kbk96,cgk98}.}
\label{tab:pbdot}
\begin{tabular}{cccccccc}
\hline
PSR JName    &  $D_{\pi}$  &  $\dot{P_b}$  &  $\dot{P_b}_{\_kin}$  & $\dot{P_b}_{\_kz}$  & $\dot{P_b}_{\_dgr}$  &  $\dot{P_b}_{\_GR}$  & $D_{\dot{P_b}}$    \\
	     &  (kpc)      &  $(\times 10^{-13})$  & $(\times 10^{-13})$  &  $(\times 10^{-13})$ &  $(\times 10^{-13})$  &  $(\times 10^{-13})$&   (kpc)\\
\hline
J0613$-$0200 & 0.78(8)	&  0.48(10)  &  +0.217(22) & -0.0133(14) & +0.034(4)   & -0.03$^{\dag}$     & 1.68(33)     \\
J0751$+$1807 & 1.07(24)   & -0.350(25) &  +0.110(25) & -0.0104(12) & +0.0125(28) &  ---          &   ---   \\
J1012$+$5307 & 1.15(24) &  0.61(4)   &  +0.96(20)  & -0.076(7)   & +0.0157(33) & -0.112(29)$^{\ddag}$ & 0.94(3) \\
J1909$-$3744 & 1.15(3)  &  5.03(5)   &  +5.07(13)  & -0.1092(29) & +0.139(12)  & -0.0291(7) & 1.140(11) \\
\hline
\end{tabular}
\end{minipage}
\end{table*}

 For nine pulsars, an independent estimate of
the distance from the parallax measurement is available. For all nine pulsars
but PSR J1024$-$0719, the parallax distance is consistent with the upper limit
$D_{\dot{P}}$.
PSR J1024$-$0719 has $D_{\dot{P}} < 0.42$ kpc but a reported Lutz-Kelker-corrected
distance $D_{\pi} = 1.08_{-0.16}^{+0.28}$ kpc, $\sim4\sigma$ away above the
upper limit $D_{\dot{P}}$.
To explain this discrepancy (also discussed in \citet{egc+13,aaa+13}), we argue that PSR J1024$-$0719 must be subject to
a minimum relative acceleration $a$ along the line of sight,
\begin{equation} 
 a = \frac{ \left| \dot{P} - \dot{P}_{\text{int}} \right| }{P} \times c = 1.7 \times 10^{-9} \text{m s}^{-2}.
\end{equation}
A possible explanation for this acceleration is the presence of a nearby star,
orbiting PSR J1024$-$0719 in a very long period. A possible companion has been
identified by \citet{srr+03}.

The same reasoning behind the corrections of Eq.~\ref{eq:pdot} also apply to the
observed orbital period derivative $\dot{P_b}$.  In addition to the previous
terms, we also consider the contribution due to gravitational radiation
assuming GR, $\dot{P_b}_{\_GR}$ but neglect the contributions from mass loss in the
binary, tidal interactions or changes in the gravitational constant $G$.
 $\dot{P_b}_{\_GR}$ is therefore the only
contribution independent of the distance to the pulsar system but requires an
estimate of the masses of the binary.

 As we measured the orbital period derivative for four pulsars
(PSRs J0613$-$0200, J0751$+$1807, J1012$+$5307 and J1909$-$3744), we investigate here
the possible bias in those measurements assuming the parallax distances from
Table~\ref{tab:distances}.  Conversely, \citet{bb96} (hereafter BB96) pointed out that the
measurement of $\dot{P_b}$ would potentially lead to more accurate distance
than the annual parallax. Hence, we also present a new distance estimate,
$D_{\dot{P_b}}$, assuming the observed $\dot{P_b}$ is the sum of all four
contributions described above.  These results are shown in Table~\ref{tab:pbdot}. 

 To estimate the gravitational radiation contribution to $\dot{P_b}$ for PSR
J0613$-$0200 without a mass measurement, we assumed $m_p=1.4$ M$_\odot$
and $i=60^\circ$. The resulting distance estimate is  $D_{\dot{P_b}} = 1.68\pm0.33$ kpc. This result is
2.2$\sigma$ consistent with the parallax distance and currently limited by the precision
on the measured $\dot{P_b}$. Continued timing of this pulsar will greatly
improve this test as the uncertainty on $\dot{P_b}$ decrease as $t^{-2.5}$.
For PSR J0751+1807, we measure a negative orbital period derivative,
$\dot{P_b}=(-3.50\pm0.25) \times 10^{-14}$, meaning the Shklovskii effect is not
the dominant contribution to  $\dot{P_b}$ in this system. We also note that our
measured composite proper motion is 3.3$\sigma$ higher than the value in \citet{nss+05}
resulting in a Shklovskii contribution to $\dot{P_b}$ that is five times larger than the one quoted
in \citet{nss+05}.
In the next section, we will combine the corrected orbital period derivative from
acceleration bias  $\dot{P_b}_{\text{corr}} =  \dot{P_b} -
\dot{P_b}_{\_\text{kin}} -
\dot{P_b}_{\_\text{kz}} - \dot{P_b}_{\_\text{dgr}} = (-4.6\pm0.4) \times 10^{-14} $  with the measurement of the Shapiro delay to constrain the masses of the two stars.

For PSR J1012$+$5307, we measured the orbital period derivative
$\dot{P_b}=(6.1\pm0.4) \times 10^{-14}$, a value similar to the one reported by \citet{lwj+09}. We
also find the contributions to $\dot{P_b}$ to be consistent with their work.
 After taking into account the companion mass and
inclination angle from \citet{kbk96,cgk98} to compute $\dot{P_b}_{\_\text{GR}}$, we find
$D_{\dot{P_b}} = 940\pm30$ pc, in very good agreement with the optical
\citep{kbk96,cgk98} and parallax distance, but more precise by a factor three and
eight, respectively.

The bias in the orbital period derivative measured for PSR J1909$-$3744 is
almost solely due to the Shklovskii effect. We get $D_{\dot{P_b}} = 1140\pm11$
pc. This result with a fractional uncertainty of only 1\% is also in very good
agreement with the parallax distance.

Twenty years ago, BB96 predicted that after only 10 years, several of the MSPs included in
this paper would have a better determination of the distance  through the measurement of
the Shklovskii contribution to $\dot{P_b}$ compared to the annual parallax.
However  we achieved a better distance estimate from
$\dot{P_b}$ than the parallax for only two pulsars so far.

We investigate here the pulsars for which we should have detected $\dot{P_b}$
based on the work by BB96 (i.e. PSRs J1455$-$3330, J2019+2425
J2145$-$0750 and J2317+1439). 
In their paper, BB96 assumed a transverse velocity of 69 km~s$^{-1}$
for pulsars where the proper motion was not measured and adopted the distance to the
pulsar based on the \citet{tc93} Galactic electron density model.

In the case of PSR J2019+2425, our measured proper motion is similar to the
value used by BB96 and the time span of our data is nine years.
  The peak-to-peak timing signature of the Shklovskii contribution
to $\dot{P_b}$ (see Eq.~1 of BB96) is $\Delta T_{pm} = 6\pm5~\mu s$,
with the large uncertainty coming from the NE2001 distance assumed.
For the three remaining pulsars, no proper-motion measurement was available at
the time and BB96 assumed in those cases a transverse velocity of 69
km~s$^{-1}$. However our new results 
reported in Table~\ref{tab:bin_velocities} show much smaller transverse
velocities for PSRs J1455$-$3330, J2145$-$0750 and J2317+1439, with
$31\pm12$ km s$^{-1}$, $40\pm3$ km s$^{-1}$,
$17\pm6$ km s$^{-1}$ respectively, resulting in a much lower Shklovskii
contribution to $\dot{P_b}$ than predicted by BB96, explaining the non-detection of this parameter after 10 to 17 years
of data with our current timing precision.

\subsection{Shapiro delay and mass measurement}
\label{sec:dis_mass}

\begin{table*}
\begin{minipage}{150mm}
\centering
\caption{Table of pulsar and companion masses. The columns indicate the pulsar
name, the previously published pulsar and companion mass (Prev. $m_p$ and Prev. $m_c$)
with the corresponding publication. The last 2 columns show our new
measurements, $m_p$ and $m_c$. $^\dag$ \citet{nsk08} did not report on the companion mass in
their proceedings. $^\ddag$ The pulsar masses were not reported by \citet{vbc+09} so
we quote the pulsar mass value based on the mass function and their companion
mass.}
\label{tab:mass}
\begin{tabular}{cccccc}
\hline
PSR JName &    Prev. $m_p$    &   Prev. $m_c$    &  Ref. &     $m_p$   &   $m_c$   \\
          &    (M$_{\odot}$)    &   (M$_{\odot}$)    &       & (M$_{\odot}$) & (M$_{\odot}$)   \\
\hline
J0751$+$1807 & $1.26\pm0.14$ & ---$^\dag$ & \citet{nss01,nsk08} & $1.64_{-0.15}^{+0.15}$ & $0.16_{-0.01}^{+0.01}$ \\
J1600$-$3053 & $0.87^\ddag$	& $0.6\pm0.7$	& \citet{vbc+09} & $1.22_{-0.35}^{+0.50}$ & $0.21_{-0.043}^{+0.06}$	\\
J1713$+$0747 & $1.31\pm0.11 $ & $0.286\pm0.012$ & \citet{zsd+15} & $1.33_{-0.08}^{+0.09}$ & $0.289_{0.011}^{+0.013}$ \\
J1802$-$2124 & $1.24\pm0.11$ & $0.78\pm0.04$ & \citet{fsk+10} &  $1.25_{-0.4}^{+0.6}$ & $0.80_{-0.16}^{+0.21}$ \\
J1857$+$0943 & $1.61^\ddag$	& $0.270\pm0.015$ & \citet{vbc+09} & $1.59_{-0.18}^{+0.21}$ & $0.268_{-0.019}^{+0.022}$\\
J1909$-$3744 & $1.53^\ddag$	& $0.212\pm0.002$ & \citet{vbc+09} & $1.54_{-0.027}^{+0.027}$ & $0.213_{-0.0024}^{+0.0024}$ \\
J1918$-$0642 & ---	&  ---  & ---	& $1.25_{-0.38}^{+0.61}$	& $0.227_{-0.046}^{+0.066}$ \\
\hline
\end{tabular}
\end{minipage}
\end{table*}

\begin{figure*}
\centering
\begin{minipage}{150mm}
\includegraphics[width=150mm]{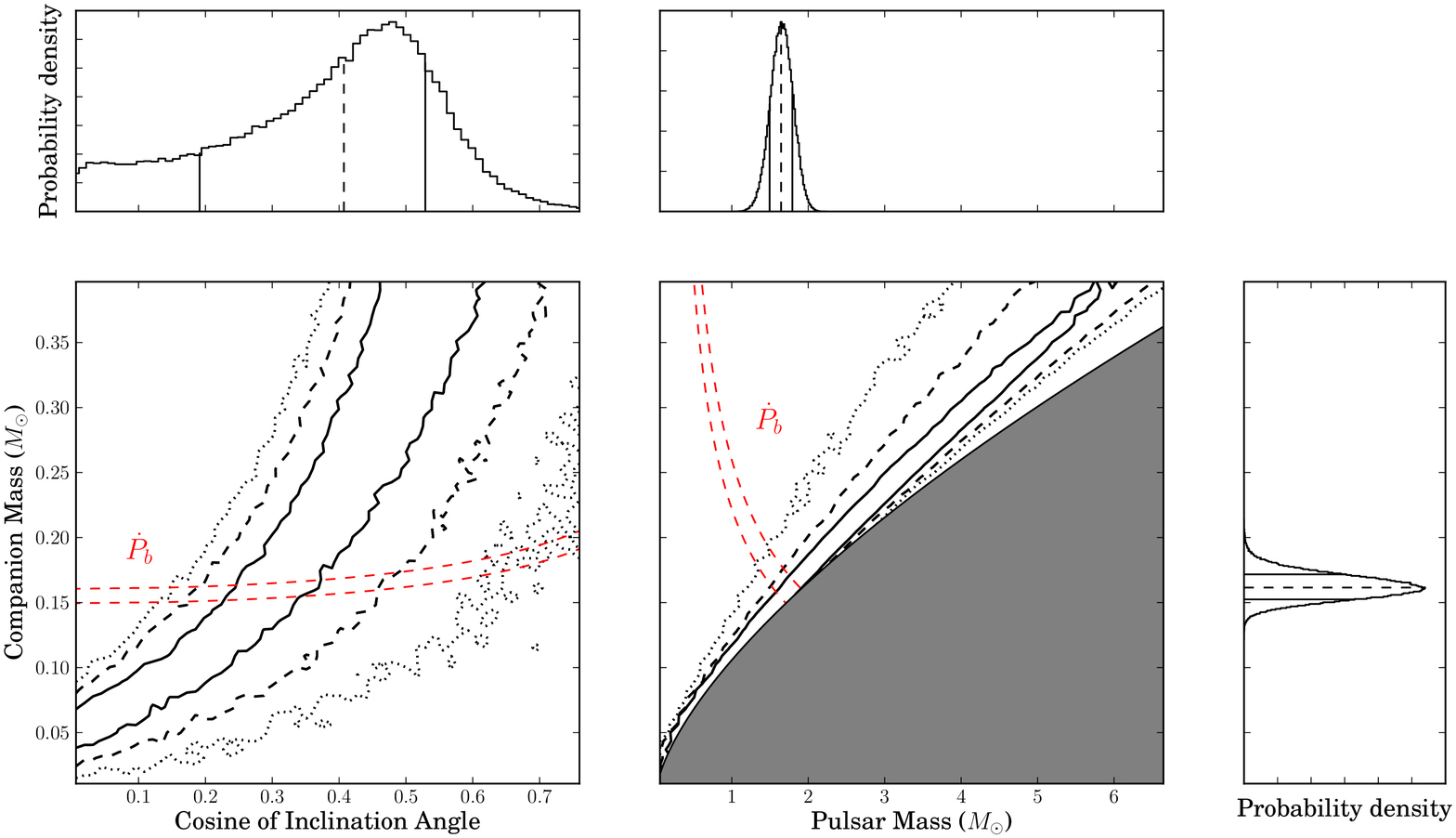}
\caption{Constraints on PSR J0751$+$1807 parameters  from the
measurement of Shapiro delay and orbital period derivative $\dot{P_b}$. The bottom left plot shows the $\cos i - m_c$
plane. The bottom right plot shows the $m_p - m_c$ plane.  The continuous black
line, the dashed line and the dotted line represent, respectively, the 68.3\%,
95.4\% and 99.7\% confidence levels of the 2-D probability density function.
The grey area is
excluded by the mass function with the condition $\sin i \leq  1$. The red
curves indicate the 1-$\sigma$ constraint required by $\dot{P_b}$ assuming GR.  The other
three panels show the projected 1-D distributions based on $\dot{P_b}$ and the
inclination angle (given by $\varsigma$). The dashed lines
indicate the median value and the continuous lines the 1-$\sigma$ contours.}
\label{fig:0751_mass}
\end{minipage}
\end{figure*}

\begin{figure*}
\centering
\begin{minipage}{150mm}
\includegraphics[width=150mm]{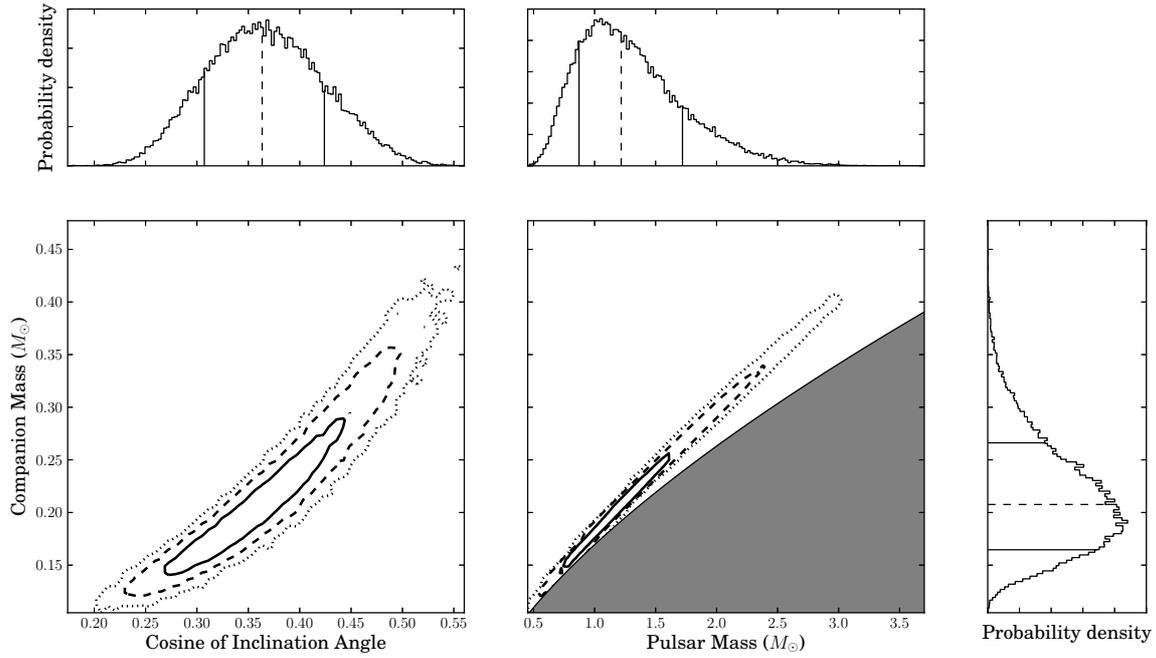}
\caption{Constraints on PSR J1600$-$3053 parameters  from the
measurement of Shapiro delay. The bottom left plot shows the $\cos i - m_c$
plane. The bottom right plot shows the $m_p - m_c$ plane.  The continuous black
line, the dashed line and the dotted line represent, respectively, the 68.3\%,
95.4\% and 99.7\% confidence levels of the 2-D probability density function.
The grey area is
excluded by the mass function with the condition $\sin i \leq  1$. The other
three panels show the projected 1-D distributions with the dashed line
indicating the median value and the continuous lines the 1-$\sigma$ contours.}
\label{fig:1600_mass}
\end{minipage}
\end{figure*}

\begin{figure*}
\begin{minipage}{150mm}
\includegraphics[width=150mm]{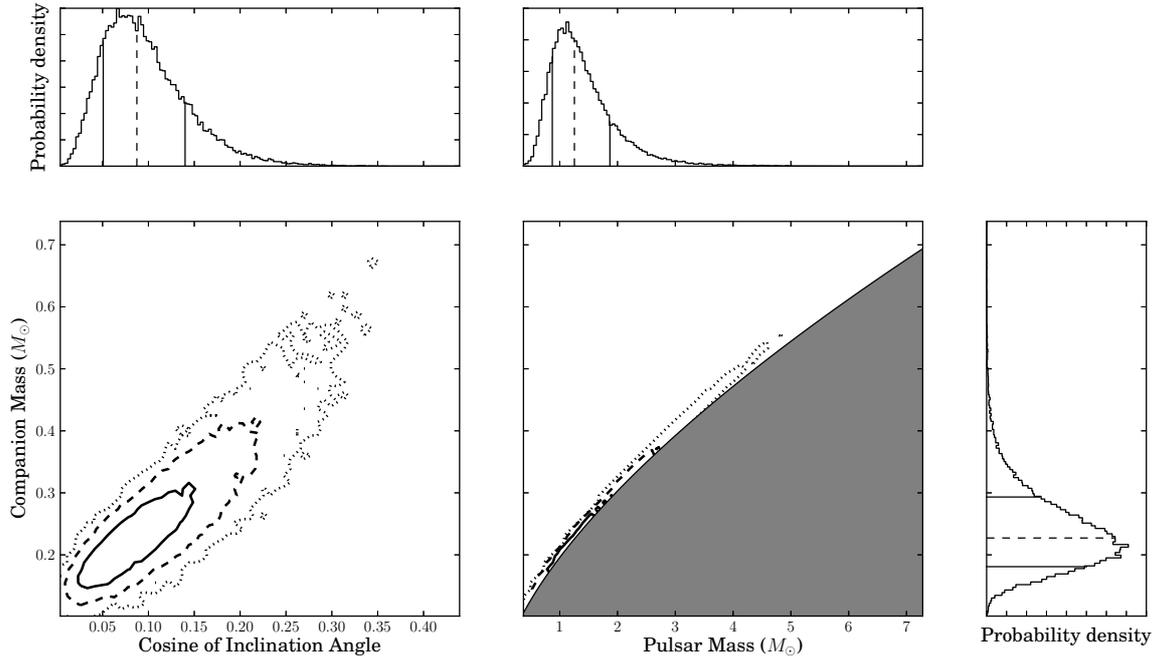}
\caption{Same caption as Fig.~\ref{fig:1600_mass} for PSR J1918$-$0642.}
\label{fig:1918_mass}
\end{minipage}
\end{figure*}

In Figures \ref{fig:0751_mass} to \ref{fig:1918_mass}, we plot, assuming GR, the
joint \mbox{2-D}
probability density function of the Shapiro delay that comes directly out of the TempoNest
analysis  for the three pulsars we achieve greatly improved mass measurements, PSRs J0751+1807, J1600$-$3053 and J1918$-$0642,
respectively. For PSR J0751+1807, we use the corrected orbital period
derivative, $\dot{P_b}_{corr} = (-4.6\pm0.4) \times 10^{-14} $, derived in the previous section to further constrain the
masses of the system.  The projection of the parameters $\dot{P_b}$ and
$\varsigma$ gives the following 68.3\%
confidence levels:
$m_p=1.64\pm0.15$~M$_{\odot}$ and
$m_c=0.16\pm0.01$~M$_{\odot}$. The inclination angle is constrained with $\cos
i < 0.64$ (2$\sigma$). Our new pulsar mass
measurement is 1.3$\sigma$ larger from the latest mass value published in \cite{nsk08}.

In the case of PSR J1600$-$3053, the posterior results from TempoNest give
 $\cos i=0.36\pm0.06$, $m_p=1.22_{-0.35}^{+0.50}$~M$_{\odot}$ and
$m_c=0.21_{-0.04}^{0.06}$~M$_{\odot}$. We now have
an accurate mass of the companion  compared to the marginal detection
by \citet{vbc+09}. Given the eccentricity $e \sim 1.7\times10^{-4}$ of this system, a detection of
the precession of periastron is likely to happen in the near future and would
greatly improve the pulsar mass measurement.

The results for PSR J1918$-$0642 translate into a pulsar mass $m_p =
1.25_{-0.4}^{+0.6}$~M$_{\odot} $ and a companion mass $m_c =
0.23\pm0.07$~M$_{\odot}$. The cosine of the inclination angle
is $0.09_{-0.04}^{+0.05}$. Based on the mass estimates for the companions to
PSRs J1600$-$3053 and J1918$-$0642, it is expected that these are low-mass
Helium white dwarfs.

In Table~\ref{tab:mass}, we summarize all our mass measurements and compare them to the values previously published in the literature. We find that PSRs J1713+0747,
J1802$-$2124, J1857+0943 and J1909$-$3744 have a mass measurement that is in very good
agreement to the values reported in the literature \citep{zsd+15,
fsk+10, vbc+09}.

\subsection{Search for annual-orbital parallax}
\label{sec:dis_kom}
For pulsars in binary systems, any change in the direction to the orbit
naturally leads to apparent variations in two of the Keplerian parameters, the intrinsic
projected semi-major axis $x_{\text{int}}$ and longitude of periastron
$\omega_{\text{int}}$.
In the case of nearby binary pulsars in wide orbits, a small periodic variation
of $x$ and $\omega$ due to the annual motion of the Earth around the Sun as
well as the orbital motion of the pulsar itself can be measured. This effect,
known as the annual-orbital parallax, can be expressed as \citep{kop95}: 

\begin{equation}
  x = x_{\text{int}} \left\lbrace 1 + \frac{\cot i }{d} (\Delta _{I_0} \sin \Omega  - \Delta _{J_0} \cos \Omega) \right\rbrace 
\end{equation}
and 
\begin{equation}
  \omega = \omega_{\text{int}} - \frac{\csc i }{d} (\Delta _{I_0} \cos \Omega  + \Delta _{J_0} \sin \Omega), 
\end{equation}
where $\Omega$ is the longitude of the ascending node. $\Delta_{I_0}$ and
$\Delta_{J_0}$ are defined in \citet{kop95} as:
\begin{equation}
\Delta_{I_0} = -X \sin \alpha + Y \cos \alpha
\end{equation}
and
\begin{equation}
\Delta_{J_0} = -X \sin \delta \cos \alpha - Y \sin \delta \sin \alpha + Z \cos \delta,
\end{equation}
where  $\mathbf{r} = (X,Y,Z)$ is the position vector of the Earth in the SSB
coordinate system.

The proper motion of the binary system also changes the apparent viewing
geometry of the orbit by \citep{ajr+96,kop96}:
\begin{equation}
\label{eq:hot}
  x = x_{\text{int}} \left\lbrace 1 + \frac{1}{\tan i} (-\mu_{\alpha} \sin \Omega  +
\mu_{\delta} \cos \Omega ) ( t - t_0)  \right\rbrace ,
\end{equation}

\begin{equation}
  \omega = \omega_{\text{int}} + \frac{1}{\sin i} (\mu_{\alpha} \cos \Omega  +
\mu_{\delta} \sin \Omega ) ( t - t_0) . 
\end{equation}
The time derivative of Eq.~\ref{eq:hot} can be expressed as
\begin{equation}
\label{eq:xdot}
\frac{\dot{x}}{x} = \mu \cot i \sin (\theta_\mu - \Omega),
\end{equation}
where $\theta_\mu$ is the position angle of the proper motion on the sky. If the inclination
angle, $i$, can be measured through, e.g., the detection of Shapiro delay, then a
measurement of $\dot{x}$ can constrain the longitude of ascending node $\Omega$.
These apparent variations in $x$ and $\omega$ are taken into account in Tempo2's
T2 binary model with the KOM and KIN parameters, corresponding to the
position angle
of the ascending node $\Omega$ and inclination angle $i$ (without the 90$^{\circ}$
ambiguity inherent to the Shapiro delay measurement). Therefore the parameter
$s \equiv \sin i$ of the Shapiro delay has to become a function of KIN.

Even a null  $\dot{x}$ can, if measured precisely enough, be useful.
According to Eq.~\ref{eq:xdot}, the maximum value for $\left|\dot{x}\right|$ is
$\dot{x}_{\max} = |x \mu \cot i|$ (obtained using the inequality $| \sin
(\theta_{\mu} - \Omega) | \leq 1$). Thus whenever the observed value and
uncertainty represent a small fraction of the interval from $- \dot{x}_{\max}$
to $\dot{x}_{\max}$, they are placing a direct constraint on $\sin
(\theta_{\mu} - \Omega)$


In our dataset, we measured the apparent variation of $\dot{x}$ for
13 pulsars, among which six are new measurements (PSRs J0751+1807,
J1455$-$3330, J1640+2224, J1751$-$2857, J1857+0943 and J1955+2908). 

\begin{figure}
\includegraphics[width=84mm]{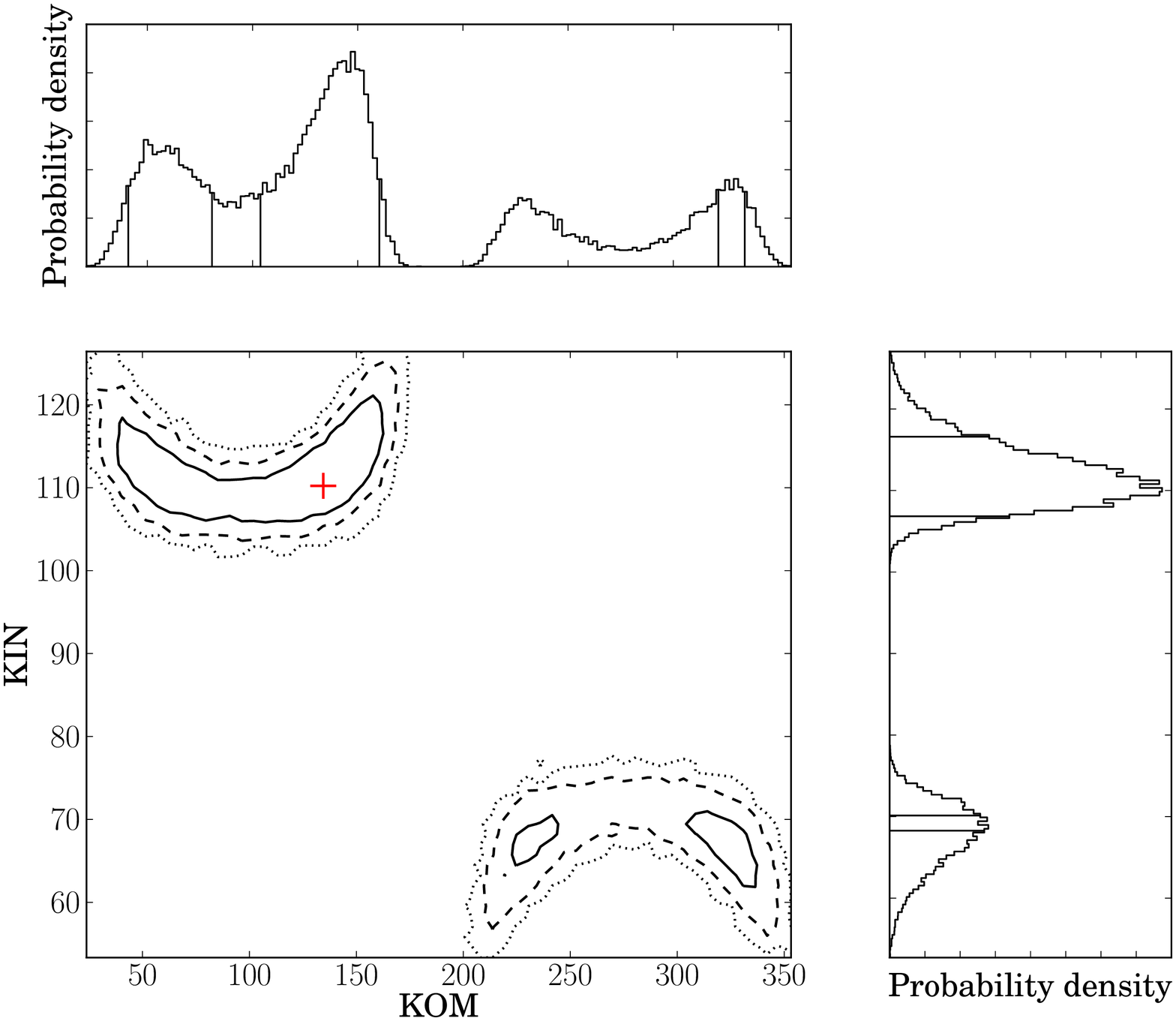}
\caption{One and two-dimensional marginalized posterior distributions of the
longitude of ascending node $\Omega $ and inclination angle $i$ for PSR
J1600$-$3053. The continuous black line, the dashed line and the dotted line
represent, respectively, the 68.3\%, 95.4\% and 99.7\% confidence levels of the
2-D probability density function. The red cross indicates the maximum
likelihood location. The continuous lines in the panels of the projected 1-D
distributions of KOM and KIN show the 68.3\% confidence levels for each
parameter.}
\label{fig:1600_komkin}
\end{figure}

\begin{figure}
\includegraphics[width=84mm]{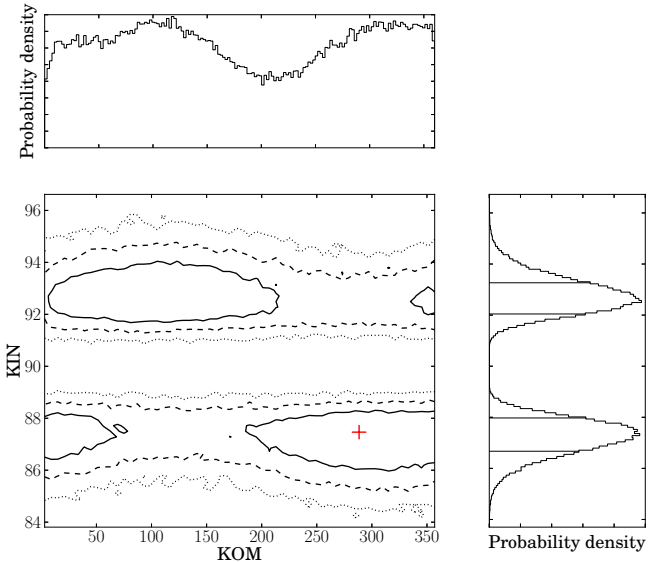}
\caption{Same caption as Fig.~\ref{fig:1600_komkin} for PSR J1857$+$0943.}
\label{fig:1857_komkin}
\end{figure}

\begin{figure}
\includegraphics[width=84mm]{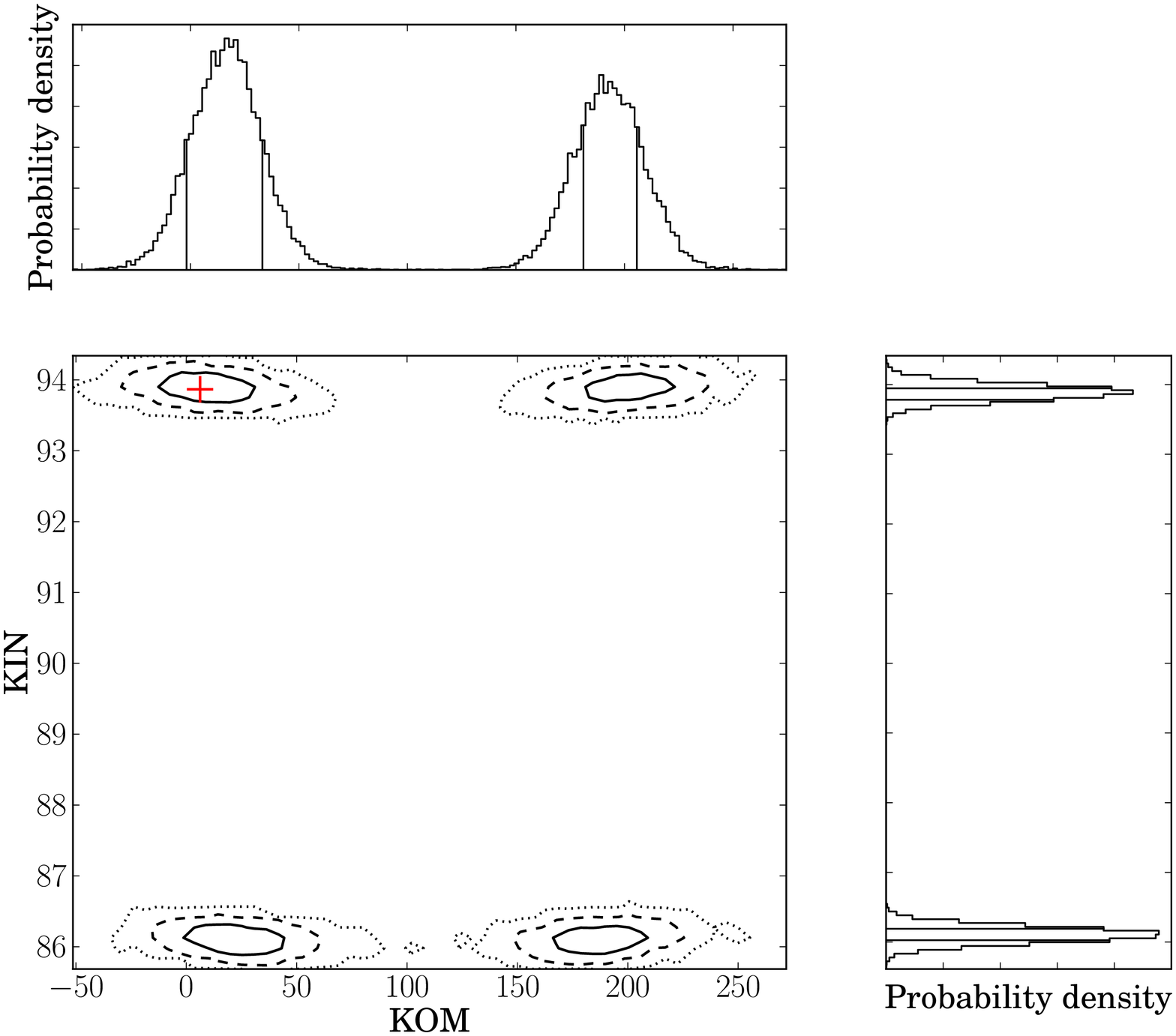}
\caption{Same caption as Fig.~\ref{fig:1600_komkin} for PSR J1909$-$3744.}
\label{fig:1909_komkin}
\end{figure}

For the three pulsars where we measured both the Shapiro delay and the variation
of  the semi-major axis (i.e. PSRs J0751+1807, J1600$-$3053 and
J1857+0943) and PSR J1909$-$3744 (where $\dot{x} =0.6\pm1.7 \times 10^{-16}$
 and $x \mu \cot i = 1.08\times 10^{-14}$), we map the KOM-KIN space with TempoNest
using the following procedure. First, we reduce the dimensionality of the Bayesian
analysis by fixing the set of white noise parameters to their maximum
likelihood values from the
timing analysis. We also choose to marginalize analytically over the astrometric and spin
parameters. Then we manually set the priors on KOM, KIN and M2 to encompass any
physical range of solution. Finally we perform the
sampling with TempoNest with the constant efficiency option turned off, in
order to more carefully explore the complex multi-modal parameter space. Because of the strong
correlation between the companion mass and  the inclination angle in the
case of PSR J0751+1807, (see Fig.~\ref{fig:0751_mass}), we do not report our
measurements as they were not constrained enough. The results are shown in
Figures \ref{fig:1600_komkin} to \ref{fig:1909_komkin} for the other three pulsars. 

For PSR J1600$-$3053, the 1-$\sigma$ contours of the 2-D posterior distribution
(Fig.~\ref{fig:1600_komkin}) give three solutions for
($\Omega, i$): $219^\circ < \Omega < 244^\circ$ and $63^\circ < i < 71^\circ$ or
$303^\circ < \Omega < 337^\circ$ and $61^\circ < i< 72^\circ$ and the preferred
solution, $37^\circ < \Omega < 163^\circ$ and $105^\circ < i < 122^\circ$.
The 2.5$\sigma$ detection of $\dot{x}$ in the PSR J1857+0943 binary system
still limit the constraints that can be set on $\Omega$ (see
Fig.~\ref{fig:1857_komkin}). Even though we do not detect  $\dot{x}$ for PSR
J1909$-$3744, we can constrain $\Omega$ (see Fig.~\ref{fig:1909_komkin}) to  $-2^\circ < \Omega < 33 ^\circ$ or $181^\circ <
\Omega < 206 ^\circ$. The preferred solution is $-2^\circ < \Omega < 33 ^\circ$
and  $93.78^\circ < i < 93.95^\circ$.
However, with this EPTA dataset,  we still have no statistical evidence for the
detection of annual-orbital parallax as we cannot distinguish between the symmetric solutions of
the pulsar orbits in these three pulsars.

\subsection{Comparison with the latest NANOGrav and PPTA results}
While this work was under review, similar analysis by NANOGrav and the PPTA were
published elsewhere \citep[][hereafter \citetalias{abb+15} and
\citetalias{rhc+16}, respectively]{abb+15,rhc+16}. \citetalias{abb+15} presents a thorough
description of their data analysis and \citet[][hereafter
\citetalias{mnf+16}]{mnf+16} report on the  study of astrometric parameters.
Other timing results and their interpretations (e.g. pulsar mass measurements) will be presented in a
series of upcoming papers. Hence, we briefly summarize here the similarities and
differences between our work and the ones by \citetalias{rhc+16} and \citetalias{mnf+16}.

\citetalias{rhc+16} used Tempo2 linearized, least-squares fitting methods  to present timing models for a set of 20 MSPs. White noise, DM variations and red
noise are included in the timing analysis and modeled with completely independent techniques from the ones
described in Section~\ref{sec:timing}. 
For all 13 pulsars observed commonly by the EPTA and the PPTA, both PTAs
achieve the detection of the parallax  with consistent results
(within 1.5$\sigma$). We note here that the parallax value of PSR J1909$-$3744 should
read $\pi=0.81\pm0.03$~mas in  \citetalias{rhc+16} (Reardon, private communication).
Also, the seven new proper motions values reported in this paper are for pulsars that are not
observed by \citetalias{rhc+16}.
We obtain similar results for the pulsar and companion masses to the
values reported in \citetalias{rhc+16}, albeit with much greater precision
in the case of PSRs J1600$-$3053. Furthermore, all our measurements of $\dot{x}$ agree with \citetalias{rhc+16}.
While  \citetalias{rhc+16} measure $\dot{P_b}$ in PSR J1022$+$1001 for the
first time, the EPTA achieve the detection of $\dot{P_b}$ for another MSP
(PSR J0613$-$0200), allowing us to get an independent distance estimate for
these systems.

\citetalias{mnf+16} report on the astrometric results for a set of 37 MSPs analyzed
with the linearized least-squares fitting package
Tempo\footnote{http://tempo.sourceforge.net/}. More details on the DM and red
noise models included in their analysis can be found in \citetalias{abb+15}.
All 14 parallax measurements for the pulsars presented commonly in this work
and in \citetalias{mnf+16} are consistent at the 2-$\sigma$ level. 
In addition, \citetalias{mnf+16} show a new parallax measurement for PSR J1918$-$0642
that was not detected with our dataset.
\citetalias{mnf+16} also present updated proper motions for 35 MSPs and derived the pulsar
velocities in galactocentric coordinates.
The new proper-motion measurement for PSR J2010$-$1323 reported in our work is
consistent at the 2-$\sigma$ level with the independent measurement from
\citetalias{mnf+16}.

Finally, \citetalias{mnf+16}  discuss in detail the same discrepancy reported
in   Section~\ref{sec:dis_shk} between
their measured parallax distance for PSR J1024$-$0719 and its constraint from $D_{\dot{P}}$.

\section{Conclusions}
\label{sec:conclusions}

We studied an ensemble of 42 MSPs from the EPTA, combining multifrequency
datasets from four different observatories, with data spanning more than 15
years for almost half of our sample. The analysis was performed with TempoNest
allowing the simultaneous determination of the white noise parameters and
modeling of the stochastic DM and red noise signals. We achieved the detection
of several new parameters: seven parallaxes,  nine proper motions and six apparent
changes in the orbital semi-major axis. We also measured Shapiro delay in
two systems, PSRs J1600$-$3053 and J1918$-$0642, with low-mass Helium white dwarf
companions.
Further observations of PSR J1600$-$3053 will likely yield the detection of the advance of
periastron, dramatically improving the mass measurement of this system and
improving the constraints on the geometry of the system. We presented an
updated mass measurement for PSR J0751+1807,
roughly consistent with the previous work by \citet{nsk08}.
We searched for the presence of annual-orbital parallax  in three systems, PSRs J1600$-$3053, J1857$+$0943 and J1909$-$3744.
However we could only set constraints on the
longitude of ascending node in PSRs J1600$-$3053 and
J1909$-$3744 with marginal evidence of annual-orbital parallax in PSR J1600$-$3053.

With an improved set of parallax distances, we investigated the difference
between the predictions from the NE2001 Galactic electron density model and the
LK-corrected parallax distances. On average we found an error of $\sim$ 80\% in
the NE2001 distances, this error increasing further at high Galactic latitudes.
Despite its flaws for high galactic latitude lines-of-sight, we find NE2001 to
still predict more accurate distances than two recent models, M2 and M3, proposed by
\citet{sch12}, based respectively on the TC93 and NE2001 models with an extended thick disk. 
We showed that a change in the scale height of the thick disk of
 the current electron density models also dramatically affects the pulsars that
are located in the Galactic plane. Our updated set of parallaxes presented here
will likely contribute to improving
on any future model of the Galactic electron density model

A comparison of the 2-D velocity distribution between isolated and binary MSPs
with a sample two times larger than the last published study \citep{gsf+11}
still shows no statistical difference, arguing that both populations originate from the same underlying population.
Through precision measurement of the orbital period derivative, we
achieved better constraints on the distance to two pulsars, PSRs J1012+5307 and
J1909$-$3744, than is possible via the detection of the annual parallax.

Based on the timing results presented in this paper and the red noise
properties of the pulsars discussed in \citet{cll+15}, we will
revisit and potentially remove some MSPs from the EPTA observing list.
The EPTA is also continuously adding more sources to its observing list, especially in
the last five years, as more MSPs are discovered through the targeted survey of
\textit{Fermi} sources \citep{rap+12} and large-scale pulsar surveys
\citep[e.g. the PALFA, HTRU and GBNCC collaborations;
][]{lbh+15,bck+13,nbb+14,slr+14}. Over 60 MSPs
are now being regularly monitored as part of the EPTA effort.

Recent progress in digital processing, leading in some cases to an increase of
the processed bandwidth by a factor of $2-4\times$, allowed new wide-band coherent
dedispersion backends to be commissioned at all EPTA sites in the last few
years \citep[see e.g.][]{kss08,dbc+11}. These new instruments provide TOAs with
improved precision that will be included in a future release of the EPTA
dataset.
 The long baselines of MSPs timing data presented here, especially when
recorded with a single
backend, are of great value, not only for the detection of the GWB but also to
a wide range of astrophysics as shown in this paper.

\section*{Acknowledgments}
The authors would like to thank D.~Schnitzeler for providing us with the M2 and M3
distances used in this work, P.~Freire, M.~Bailes, T.~Tauris and N.~Wex for
useful discussions, P.~Demorest for his contribution to the pulsar
instrumentation at the NRT.

Part of this work is based on observations with the 100-m telescope of the
Max-Planck-Institut f\"ur Radioastronomie (MPIfR) at Effelsberg in Germany. Pulsar
research at the Jodrell Bank Centre for Astrophysics and the observations using
the Lovell Telescope are supported by a consolidated grant from the STFC in the
UK. The Nan{\c c}ay radio observatory is operated by the Paris Observatory,
associated to the French Centre National de la Recherche Scientifique (CNRS).
We acknowledge financial support from `Programme National de Cosmologie et
Galaxies' (PNCG) of CNRS/INSU, France. The Westerbork Synthesis Radio Telescope
is operated by the Netherlands Institute for Radio Astronomy (ASTRON) with
support from the Netherlands Foundation for Scientific Research (NWO).

CGB, GHJ, RK, KL, KJL, DP acknowledge the support from the `LEAP' ERC Advanced
Grant (337062).
RNC acknowledges the support of the International Max Planck Research School
Bonn/Cologne and the Bonn-Cologne Graduate School.
JG and AS are supported by the Royal Society.
JWTH acknowledges funding from an NWO Vidi fellowship and CGB, JWTH acknowledge
the support from the ERC Starting Grant `DRAGNET' (337062).
KJL is supported by the National Natural Science Foundation of China (Grant No.11373011).
PL acknowledges the support of the International Max Planck Research School Bonn/Cologne.
CMFM was supported by a Marie Curie International Outgoing Fellowship within
the 7th European Community Framework Programme.
SO is supported by the Alexander von Humboldt Foundation.
This research was in part supported by ST's appointment to the NASA
Postdoctoral Program at the Jet Propulsion Laboratory, administered by Oak
Ridge Associated Universities through a contract with NASA.
RvH is supported by NASA Einstein Fellowship grant PF3-140116.

The authors acknowledge the use of the Hydra and Hercules computing cluster from
Rechenzentrum Garching. This research has made extensive use of NASA's Astrophysics Data
System, the ATNF Pulsar Catalogue and the Python Uncertainties package,
http://pythonhosted.org/uncertainties/.

\bibliographystyle{mnras}
\bibliography{epta}

\appendix
\section{Data description}
\begin{figure*}
\includegraphics[height=180mm,angle=-90]{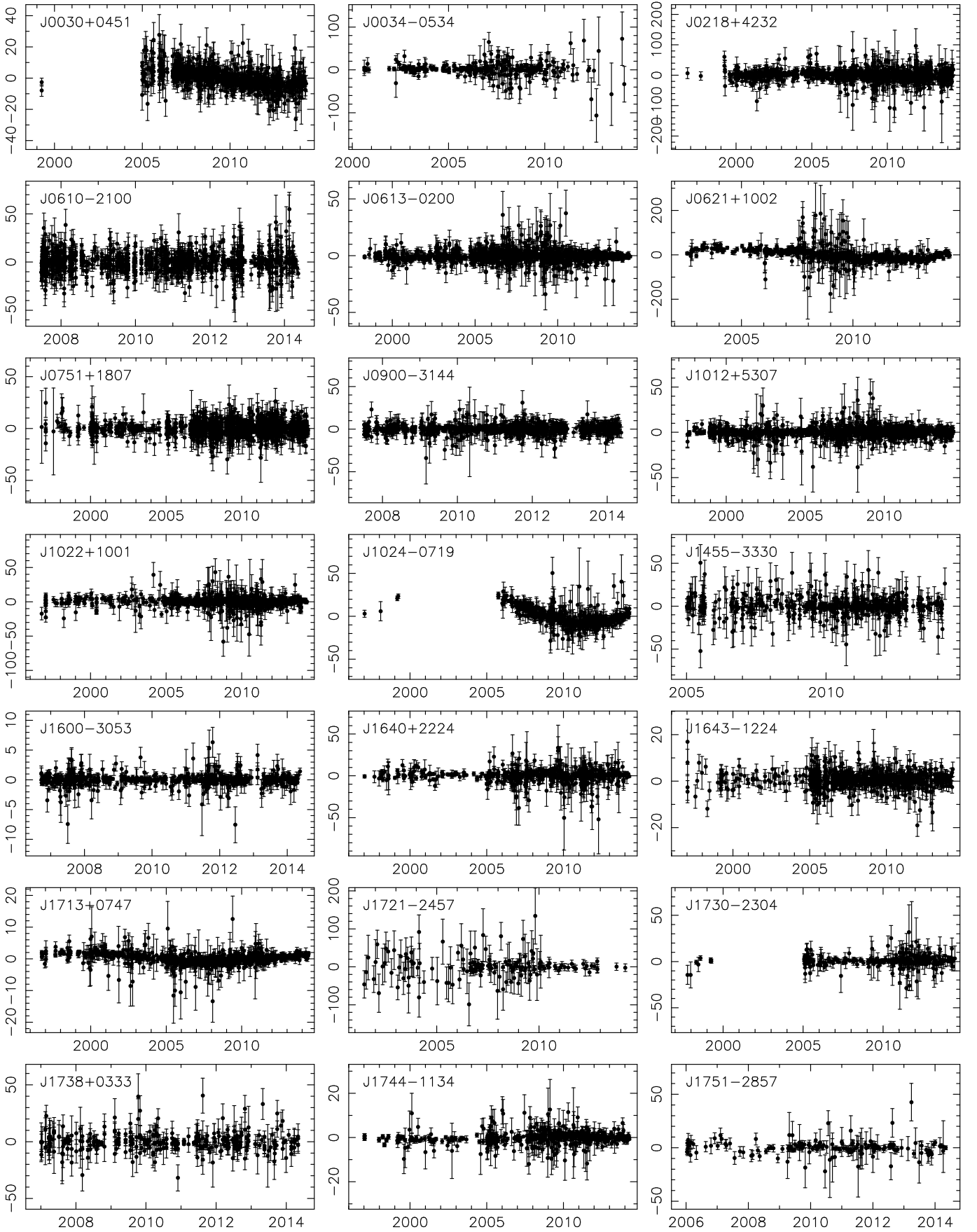}
\caption{Timing residuals in microseconds (y-axis) for the first 21 pulsars as
a function of time in years (x-axis). The plots show the multifrequency
residuals after subtracting the contribution from the DM model. The red noise
seen in the timing residuals of PSRs J0030+0451 and J1024$-$0719 will be
discussed by \citet{cll+15}.}
\label{plot:residuals-1}
\end{figure*}

\begin{figure*}
\includegraphics[height=180mm,angle=-90]{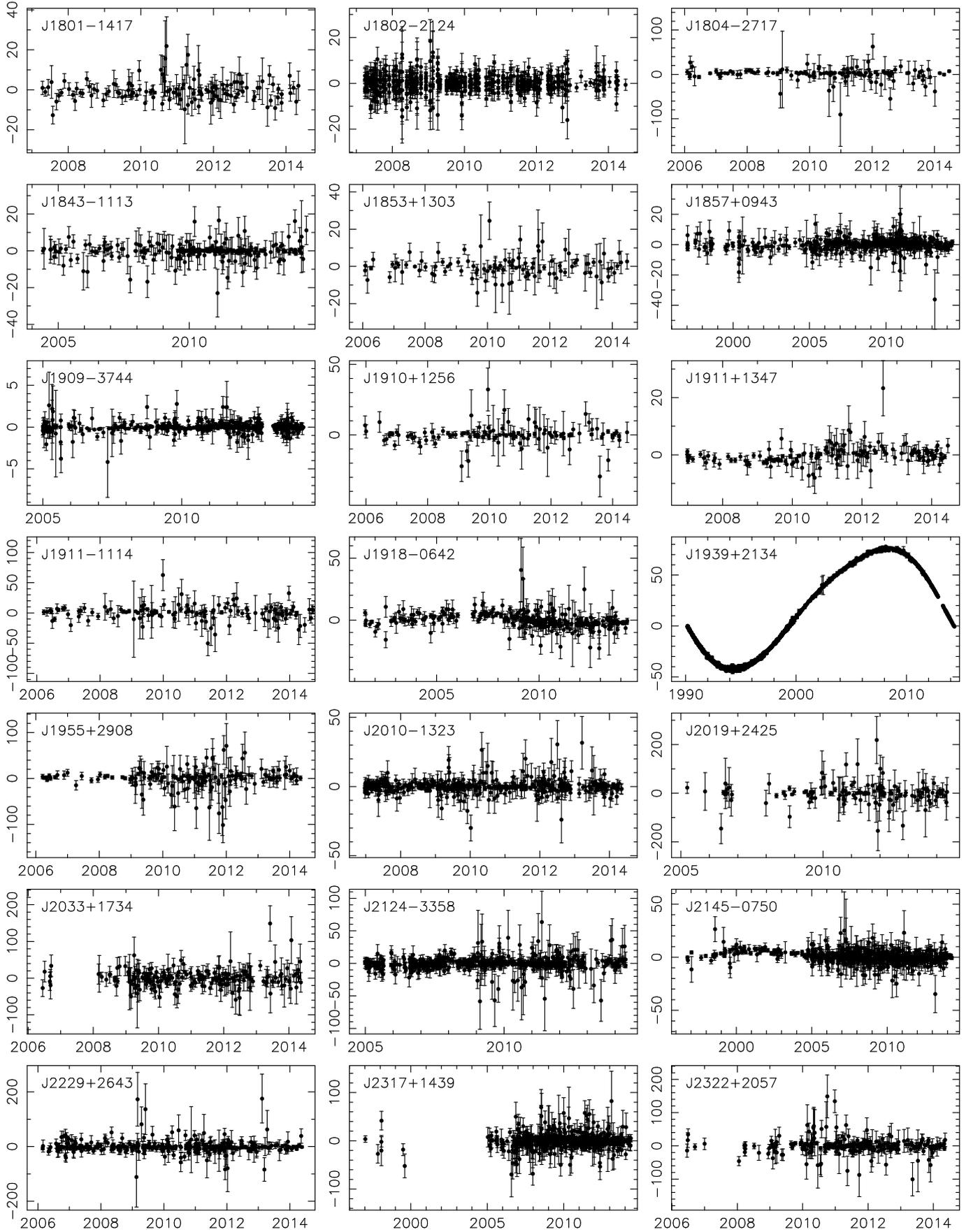}
\caption{Same caption as Fig. \ref{plot:residuals-1} for the last 21 MSPs. The
large amount of red noise 
seen in the timing residuals of PSR J1939+2134 will be 
discussed by \citet{cll+15}.}
\label{plot:residuals-2}
\end{figure*}

\begin{figure*}
\includegraphics[height=180mm,angle=-90]{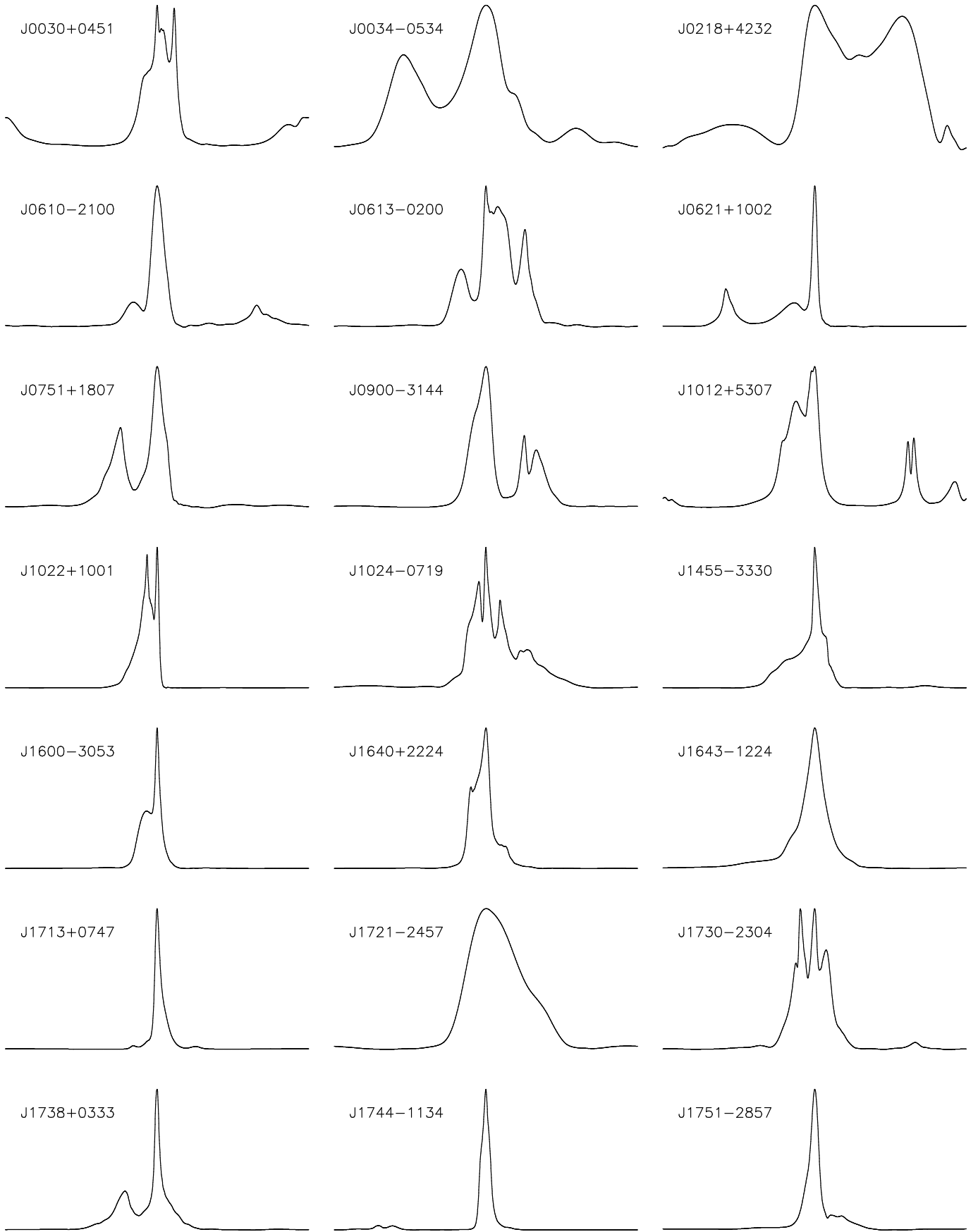}
\caption{Reference profiles of total intensity I for the first 21 MSPs observed
at 1400 MHz with the NRT. The profiles are centered with respect to the peak
maximum. For each pulsar, the full pulse phase is shown and the intensity is in
arbitrary units.}
\label{plot:templates-1}
\end{figure*}

\begin{figure*}
\includegraphics[height=180mm,angle=-90]{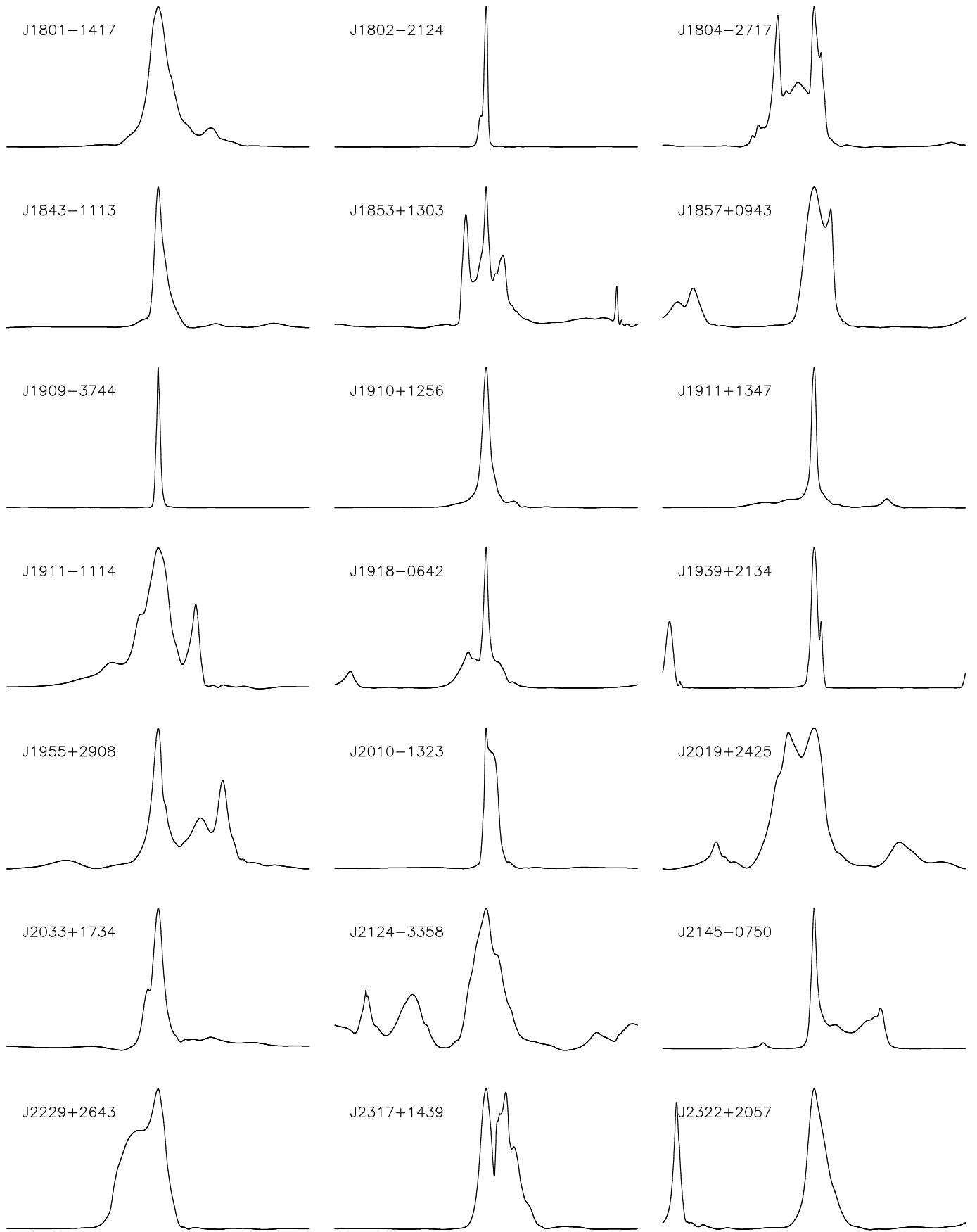}
\caption{Same caption as Fig. \ref{plot:templates-1} for the last 21 MSPs.}
\label{plot:templates-2}
\end{figure*}

\onecolumn
\begin{longtable}{lcrrrrcccrrr}
\caption{Summary of the EPTA dataset. The columns indicate respectively the
pulsar name, the number of JUMPs (N$_{j}$) included in the timing solution, the
maximum likelihood results for the red noise (RN) and DM models (dimensionless amplitude $A$ and
spectral index $\gamma$ for each model), the
observatory, frequency band (in MHz), data span in years, number of TOAs along with the maximum
likelihood values for the EFAC ($E_f$) and EQUAD ($E_q$ in units of seconds) parameters (EQUAD
shown in log10-base).} \label{tab:data} \\
\hline
PSR JName   & N$_{j}$   & $A_{\text{RN}}$ & $\gamma_{\text{RN}}$ & $A_{\text{DM}}$  & $\gamma_{\text{DM}}$ & Observatory & Frequency & Year Range  & N$_{\text{TOAs}}$ & $E_f$ & $E_q$ \\
\hline
\endfirsthead
\multicolumn{12}{c}%
{\tablename\ \thetable\ -- \textit{Continued from previous page}} \\
\hline
PSR JName   & N$_{j}$ & $A_{\text{RN}}$ & $\gamma_{\text{RN}}$ & $A_{\text{DM}}$  & $\gamma_{\text{DM}}$   & Observatory & Frequency & Year Range  & N$_{\text{TOAs}}$ & $E_f$ & $E_q$ \\
\hline
\endhead
\hline 
J0030$+$0451  & 6  & -14.65 & 5.43 & -19.06 & 3.27\\
            &   &    &   & &   & EFF\dotfill & 1360 & $2009.7-2013.5$ & 33 & 1.17 & -9.17  \\
            &   &    &   & &   &              & 1410 & $1999.3-2009.6$ & 10 & 0.62 & -9.00  \\
            &   &    &   & &   &              & 2639 & $2008.4-2013.4$ & 34 & 1.03 & -6.80  \\
            &   &    &   & &   & JBO\dotfill & 1520 & $2012.5-2014.3$ & 50 & 1.01 & -8.87  \\
            &   &    &   & &   & NRT\dotfill & 1400 & $2005.0-2011.9$ & 552 & 1.18 & -6.13  \\
            &   &    &   & &   &              & 1600 & $2009.0-2014.3$ & 138 & 1.59 & -9.86  \\
            &   &    &   & &   &              & 2000 & $2006.9-2014.0$ & 90 & 1.32 & -6.07  \\
\\
J0034$-$0534  & 5  & -14.93 & 1.27 & -12.43 & 2.84\\
            &   &    &   & &   & NRT\dotfill & 1400 & $2005.9-2011.7$ & 56 & 0.94 & -5.11  \\
            &   &    &   & &   &              & 1600 & $2012.0-2014.1$ & 10 & 2.00 & -9.50  \\
            &   &    &   & &   & WSRT\dotfill & 328 & $2000.6-2010.5$ & 112 & 1.22 & -8.45  \\
            &   &    &   & &   &              & 382 & $2000.6-2010.5$ & 98 & 0.75 & -7.96  \\
\\
J0218$+$4232  & 12  & -13.32 & 2.78 & -11.14 & 2.09\\
            &   &    &   & &   & EFF\dotfill & 1360 & $2009.6-2013.5$ & 34 & 1.17 & -6.75  \\
            &   &    &   & &   &              & 1410 & $1996.8-2009.3$ & 178 & 1.21 & -8.52  \\
            &   &    &   & &   & JBO\dotfill & 1400 & $2009.1-2009.4$ & 13 & 1.50 & -6.08  \\
            &   &    &   & &   &              & 1520 & $2009.6-2014.4$ & 97 & 1.10 & -6.68  \\
            &   &    &   & &   & NRT\dotfill & 1400 & $2005.6-2012.2$ & 406 & 1.17 & -7.45  \\
            &   &    &   & &   &              & 1600 & $2009.0-2014.3$ & 157 & 1.54 & -6.59  \\
            &   &    &   & &   &              & 2000 & $2006.9-2013.6$ & 14 & 1.46 & -5.71  \\
            &   &    &   & &   & WSRT\dotfill & 1380 & $1999.8-2010.5$ & 49 & 1.55 & -6.32  \\
            &   &    &   & &   &              & 328 & $2000.1-2010.5$ & 125 & 1.21 & -6.51  \\
            &   &    &   & &   &              & 382 & $1999.6-2010.5$ & 123 & 0.98 & -8.46  \\
\\
J0610$-$2100  & 2  & -12.85 & 0.23 & -13.74 & 5.46\\
            &   &    &   & &   & JBO\dotfill & 1520 & $2010.5-2014.4$ & 179 & 0.85 & -8.44  \\
            &   &    &   & &   & NRT\dotfill & 1400 & $2007.5-2011.7$ & 631 & 1.62 & -9.29  \\
            &   &    &   & &   &              & 1600 & $2011.2-2014.3$ & 224 & 1.26 & -8.48  \\
\\
J0613$-$0200  & 13  & -16.15 & 6.88 & -11.71 & 1.07\\
            &   &    &   & &   & EFF\dotfill & 1360 & $2008.0-2013.5$ & 42 & 1.01 & -6.80  \\
            &   &    &   & &   &              & 1410 & $1998.3-2009.3$ & 241 & 1.10 & -8.95  \\
            &   &    &   & &   &              & 2639 & $2006.6-2013.5$ & 64 & 1.01 & -7.79  \\
            &   &    &   & &   & JBO\dotfill & 1400 & $2009.0-2009.4$ & 24 & 1.35 & -8.44  \\
            &   &    &   & &   &              & 1520 & $2009.6-2014.3$ & 191 & 0.94 & -5.96  \\
            &   &    &   & &   & NRT\dotfill & 1400 & $2005.0-2011.8$ & 334 & 1.02 & -6.35  \\
            &   &    &   & &   &              & 1600 & $2009.0-2014.4$ & 84 & 1.00 & -9.77  \\
            &   &    &   & &   &              & 2000 & $2006.9-2012.8$ & 51 & 1.10 & -6.41  \\
            &   &    &   & &   & WSRT\dotfill & 1380 & $1999.6-2010.5$ & 171 & 1.00 & -9.59  \\
            &   &    &   & &   &              & 328 & $2000.6-2010.5$ & 87 & 0.75 & -6.32  \\
            &   &    &   & &   &              & 382 & $2000.6-2010.5$ & 80 & 0.92 & -5.69  \\
\\
J0621$+$1002  & 9  & -12.04 & 2.50 & -16.98 & 3.80\\
            &   &    &   & &   & EFF\dotfill & 1360 & $2009.6-2013.5$ & 42 & 0.83 & -9.46  \\
            &   &    &   & &   &              & 1410 & $2002.6-2009.3$ & 88 & 0.63 & -7.27  \\
            &   &    &   & &   &              & 2639 & $2006.0-2013.4$ & 47 & 0.43 & -4.39  \\
            &   &    &   & &   & JBO\dotfill & 1400 & $2009.0-2009.4$ & 18 & 1.08 & -7.86  \\
            &   &    &   & &   &              & 1520 & $2009.6-2014.3$ & 140 & 1.09 & -9.34  \\
            &   &    &   & &   & NRT\dotfill & 1400 & $2006.1-2011.9$ & 168 & 1.23 & -6.58  \\
            &   &    &   & &   &              & 1600 & $2009.0-2014.3$ & 33 & 1.31 & -9.74  \\
            &   &    &   & &   & WSRT\dotfill & 1380 & $2006.0-2010.5$ & 68 & 1.18 & -8.23  \\
            &   &    &   & &   &              & 323 & $2007.5-2010.5$ & 34 & 2.13 & -6.47  \\
            &   &    &   & &   &              & 367 & $2007.5-2010.5$ & 35 & 1.72 & -8.12  \\
\\
J0751$+$1807  & 8  & -19.50 & 4.86 & -11.77 & 2.83\\
            &   &    &   & &   & EFF\dotfill & 1360 & $2009.6-2013.4$ & 29 & 0.91 & -5.34  \\
            &   &    &   & &   &              & 1410 & $1996.8-2004.6$ & 159 & 0.98 & -5.74  \\
            &   &    &   & &   &              & 2639 & $1999.2-2013.5$ & 64 & 1.15 & -5.36  \\
            &   &    &   & &   & JBO\dotfill & 1400 & $2009.0-2009.4$ & 28 & 1.03 & -8.01  \\
            &   &    &   & &   &              & 1520 & $2009.6-2014.3$ & 179 & 1.00 & -6.09  \\
            &   &    &   & &   & NRT\dotfill & 1400 & $2005.0-2011.9$ & 598 & 1.19 & -6.00  \\
            &   &    &   & &   &              & 1600 & $2011.2-2014.4$ & 362 & 1.07 & -7.50  \\
            &   &    &   & &   &              & 2000 & $2007.5-2013.4$ & 40 & 0.88 & -5.72  \\
            &   &    &   & &   & WSRT\dotfill & 1380 & $2007.4-2010.5$ & 32 & 1.55 & -8.63  \\
\\
J0900$-$3144  & 4  & -15.58 & 5.04 & -11.55 & 3.05\\
            &   &    &   & &   & JBO\dotfill & 1400 & $2009.1-2009.4$ & 9 & 1.09 & -9.26  \\
            &   &    &   & &   &              & 1520 & $2009.7-2014.3$ & 99 & 1.02 & -7.47  \\
            &   &    &   & &   & NRT\dotfill & 1400 & $2007.5-2012.1$ & 321 & 1.04 & -9.21  \\
            &   &    &   & &   &              & 1600 & $2009.0-2014.4$ & 329 & 1.04 & -5.64  \\
            &   &    &   & &   &              & 2000 & $2007.7-2014.2$ & 117 & 1.28 & -8.07  \\
\\
J1012$+$5307  & 14  & -13.07 & 1.52 & -17.57 & 3.46\\
            &   &    &   & &   & EFF\dotfill & 1360 & $2009.6-2013.4$ & 37 & 0.56 & -5.47  \\
            &   &    &   & &   &              & 1410 & $1997.5-2009.3$ & 404 & 0.95 & -6.47  \\
            &   &    &   & &   &              & 2639 & $2006.6-2013.5$ & 88 & 1.02 & -7.43  \\
            &   &    &   & &   & JBO\dotfill & 1400 & $2009.0-2009.4$ & 12 & 1.13 & -7.83  \\
            &   &    &   & &   &              & 1520 & $2009.8-2014.3$ & 96 & 0.96 & -5.86  \\
            &   &    &   & &   & NRT\dotfill & 1400 & $2005.5-2011.7$ & 239 & 1.21 & -9.32  \\
            &   &    &   & &   &              & 1600 & $2009.0-2014.4$ & 234 & 1.19 & -6.06  \\
            &   &    &   & &   &              & 2000 & $2007.5-2014.1$ & 18 & 1.18 & -9.79  \\
            &   &    &   & &   & WSRT\dotfill & 328 & $2000.9-2010.5$ & 87 & 1.15 & -5.52  \\
            &   &    &   & &   &              & 382 & $2000.9-2010.5$ & 82 & 1.07 & -7.66  \\
            &   &    &   & &   &              & 1380.1 & $1999.6-2001.2$ & 26 & 1.14 & -8.21  \\
            &   &    &   & &   &              & 1380.2 & $2001.2-2010.5$ & 136 & 0.90 & -6.96  \\
\\
J1022$+$1001  & 9  & -13.08 & 1.70 & -11.63 & 1.10\\
            &   &    &   & &   & EFF\dotfill & 1360 & $2008.1-2013.5$ & 76 & 1.02 & -5.84  \\
            &   &    &   & &   &              & 1410 & $1996.8-2009.3$ & 164 & 0.65 & -5.43  \\
            &   &    &   & &   &              & 2639 & $2006.2-2013.5$ & 88 & 2.09 & -7.32  \\
            &   &    &   & &   & JBO\dotfill & 1400 & $2009.0-2011.0$ & 40 & 1.03 & -6.29  \\
            &   &    &   & &   &              & 1520 & $2009.6-2014.3$ & 187 & 1.30 & -5.66  \\
            &   &    &   & &   & NRT\dotfill & 1400 & $2005.0-2008.4$ & 127 & 1.23 & -5.84  \\
            &   &    &   & &   &              & 1600 & $2011.9-2014.3$ & 44 & 1.30 & -7.12  \\
            &   &    &   & &   & WSRT\dotfill & 1380 & $2006.0-2010.5$ & 58 & 0.82 & -6.01  \\
            &   &    &   & &   &              & 323 & $2007.5-2010.5$ & 26 & 1.67 & -4.81  \\
            &   &    &   & &   &              & 367 & $2007.5-2010.0$ & 17 & 0.79 & -4.96  \\
\\
J1024$-$0719  & 8  & -13.69 & 3.17 & -12.96 & 6.12\\
            &   &    &   & &   & EFF\dotfill & 1360 & $2008.1-2013.5$ & 33 & 1.05 & -9.41  \\
            &   &    &   & &   &              & 1410 & $1997.0-2009.3$ & 27 & 0.75 & -9.76  \\
            &   &    &   & &   &              & 2639 & $2006.0-2013.5$ & 53 & 1.23 & -9.75  \\
            &   &    &   & &   & JBO\dotfill & 1400 & $2009.0-2009.4$ & 32 & 1.10 & -8.08  \\
            &   &    &   & &   &              & 1520 & $2009.6-2014.3$ & 127 & 1.19 & -8.53  \\
            &   &    &   & &   & NRT\dotfill & 1400 & $2006.0-2011.7$ & 176 & 1.24 & -8.60  \\
            &   &    &   & &   &              & 1600 & $2009.0-2014.3$ & 77 & 0.86 & -5.88  \\
            &   &    &   & &   &              & 2000 & $2006.9-2010.5$ & 12 & 0.89 & -6.25  \\
            &   &    &   & &   & WSRT\dotfill & 1380 & $2007.4-2010.2$ & 24 & 1.12 & -9.43  \\
\\
J1455$-$3330  & 2  & -16.31 & 4.03 & -11.38 & 2.44\\
            &   &    &   & &   & JBO\dotfill & 1520 & $2009.8-2014.3$ & 25 & 1.01 & -5.80  \\
            &   &    &   & &   & NRT\dotfill & 1400 & $2005.0-2011.9$ & 338 & 1.23 & -9.53  \\
            &   &    &   & &   &              & 1600 & $2009.0-2014.2$ & 161 & 1.04 & -8.98  \\
\\
J1600$-$3053  & 3  & -16.56 & 2.71 & -11.64 & 1.41\\
            &   &    &   & &   & JBO\dotfill & 1520 & $2011.5-2014.3$ & 44 & 1.43 & -6.19  \\
            &   &    &   & &   & NRT\dotfill & 1400 & $2006.7-2011.7$ & 230 & 1.15 & -7.62  \\
            &   &    &   & &   &              & 1600 & $2010.8-2014.4$ & 151 & 0.99 & -9.27  \\
            &   &    &   & &   &              & 2000 & $2006.9-2014.2$ & 106 & 1.12 & -6.58  \\
\\
J1640$+$2224  & 7  & -13.11 & 0.12 & -16.87 & 0.75\\
            &   &    &   & &   & EFF\dotfill & 1360 & $2008.1-2013.5$ & 81 & 0.89 & -6.64  \\
            &   &    &   & &   &              & 1410 & $1997.0-2009.7$ & 122 & 0.93 & -7.96  \\
            &   &    &   & &   &              & 2639 & $2006.2-2013.5$ & 67 & 1.02 & -6.86  \\
            &   &    &   & &   & JBO\dotfill & 1400 & $2009.1-2009.4$ & 10 & 0.88 & -8.83  \\
            &   &    &   & &   &              & 1520 & $2009.6-2014.3$ & 148 & 1.32 & -8.92  \\
            &   &    &   & &   & NRT\dotfill & 1400 & $2005.0-2011.9$ & 103 & 1.22 & -8.46  \\
            &   &    &   & &   &              & 1600 & $2010.8-2013.8$ & 24 & 1.11 & -9.45  \\
            &   &    &   & &   & WSRT\dotfill & 1380 & $2006.0-2010.2$ & 40 & 1.28 & -8.82  \\
\\
J1643$-$1224  & 8  & -19.04 & 3.44 & -10.99 & 1.70\\
            &   &    &   & &   & EFF\dotfill & 1360 & $2009.6-2013.4$ & 27 & 1.08 & -8.64  \\
            &   &    &   & &   &              & 1410 & $1997.0-2009.7$ & 94 & 1.11 & -9.77  \\
            &   &    &   & &   &              & 2639 & $2006.6-2013.5$ & 43 & 0.69 & -5.37  \\
            &   &    &   & &   & JBO\dotfill & 1400 & $2009.1-2009.4$ & 11 & 0.95 & -8.85  \\
            &   &    &   & &   &              & 1520 & $2009.7-2014.3$ & 76 & 0.74 & -5.61  \\
            &   &    &   & &   & NRT\dotfill & 1400 & $2005.0-2011.7$ & 334 & 1.24 & -7.19  \\
            &   &    &   & &   &              & 1600 & $2009.0-2014.3$ & 71 & 1.35 & -8.50  \\
            &   &    &   & &   &              & 2000 & $2006.9-2013.6$ & 49 & 1.05 & -8.45  \\
            &   &    &   & &   & WSRT\dotfill & 1380 & $2006.0-2010.5$ & 54 & 1.29 & -9.87  \\
\\
J1713$+$0747  & 13  & -15.29 & 5.62 & -11.98 & 1.47\\
            &   &    &   & &   & EFF\dotfill & 1360 & $2008.1-2013.4$ & 40 & 0.44 & -6.08  \\
            &   &    &   & &   &              & 1410 & $1996.8-2009.7$ & 164 & 0.98 & -6.42  \\
            &   &    &   & &   &              & 2639 & $2006.6-2013.5$ & 61 & 1.08 & -6.64  \\
            &   &    &   & &   & JBO\dotfill & 1400 & $2009.1-2011.0$ & 18 & 1.27 & -6.22  \\
            &   &    &   & &   &              & 1520 & $2009.6-2012.0$ & 53 & 1.78 & -6.54  \\
            &   &    &   & &   & NRT\dotfill & 1400 & $2005.0-2011.6$ & 354 & 1.24 & -6.81  \\
            &   &    &   & &   &              & 1600 & $2009.0-2014.4$ & 173 & 0.97 & -7.48  \\
            &   &    &   & &   &              & 2000 & $2005.2-2013.8$ & 97 & 1.23 & -7.15  \\
            &   &    &   & &   & WSRT\dotfill & 840 & $1999.5-2007.8$ & 53 & 0.90 & -8.81  \\
            &   &    &   & &   &              & 1380 & $1999.7-2001.1$ & 22 & 0.55 & -7.98  \\
            &   &    &   & &   &              & 1380 & $2001.1-2010.5$ & 114 & 0.53 & -8.74  \\
            &   &    &   & &   &              & 2273 & $2006.9-2010.4$ & 39 & 1.05 & -7.87  \\
\\
J1721$-$2457  & 3  & -16.50 & 6.68 & -10.04 & 1.22\\
            &   &    &   & &   & NRT\dotfill & 1400 & $2006.3-2011.8$ & 58 & 0.33 & -5.04  \\
            &   &    &   & &   &              & 1600 & $2009.1-2014.2$ & 13 & 0.77 & -5.21  \\
            &   &    &   & &   & WSRT\dotfill & 1380 & $2001.5-2010.5$ & 79 & 2.38 & -5.13  \\
\\
J1730$-$2304  & 7  & -16.33 & 0.11 & -11.42 & 2.37\\
            &   &    &   & &   & EFF\dotfill & 1360 & $2010.9-2013.5$ & 19 & 0.71 & -7.94  \\
            &   &    &   & &   &              & 1410 & $1997.8-1999.3$ & 8 & 0.72 & -8.07  \\
            &   &    &   & &   &              & 2639 & $2011.0-2013.2$ & 9 & 1.31 & -6.32  \\
            &   &    &   & &   & JBO\dotfill & 1400 & $2009.0-2009.4$ & 5 & 1.63 & -6.91  \\
            &   &    &   & &   &              & 1520 & $2009.7-2014.5$ & 83 & 1.39 & -6.90  \\
            &   &    &   & &   & NRT\dotfill & 1400 & $2005.1-2011.8$ & 106 & 1.01 & -6.75  \\
            &   &    &   & &   &              & 1600 & $2011.0-2014.4$ & 29 & 1.07 & -8.43  \\
            &   &    &   & &   &              & 2000 & $2007.4-2011.7$ & 9 & 1.30 & -7.73  \\
\\
J1738$+$0333  & 2  & -15.34 & 0.36 & -12.09 & 1.89\\
            &   &    &   & &   & JBO\dotfill & 1520 & $2011.5-2014.3$ & 56 & 1.06 & -5.89  \\
            &   &    &   & &   & NRT\dotfill & 1400 & $2007.0-2011.7$ & 199 & 1.14 & -9.60  \\
            &   &    &   & &   &              & 1600 & $2011.2-2014.3$ & 63 & 1.14 & -9.24  \\
\\
J1744$-$1134  & 8  & -13.85 & 2.90 & -17.65 & 4.57\\
            &   &    &   & &   & EFF\dotfill & 1360 & $2009.6-2013.5$ & 22 & 0.58 & -6.17  \\
            &   &    &   & &   &              & 1410 & $1997.0-2009.7$ & 100 & 1.03 & -6.14  \\
            &   &    &   & &   &              & 2639 & $2007.1-2013.4$ & 42 & 0.88 & -6.14  \\
            &   &    &   & &   & JBO\dotfill & 1520 & $2009.6-2014.3$ & 68 & 0.77 & -6.07  \\
            &   &    &   & &   & NRT\dotfill & 1400 & $2005.0-2011.7$ & 141 & 1.43 & -6.65  \\
            &   &    &   & &   &              & 1600 & $2010.9-2014.3$ & 73 & 1.24 & -6.51  \\
            &   &    &   & &   &              & 2000 & $2009.9-2012.7$ & 27 & 1.14 & -7.42  \\
            &   &    &   & &   & WSRT\dotfill & 323 & $2007.6-2010.2$ & 32 & 0.90 & -5.73  \\
            &   &    &   & &   &              & 367 & $2007.6-2010.2$ & 31 & 1.04 & -9.40  \\
\\
J1751$-$2857  & 2  & -19.67 & 6.32 & -16.35 & 4.08\\
            &   &    &   & &   & JBO\dotfill & 1520 & $2009.3-2014.3$ & 37 & 1.56 & -7.05  \\
            &   &    &   & &   & NRT\dotfill & 1400 & $2006.0-2011.8$ & 75 & 1.57 & -6.79  \\
            &   &    &   & &   &              & 1600 & $2011.2-2014.3$ & 32 & 1.14 & -9.56  \\
\\
J1801$-$1417  & 2  & -17.96 & 6.45 & -10.84 & 2.28\\
            &   &    &   & &   & JBO\dotfill & 1520 & $2009.7-2014.3$ & 55 & 0.96 & -8.48  \\
            &   &    &   & &   & NRT\dotfill & 1400 & $2007.3-2011.8$ & 49 & 1.68 & -7.12  \\
            &   &    &   & &   &              & 1600 & $2009.0-2014.1$ & 22 & 1.47 & -8.21  \\
\\
J1802$-$2124  & 3  & -19.55 & 6.88 & -10.79 & 1.73\\
            &   &    &   & &   & JBO\dotfill & 1520 & $2011.4-2014.5$ & 26 & 1.01 & -9.85  \\
            &   &    &   & &   & NRT\dotfill & 1400 & $2007.2-2011.8$ & 354 & 1.04 & -9.28  \\
            &   &    &   & &   &              & 1600 & $2009.0-2014.2$ & 105 & 1.07 & -6.94  \\
            &   &    &   & &   &              & 2000 & $2008.3-2009.9$ & 37 & 1.12 & -7.02  \\
\\
J1804$-$2717  & 2  & -18.45 & 4.18 & -17.42 & 0.76\\
            &   &    &   & &   & JBO\dotfill & 1520 & $2009.1-2014.5$ & 53 & 1.14 & -6.76  \\
            &   &    &   & &   & NRT\dotfill & 1400 & $2006.1-2011.8$ & 50 & 0.80 & -5.94  \\
            &   &    &   & &   &              & 1600 & $2009.1-2014.2$ & 13 & 1.10 & -9.67  \\
\\
J1843$-$1113  & 4  & -17.38 & 5.43 & -10.94 & 1.45\\
            &   &    &   & &   & JBO\dotfill & 1520 & $2009.6-2014.5$ & 47 & 0.72 & -5.25  \\
            &   &    &   & &   & NRT\dotfill & 1400 & $2008.0-2011.8$ & 63 & 0.86 & -9.34  \\
            &   &    &   & &   &              & 1600 & $2010.8-2014.3$ & 47 & 1.01 & -9.11  \\
            &   &    &   & &   & WSRT\dotfill & 1380 & $2004.4-2010.4$ & 67 & 1.33 & -7.09  \\
\\
J1853$+$1303  & 2  & -15.59 & 5.83 & -18.67 & 1.11\\
            &   &    &   & &   & JBO\dotfill & 1520 & $2009.6-2014.5$ & 34 & 1.03 & -6.18  \\
            &   &    &   & &   & NRT\dotfill & 1400 & $2006.1-2011.8$ & 49 & 0.96 & -7.79  \\
            &   &    &   & &   &              & 1600 & $2009.1-2014.3$ & 18 & 0.79 & -5.82  \\
\\
J1857$+$0943  & 8  & -13.37 & 2.53 & -17.42 & 5.06\\
            &   &    &   & &   & EFF\dotfill & 1360 & $2008.1-2013.4$ & 25 & 0.26 & -5.56  \\
            &   &    &   & &   &              & 1410 & $1997.0-2009.7$ & 106 & 0.80 & -8.75  \\
            &   &    &   & &   &              & 2639 & $2008.2-2013.5$ & 43 & 1.05 & -8.27  \\
            &   &    &   & &   & JBO\dotfill & 1400 & $2010.9-2011.0$ & 7 & 1.14 & -7.37  \\
            &   &    &   & &   &              & 1520 & $2009.6-2012.0$ & 31 & 0.58 & -5.71  \\
            &   &    &   & &   & NRT\dotfill & 1400 & $2005.1-2011.8$ & 102 & 0.77 & -6.08  \\
            &   &    &   & &   &              & 1600 & $2011.0-2014.3$ & 58 & 1.18 & -6.81  \\
            &   &    &   & &   &              & 2000 & $2010.1-2013.5$ & 13 & 1.25 & -7.40  \\
            &   &    &   & &   & WSRT\dotfill & 1380 & $2006.2-2010.5$ & 59 & 1.29 & -6.09  \\
\\
J1909$-$3744  & 2  & -14.18 & 2.17 & -16.84 & 6.70\\
            &   &    &   & &   & NRT\dotfill & 1400 & $2005.0-2011.8$ & 156 & 1.13 & -7.89  \\
            &   &    &   & &   &              & 1600 & $2010.7-2014.4$ & 219 & 0.97 & -7.17  \\
            &   &    &   & &   &              & 2000 & $2005.2-2013.8$ & 50 & 1.15 & -7.39  \\
\\
J1910$+$1256  & 2  & -16.72 & 0.09 & -19.73 & 3.84\\
            &   &    &   & &   & JBO\dotfill & 1520 & $2009.1-2014.5$ & 46 & 0.79 & -8.57  \\
            &   &    &   & &   & NRT\dotfill & 1400 & $2006.0-2011.8$ & 52 & 1.12 & -7.59  \\
            &   &    &   & &   &              & 1600 & $2012.0-2014.2$ & 14 & 0.66 & -7.84  \\
\\
J1911$+$1347  & 2  & -14.84 & 6.85 & -12.89 & 4.27\\
            &   &    &   & &   & JBO\dotfill & 1520 & $2009.3-2014.5$ & 69 & 0.83 & -6.12  \\
            &   &    &   & &   & NRT\dotfill & 1400 & $2007.0-2011.8$ & 44 & 0.86 & -6.38  \\
            &   &    &   & &   &              & 1600 & $2009.1-2014.4$ & 27 & 1.17 & -8.05  \\
\\
J1911$-$1114  & 3  & -18.94 & 3.55 & -14.02 & 1.19\\
            &   &    &   & &   & JBO\dotfill & 1400 & $2009.1-2009.4$ & 5 & 0.90 & -7.45  \\
            &   &    &   & &   &              & 1520 & $2009.6-2015.0$ & 59 & 1.10 & -8.98  \\
            &   &    &   & &   & NRT\dotfill & 1400 & $2006.2-2011.8$ & 52 & 1.23 & -9.21  \\
            &   &    &   & &   &              & 1600 & $2012.0-2014.2$ & 14 & 2.53 & -8.35  \\
\\
J1918$-$0642  & 5  & -14.07 & 4.57 & -18.27 & 3.41\\
            &   &    &   & &   & JBO\dotfill & 1400 & $2009.1-2009.6$ & 12 & 1.21 & -6.44  \\
            &   &    &   & &   &              & 1520 & $2009.6-2014.3$ & 97 & 0.98 & -6.90  \\
            &   &    &   & &   & NRT\dotfill & 1400 & $2006.8-2011.8$ & 57 & 0.92 & -5.95  \\
            &   &    &   & &   &              & 1600 & $2010.9-2014.3$ & 26 & 0.70 & -7.61  \\
            &   &    &   & &   & WSRT\dotfill & 1380 & $2001.5-2010.1$ & 86 & 0.99 & -9.25  \\
\\
J1939$+$2134  & 11  & -14.86 & 6.89 & -11.21 & 2.57\\
            &   &    &   & &   & EFF\dotfill & 1360 & $2009.6-2011.5$ & 32 & 2.05 & -6.55  \\
            &   &    &   & &   &              & 1410 & $1996.8-2009.4$ & 223 & 1.31 & -6.32  \\
            &   &    &   & &   & JBO\dotfill & 1520 & $2009.7-2012.0$ & 54 & 0.70 & -6.63  \\
            &   &    &   & &   & NRT\dotfill & 1400 & $2005.0-2011.8$ & 249 & 2.67 & -8.32  \\
            &   &    &   & &   &              & 1600 & $2005.0-2014.3$ & 202 & 2.13 & -8.65  \\
            &   &    &   & &   &              & 2000 & $2005.0-2013.4$ & 119 & 1.76 & -6.36  \\
            &   &    &   & &   &              & 1400 & $1990.2-1999.8$ & 2058 & 1.16 & -6.56  \\
            &   &    &   & &   & WSRT\dotfill & 1380 & $1999.6-2010.5$ & 148 & 2.63 & -8.67  \\
            &   &    &   & &   &              & 2273 & $2006.9-2010.5$ & 37 & 1.04 & -9.79  \\
            &   &    &   & &   &              & 840 & $2000.3-2007.9$ & 52 & 1.44 & -8.71  \\
\\
J1955$+$2908  & 3  & -17.54 & 4.95 & -16.27 & 1.38\\
            &   &    &   & &   & JBO\dotfill & 1400 & $2009.0-2009.4$ & 10 & 1.11 & -7.53  \\
            &   &    &   & &   &              & 1520 & $2009.6-2014.2$ & 80 & 1.01 & -8.45  \\
            &   &    &   & &   & NRT\dotfill & 1400 & $2006.2-2011.8$ & 47 & 1.14 & -7.23  \\
            &   &    &   & &   &              & 1600 & $2011.2-2014.3$ & 20 & 1.38 & -8.80  \\
\\
J2010$-$1323  & 4  & -17.75 & 2.26 & -11.66 & 3.44\\
            &   &    &   & &   & JBO\dotfill & 1400 & $2009.1-2009.6$ & 13 & 0.84 & -9.10  \\
            &   &    &   & &   &              & 1520 & $2009.6-2014.3$ & 87 & 1.12 & -9.17  \\
            &   &    &   & &   & NRT\dotfill & 1400 & $2007.0-2011.8$ & 177 & 1.23 & -7.66  \\
            &   &    &   & &   &              & 1600 & $2010.8-2014.4$ & 77 & 1.11 & -7.74  \\
            &   &    &   & &   &              & 2000 & $2007.3-2012.7$ & 36 & 1.34 & -7.43  \\
\\
J2019$+$2425  & 2  & -15.49 & 2.07 & -17.80 & 3.18\\
            &   &    &   & &   & JBO\dotfill & 1520 & $2009.6-2014.4$ & 59 & 1.41 & -9.85  \\
            &   &    &   & &   & NRT\dotfill & 1400 & $2005.2-2011.8$ & 44 & 1.28 & -6.02  \\
            &   &    &   & &   &              & 1600 & $2011.9-2014.3$ & 27 & 1.07 & -8.00  \\
\\
J2033$+$1734  & 3  & -19.52 & 0.19 & -12.39 & 2.13\\
            &   &    &   & &   & JBO\dotfill & 1400 & $2009.1-2009.6$ & 14 & 1.38 & -9.98  \\
            &   &    &   & &   &              & 1520 & $2009.6-2014.4$ & 86 & 0.85 & -8.78  \\
            &   &    &   & &   & NRT\dotfill & 1400 & $2006.4-2011.9$ & 58 & 1.19 & -8.97  \\
            &   &    &   & &   &              & 1600 & $2011.8-2014.3$ & 36 & 1.32 & -7.72  \\
\\
J2124$-$3358  & 4  & -16.98 & 6.07 & -14.42 & 1.20\\
            &   &    &   & &   & JBO\dotfill & 1400 & $2009.1-2009.3$ & 7 & 1.25 & -9.81  \\
            &   &    &   & &   &              & 1520 & $2009.6-2014.3$ & 51 & 1.14 & -9.14  \\
            &   &    &   & &   & NRT\dotfill & 1400 & $2005.0-2011.8$ & 339 & 1.30 & -8.47  \\
            &   &    &   & &   &              & 1600 & $2009.0-2014.4$ & 97 & 1.69 & -6.15  \\
            &   &    &   & &   &              & 2000 & $2006.9-2012.9$ & 50 & 1.01 & -5.14  \\
\\
J2145$-$0750  & 11  & -14.29 & 4.83 & -11.79 & 1.33\\
            &   &    &   & &   & EFF\dotfill & 1360 & $2009.6-2013.4$ & 30 & 0.81 & -9.25  \\
            &   &    &   & &   &              & 1410 & $1996.8-2009.6$ & 117 & 0.65 & -5.92  \\
            &   &    &   & &   &              & 2639 & $2006.9-2013.5$ & 51 & 0.81 & -5.62  \\
            &   &    &   & &   & JBO\dotfill & 1400 & $2009.1-2009.4$ & 9 & 1.47 & -7.57  \\
            &   &    &   & &   &              & 1520 & $2009.6-2014.3$ & 82 & 1.05 & -6.06  \\
            &   &    &   & &   & NRT\dotfill & 1400 & $2005.0-2011.8$ & 237 & 1.11 & -6.23  \\
            &   &    &   & &   &              & 1600 & $2010.8-2013.9$ & 125 & 1.26 & -7.60  \\
            &   &    &   & &   &              & 2000 & $2007.3-2013.8$ & 47 & 1.26 & -6.58  \\
            &   &    &   & &   & WSRT\dotfill & 1380 & $2006.0-2010.1$ & 41 & 1.13 & -9.02  \\
            &   &    &   & &   &              & 2273 & $2006.9-2007.3$ & 6 & 2.82 & -8.78  \\
            &   &    &   & &   &              & 323 & $2007.3-2010.2$ & 30 & 1.68 & -6.10  \\
            &   &    &   & &   &              & 367 & $2007.3-2010.2$ & 25 & 1.34 & -5.32  \\
\\
J2229$+$2643  & 5  & -15.69 & 4.55 & -17.19 & 1.80\\
            &   &    &   & &   & EFF\dotfill & 1360 & $2010.7-2013.5$ & 26 & 1.38 & -5.49  \\
            &   &    &   & &   &              & 2639 & $2007.3-2013.2$ & 23 & 0.54 & -5.34  \\
            &   &    &   & &   & JBO\dotfill & 1400 & $2009.0-2009.4$ & 11 & 1.91 & -5.76  \\
            &   &    &   & &   &              & 1520 & $2009.6-2014.4$ & 71 & 1.04 & -6.23  \\
            &   &    &   & &   & NRT\dotfill & 1400 & $2006.2-2011.8$ & 150 & 1.79 & -6.67  \\
            &   &    &   & &   &              & 1600 & $2010.9-2014.4$ & 35 & 1.36 & -7.95  \\
\\
J2317$+$1439  & 7  & -15.05 & 0.88 & -15.56 & 1.06\\
            &   &    &   & &   & EFF\dotfill & 1360 & $2009.6-2013.5$ & 32 & 1.78 & -8.19  \\
            &   &    &   & &   &              & 1410 & $1997.0-2009.6$ & 15 & 1.31 & -8.40  \\
            &   &    &   & &   &              & 2639 & $2007.6-2013.2$ & 41 & 1.13 & -6.63  \\
            &   &    &   & &   & JBO\dotfill & 1400 & $2009.0-2009.4$ & 9 & 1.95 & -8.22  \\
            &   &    &   & &   &              & 1520 & $2009.6-2014.3$ & 79 & 1.03 & -8.27  \\
            &   &    &   & &   & NRT\dotfill & 1400 & $2005.0-2011.9$ & 238 & 1.41 & -7.36  \\
            &   &    &   & &   &              & 1600 & $2009.0-2014.4$ & 93 & 1.48 & -6.84  \\
            &   &    &   & &   & WSRT\dotfill & 1380 & $2006.3-2010.5$ & 48 & 1.43 & -9.11  \\
\\
J2322$+$2057  & 3  & -19.91 & 6.37 & -13.89 & 3.70\\
            &   &    &   & &   & JBO\dotfill & 1400 & $2009.1-2009.4$ & 8 & 0.80 & -6.09  \\
            &   &    &   & &   &              & 1520 & $2009.6-2014.4$ & 113 & 1.40 & -6.01  \\
            &   &    &   & &   & NRT\dotfill & 1400 & $2006.5-2011.8$ & 59 & 1.58 & -9.55  \\
            &   &    &   & &   &              & 1600 & $2009.0-2014.3$ & 49 & 1.21 & -7.92  \\

\end{longtable}

\bsp	
\label{lastpage}
\end{document}